%% file: simple.tex
\documentclass[a4paper,11pt]{article}

\usepackage{amsmath, amssymb, amsthm}
\usepackage{amsopn}
\usepackage{graphicx}
\usepackage[utf8]{inputenc}
\usepackage{nicefrac}
\usepackage{enumerate}
\usepackage[usenames,dvipsnames]{xcolor}

\iftrue
\usepackage[colorlinks=true,linkcolor=Blue,citecolor=Green]{hyperref}
\hypersetup{
  pdfauthor={Heuna Kim, Günter Rote},
  pdftitle={Congruence Testing for Point Sets in 4 Dimensions},
  pdfsubject={Congruence Testing},
}
\usepackage{bookmark}
\fi

\usepackage{longtable}
\usepackage{emptypage}

\usepackage{paralist}

\usepackage{fix-cm}

\usepackage[margin=2.5cm,top=2.5cm]{geometry}
\usepackage[capitalise%,nameinlink,noabbrev
]{cleveref}
\crefname{enumi}{}{}

\graphicspath{{./Figures/}}

\newtheorem{theorem}{Theorem}%[chapter]
\newtheorem{lemma}[theorem]{Lemma}
\newtheorem{proposition}[theorem]{Proposition}

\newtheorem{corollary}[theorem]{Corollary}
\newtheorem{conjecture}[theorem]{Conjecture}

\theoremstyle{definition}
\newtheorem{definition}[theorem]{Definition}

\let\phi=\varphi

\newcommand{\reals}{\mathbb{R}}

\newcommand{\ta}{\tilde{A}}
\newcommand{\tb}{\tilde{B}}

\newcommand{\plane}{\mathcal{P}}

\newcommand{\sa}{\mathbb{S}^3}
\newcommand{\sphere}{\mathbb{S}}
\newcommand{\mt}{\mathbb{T}^2} %\mathcal{T}}
\newcommand{\ifam}{\mathcal{F}}
\newcommand{\emphi}[1]{\smallskip\noindent \textbf{#1}}

\newcommand{\symt}{\Theta}
\newcommand{\spaces}{\mathcal{S}}
\DeclareMathOperator{\sym}{Sym}

\newcommand{\eps}{\varepsilon}
\newcommand{\psfig}{predecessor-successor figure}

\newcommand{\PACK}{}
\newcommand{\DELTA}{}
\def\PACK-F.{829}
\def\DELTA0{0.0005}

\title{Congruence Testing of Point Sets in 4 Dimensions}
\author{Heuna Kim and G\"unter Rote \\
Institut f\"ur Informatik, Freie Universit\"at Berlin}
\date{March 23, 2016}

\begin{document}

%\listoftodos
%\vspace{1em}
%\input Abstract}
\maketitle

\begin{abstract}
  %We show that
\noindent
  Congruence between two sets of $n$ points in 4 dimension can be checked
  in $O(n\log n)$ time.
\end{abstract}
	
\tableofcontents

%\newpage

\input Problem_simple

\input Overview

\input Previous

%% PART 1. GEOMETRIC TOOLS
\input Rotation

\input AngleIsoclinic

\input Pluecker

\input Hopf

\input ClosestPair

\input Coxeter

\input Packing
\input Canonical

\input Congruence_Types

%\section{Building on Previous Ideas}\label{previous}
\input Dimension_Reduction

%\part[The New Algorithm in 4-space]{The New Algorithm in 4-space}\label{Part:algo}

\input Pruning

\input Overview_Algorithm

%\chapter{Five Crucial Modules}\label{sec:modules}
\input Iter

\input Orbit

\input Mirror

\input Algorithm_K

\input Plane
\input Cycle

%\input ConcRemark

%\bookmarksetup{startatroot}% this is it
%\addtocontents{toc}{\bigskip}% perhaps as well

%\part[More Implications]{More Implications}\label{Part:more}

%\input Highdim}

\input Conclusion

\addcontentsline{toc}{section}{Bibliography}
%\phantomsection
\bibliographystyle{abbrv}
\bibliography{cong}

\end{document}

% Clifford for elliptic space
% Plücker why is small dim. better

%% file: Problem_simple.tex
\section{%Introduction and 
Problem Statement}

% In the following sections, more precise definitions of the problem
% and the computational model are described.
% %and
% %previous work is summarized.
% %together with understanding the meaning of being efficient.
% %Before explaining what are new contributions,
% %ideas or simple trials for the problem 
% %that are probably not going to work are described.
% This chapter wraps up with stating contributions of this
% thesis and organization of the remaining chapters.

\paragraph{Problem Statement.}
\label{sec:probstat}
We are given two $n$-point sets $A,B\subset \mathbb{R}^d$.
Our aim is to test whether they are \emph{congruent}, 
i.\,e., whether there exists an orthogonal matrix and
a translation vector $t$ such that $B=RA+t$.

%\paragraph{Motivation.}
As well as being intuitive, congruence is a fundamental notion in
geometry.
The well-known congruence criteria for triangles go back to
300 BC, written in Euclid's Elements.
%The method of testing congruence of triangles is well-known:
%SAS, SSS, ASA, AAS, and RHS where S represents a side, A for an
%angle, R for an right angle, and H for Hypothenuse. 
%Although this method was known for dozens of centuries, 
Algorithms for testing congruence for point sets have been developed
since the  1970's \cite{Man,Ata,Hig,Sug,Atk,Alt,BK,Aku}.

\paragraph{Problem Set-Up.}

We may eliminate the translation vector $t$
by translating $A$ and
$B$ so that their centroids are moved to the origin. 

Also, we only consider orthogonal matrices $R$ of determinant $+1$ (``direct
congruences'', rotations). If we want to allow orthogonal
matrices of determinant $-1$ as well (``mirror congruence'', rotations and a reflection),
we can just repeat the algorithm with a reflected copy of $B$.

We solve this problem in $O(n\log n)$ time in $d=4$ dimensions.
This paper accompanies our conference article~\cite{RH},
and it contains all details that were omitted there for lack of space.

\section{The Computational Model}
\label{sec:model}

Because the output of problem is sensitive to numerical errors,
it is natural to consider an approximate version of the congruence
testing problem. However, 
 congruence testing with error tolerances 
is known to be NP-hard~\cite{die,iwa}.
We therefore restrict our concern to the exact case.
(Error
tolerances are discussed again in
Chapter~\ref{chap:conclusion}.)

We assume that we can perform exact arithmetic with real numbers.
Thus, we use the \emph{Real Random-Access Machine (Real-RAM) model}
(see for example \cite[Chapter 1.4]{PS}).  This assumption is common in
Computational Geometry.
In this model, we can freely use square
roots, sines and cosines,
and basic operations from linear algebra such as eigenvalues of
$2\times2$-matrices (or matrices of any constant size), 
and we assume that we obtain exact results in constant time.
Sines and cosines and their inverses can
be eliminated in favor of purely algebraic operations.

If we would restrict our model to rational inputs, this would
severely restrict the problem. For example, in the plane, the only
rotational symmetries that a point set with rational coordinates 
can have are multiples of $90^\circ$.
A three-fold symmetry is possible in 3 dimensions, but
a fivefold symmetry is impossible in any
dimension.
An input with limited symmetry, however, would not be interesting
enough to consider.
%Why symmetry makes the problem more interesting or difficult?
The previous algorithms
 for the exact congruence testing problem~\cite{Man,Ata,Hig,Sug,Atk,Alt}
 were developed around 70's--80's under the same assumption.
The only papers that explicitly mentioned the assumption
 are~\cite{Atk,Alt}.

%% file: Overview.tex
\section{Overview}

%is shown to be $\Omega( n \log n)$~\cite{Atk} in 1987
%and conjectured to be $O(n \log n)$ in any fixed dimension $d$.

In \cref{sec:prevA}, we survey the previous literature on congruence
testing, since we build on ideas from these algorithms.

Our new algorithm uses tools that exploit the geometric structure of
4-space.
The first part of the body of the paper (Sections~\ref{sec:rot}--\ref{sec:congtype})
is develops these tools.
We elaborate
 the necessary background to understand how these tools work.

As a result of symmetries, 
we will also encounter beautiful mathematical structures.
In particular, 
the structure of Hopf bundles organizes the special case of 
isoclinic planes in a pleasant way. 
We describe
the interpretation of the Hopf fibration in a very geometric and
elementary way by using four-dimensional rotations. 
This interpretation clarifies
the necessary and sufficient condition for a collection of great
circles to be in a Hopf bundle.

This part includes preliminaries about four-dimensional rotations,
 angles between a pair of planes,
Pl\"ucker coordinates, four-dimensional Coxeter groups, kissing
numbers and representing a neighborhood of a vertex or an edge in 4-space.

The second part
 (Sections~\ref{sec:1+d_red}%{sec:overview}
--\ref{sec:torus})
is devoted to
the new algorithm.
The techniques of closest pairs,
pruning, and dimension reduction
have been used for congruence testing before. 
We extend these ideas and apply them in a novel way. 
% Our algorithm uses
% closest pairs not just as a convenient tool for matching, but
% as a devise for scrutinizing an actual structure of a closest-pair graph
% (i.\,e., a graph whose vertices are points and edges indicate the
% minimum distance).
Our algorithm extracts helices around great circles on 3-sphere 
by taking advantage of the structure of the closest-pair graph.

% The auxiliary algorithm in Section~\ref{sec:plane} of pruning these
% extracted great circles is of independent interest; this algorithm
% includes the step that, given a set of great circles, selects a subset
% of great circles in the same Hopf bundle.

On top of the previous dimension reduction principle,
a new efficient dimension reduction technique, called 2+2
dimension reduction (Section~\ref{sec:torus}) 
is developed for the case 
that a given point set has rotational symmetries that
make two orthogonal planes invariant. 
This technique is simple to implement and
resolves the tricky case in Brass and Knauer's algorithm~\cite{BK} as
well.
%  We would expect that 2+2 dimension reduction can be
% extended to higher dimensions %(see Section~\ref{sec:exthd}) 
% and applied to other problems that
% require dimension reduction.

\begin{figure}[hbt]
	\centering
\includegraphics[width=0.7\textwidth]{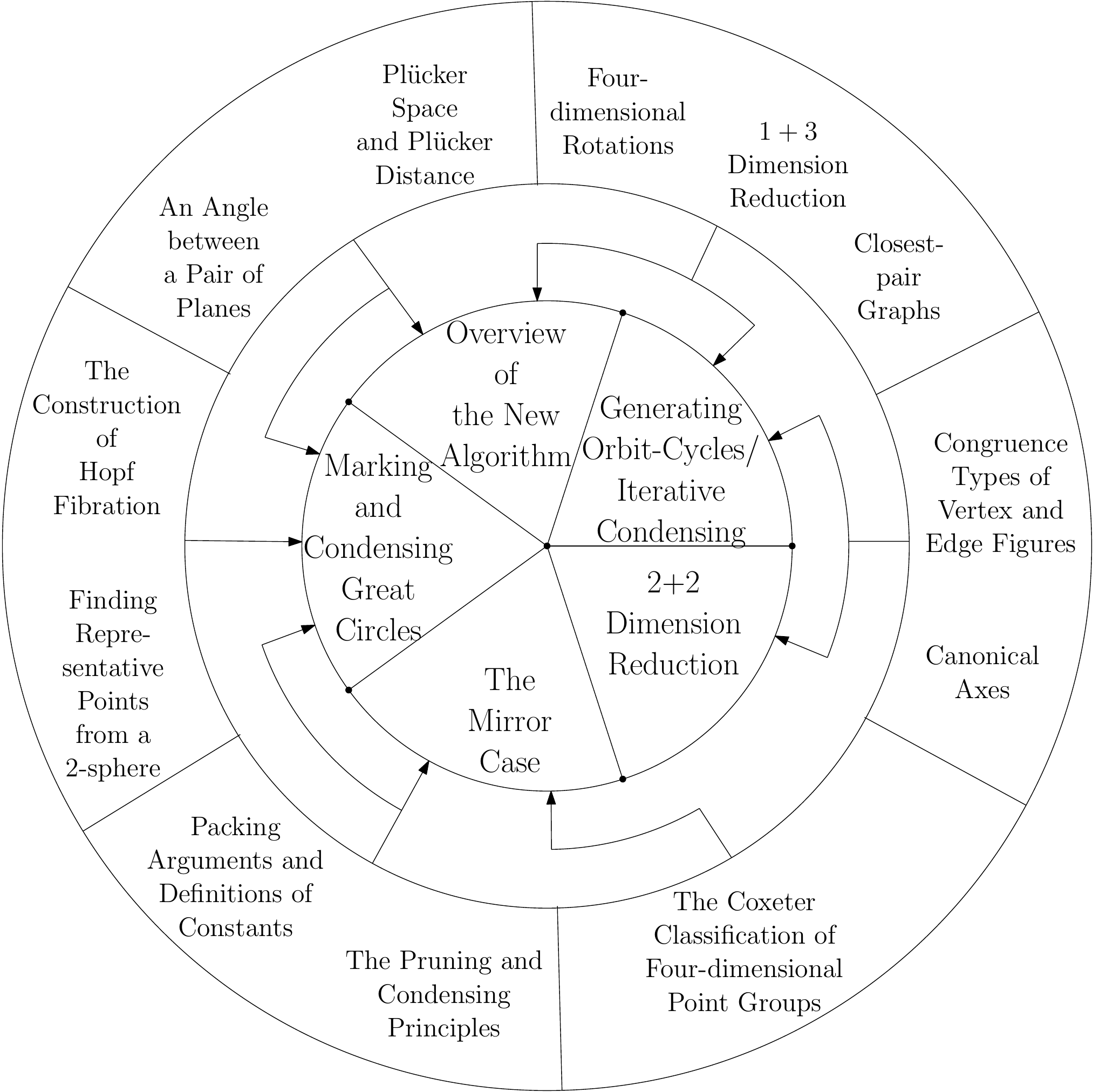}
\caption{Dependence between the first part
  (Sections~\ref{sec:rot}--\ref{sec:congtype}, in the outer ring)
 and the second
part (Sections~\ref{sec:1+d_red}--\ref{sec:torus}, in the inner disk). }
	\label{fig:section}
\end{figure}

Figure~\ref{fig:section} shows the relations between 
the sections in the two parts.

%% file: Previous.tex
\section{Previous Algorithms}
\label{sec:prevA}

It has been shown that the complexity of the problem is $\Omega(n \log n)$
independently by Atallah (1985)~\cite{Ata},
Highnam (1986)~\cite{Hig}, and Atkinson (1987)~\cite{Atk}.
This bound holds
even in one-dimensional space. It can be easily seen
by reducing the problem 
to the set equality problem in the real line.

It has been believed
that
the congruence testing problem should be fixed-parameter 
tractable with the dimension parameter $d$, that is, solvable in
time $O(f(d)n^C)$ for an arbitrary function $f(d)$ and a constant $C$
that is independent of $d$.
More precisely, is is believed
 that $O(n \log n)$ algorithms should 
exist for any fixed dimension.
 
The previous best deterministic algorithm by Brass and Knauer (2000)~\cite{BK} 
 for $d$-space for $d>3$ 
took time $O(n^{\lceil d/3 \rceil} \log
n)$. The previous best randomized algorithm by Akutsu
(1998)~\cite{Aku} takes time $O(n^{\lfloor d/2\rfloor /2}
\log n)$ for $d \geq 6$ and time $O(n^{3/2} \log n)$ for $d=4,5$.  
Therefore, the previous best result in 4-space was time $O(n^2 \log
n)$ deterministically and time $O(n\sqrt{n} \log n)$ with randomization.

We will survey some previous congruence testing algorithms, because we
will use them and build on their ideas.
A thorough survey of previous algorithms is given in Kim's thesis~\cite{my-thesis}.

\subsection{Congruence Testing in the Plane}
\label{sec:2d}

The first efficient algorithm for congruence testing
is due to Manacher (1976)~\cite{Man}.
After translating point sets so that the centroid lies on the origin $O$,
Manacher's algorithm sorts all input points by spherical coordinates
$(r,\theta)$, first by angle $\theta$, and secondly by distance $r$ to
the origin $O$ in both increasing order.
Let $\{p_i = (r_i, \theta_i)|i=1,\ldots, n\}$ be the given point set indexed by the
sorted order. Finally, the algorithm generates a cyclic sequence 
in which $r_i$ and $\angle p_iOp_{i +1} = \theta_{i+1}-\theta_i$
alternate for $1\leq i \leq n$:
$\langle r_1, \theta_2-\theta_1, r_2, \ldots,\theta_n -\theta_{n-1},
r_n \rangle.$ Point sets are congruent if and only if the
corresponding cyclic sequences are the same up to cyclic shifts.
His paper~\cite{Man} mainly discusses geometric applications of the algorithm by Knuth, Morris and Pratt
~\cite{KMP} that determines whether a string $y$ contains a string $x$ as a substring 
in linear time $O(|x|+|y|)$. It can be determined if two sequences are the same up to cyclic shifts
by duplicating one sequence and checking if the original sequence is a
subsequence of the duplicated sequence by 
using the algorithm by Knuth, Morris and Pratt in linear time. 
Manacher's algorithm takes time $O(n \log n)$ due to sorting.

\subsection{Congruence Testing In Three-Dimensional Space}
\label{sec:3d}

Two algorithms rely on graph isomorphism for planar graphs:
an algorithm of
Sugihara~\cite{Sug} for congruence testing between (not necessarily
convex) polytopes,
and an algorithm of
Alt, Mehlhorn, Wagener, and Welzl~\cite{Alt} for point sets.
They are based on the fact that the graph of a convex polytope is
planar.
Thus, with some preprocessing, they reduce congruence testing to
isomorphism testing for labeled planar graphs, which can be solved in
 time $O(n \log n)$ by applying the partitioning algorithm of Hopcroft and
Tarjan~\cite{HT}, see also~\cite[Section 4.13]{AHU}.
The total time for these algorithms is $O(n \log n)$.
We will not use this approach.

\paragraph{Atkinson (1987)~\cite{Atk}.}
Atkinson's algorithm determines congruence for two point sets in 3-space.
This algorithm begins by reducing input point sets as small as possible 
while preserving symmetries and then considers 
a bounded number of all the possible matches of the original point sets, 
obtained by the reduced sets.
The principle of this reduction procedure is different from \emph{pruning} mentioned
before (Section~\ref{sec:model} and Section~\ref{sec:pruning}) but is more
related to a \emph{canonical set procedure} (Sections~\ref{sec:cano}
and~\ref{sec:cpp}). The
first part of this algorithm can be considered a canonical set
procedure of for a 2-sphere with rotational symmetries.
This reduction procedure preserves symmetries, whereas pruning does not.
For reduction, Atkinson's algorithm constructs a closest-pair graph,
i.\,e., a graph whose vertices are points and edges are pairs of
vertices that achieve the minimum distance (see
Section~\ref{sec:clograph}).
Then, the algorithm prunes points by the congruence type of
a neighborhood of a vertex $v$, or by a \emph{vertex figure} (see
Section~\ref{sec:congtype}). 
Then, either there is a component of the centroid different from the
original centroid or the degrees of vertices are bounded 
since $v$ has at most five closest vertices by the kissing number
(see Section~\ref{sec:packing}).
The sets can be further reduced 
by traversing each component if the degree is two and by comparing faces if
the degree is three or four or five.
At the end, there are only two cases: 
\begin{enumerate}
	\item[(i)] the reduced set is a singleton set or consists of
		two antipodal points.
	\item[(ii)] the reduced set has a bounded cardinality more
		than two.
\end{enumerate}
For (ii), the algorithm finds all the rotations from a pair of
non-antipodal points in one reduced set to any pair of non-antipodal
points at the same angle in another reduced set.
Then, it checks if at least one of such rotations   
transforms one input point set to another input point set.  
For (i), the rotational axis can be identified so 
Manacher's algorithm~\cite{Man} (see the first paragraph of
Section~\ref{sec:2d}) can be applied after each point is represented by
a cylindrical coordinate regarding the common rotational axis as the
$z$-axis.

The variation of the first part of Atkinson's algorithm
will be used in the new algorithm and
discussed again in
Lemma~\ref{lem:rep} in Section~\ref{sec:prusphere}.

\subsection{Congruence Testing in Four and Higher Dimensions}
\label{sec:hd}
\paragraph{Alt, Mehlhorn, Wagener, and Welzl  (1988)~\cite{Alt}.}
This algorithm reduces a $d$-dimensional congruence testing problem to
$n$ subproblems in $(d-1)$-space.
As in the three-dimensional case of their algorithm, 
it projects all the points radially to the $(d-1)$-dimensional 
unit sphere $S$ centered at
the centroid and labels them with distances to the centroid.
Fix one point $a$ in the resulting set from $A$.
The next step is another projection to a $(d-2)$-dimensional unit
sphere $S'$ which is the intersection of $S$ and the hyperplane that
orthogonally bisects the line segment from $a$ to the centroid;
project each point $x$ in $A$ except $a$ onto $S'$ along the arc from
$x$ to $a$ on the surface of $S$. Let us denote the resulting set as
$A'$. We can obtain sets $B_1',\ldots,B_n'$ in the same manner 
by fixing all points of $B$. Two given point sets $A$ and $B$ are
congruent if and only if $A'$ is congruent to at least one of $B_i'$
for $i=1,\ldots, n$ with taking labels into account. 
As a result, we obtained $n$ subproblems
to determine congruence between $A'$ and $B'_i$ for $i=1,\ldots,n$.
Thus, the algorithm by Alt, Mehlhorn, Wagener, and Welzl takes time
$O(n^{d-2} \log n)$ in $d$-space.

The following algorithms achieve the best known deterministic and
randomized running times in high dimension:
\paragraph{Akutsu (1998)~\cite{Aku}:}
Akutsu developed
 a \emph{Monte Carlo algorithm} for
congruence testing problem in $d$-space.
The algorithm is
randomized and it is
 based on the birthday paradox. For two congruent
input sets, the algorithm is guaranteed to find the congruence only
with high probability.
Akutsu stated the running time as $O(n^{(d-1)/2} \log n)$,
and he mentioned that it can be reduced
to $O(n^{d/4+O(1)})$ by an unpublished idea of J.~Matou\v{s}ek.
Matou\v{s}ek's idea is to match 
closest \emph{pairs}, i.\,e., pairs of points that attain the minimum
distance (see Section~\ref{sec:clograph}) instead of input
\emph{points}
Since a closest pair together with
the centroid spans a 2-plane,
This idea allows to reduce the dimension by two steps at a time.
 and the orthogonal space of a 2-plane is
$(d-2)$-dimensional if the ambient space is $d$-dimensional.
This idea can be used for the 
algorithm by Alt, Mehlhorn, Wagener, and Welzl as well,
and it improves the time to
$O(n^{\lfloor \frac{d}{2} \rfloor} \log n)$
%The precise complexity of Akutsu's algorithm together with
%Matou\v{s}ek's idea is $O(n^{\lfloor d/2\rfloor /2}
%\log n)$ for $d \geq 6$ and $O(n^{3/2} \log n)$ for $d=4,5$.
%
% Akutsu's algorithm also reduces the original problem to multiple
% lower-dimensional subproblem. With randomization, however, his algorithm chooses
% $O(\sqrt{n})$ points randomly and projects original input points to a
% hyperplane $H$ by fixing one point among randomly chosen points as an
% orthogonal vector to $H$.  
% At the end, this generates two
% collections of projected sets in 3-space
% such that (a) if there is a congruent pair from each collection, the two given input
% sets need to be congruent and (b) if not, the probability that they are
% congruent is very low due to the birthday paradox.
% Instead of comparing every
% pair from each collection, by using the canonical form
% and lexicographical sorting, whether there exists a pair of two
% congruent sets can be determined efficiently in 
% time $O(n^{(d-1)/2} \log n)$ without Matou\v{s}ek's idea.
% The accurate analysis with Matou\v{s}ek's idea was not provided
% but still the algorithm has the bounded error probability 
% after modification since the
% analysis for the error probability does not depend on the dimension
% reduction but uses only the birthday paradox. Then, 
% With Matou\v{s}ek's idea,
Akutsu's algorithm takes time $O(n^{\lfloor d/2\rfloor /2}
\log n)$ for $d \geq 6$ and $O(n^{3/2} \log n)$ for $d=4,5$.
(The accurate analysis was not provided in~\cite{Aku}.)

%in $O(mn \log(mn))$ time where $m$ is
%the size of the collection of projected sets. Since $O(m\sqrt{n})$
%points are projected, it takes time $O(mn^{3/2} log (mn^{3/2}))$.
%In $i$th depth $m = O(n^{(i-1)/2})$ and without Matou\v{s}ek's idea,
%the total depth is $d-2$ and the projection step is not excuted in
%the last step so the total time complexity is $O(n^{(d-1)/2} \log n)$. 

% The projection step of Akutsu's algorithm is also different from that
% of
% Alt, Mehlhorn, Wagener and Welzl. Instead of projecting points
% radially, Akutsu's algorithm projects all points to the hyperplane
% perpendicular to the segment $\overline{cp}$ from the centroid to $p$
% where $p$ is one of the points chosen randomly before.

% \begin{wrapfigure}{r}{0.28\textwidth}
% 	\centering
% 	\includegraphics[width=0.26\textwidth]{BK}
% 	\caption{If all the closest-pairs are in orthogonal 2-planes,
% 	any triple can be mapped to other $\Omega(n^2)$ triples by rotations.}
% 	\label{fig:BK}
% \end{wrapfigure}
% During the projection step, labeling is done simultaneously  
% in a way that two points have the same label if there is an
% orientation-preserving isometry between these two points that
% fixes the
% segment $\overline{cp}$.
% A variant of this projection method is used in the new algorithm
% and discussed in Section~\ref{sec:1+d_red}.

\paragraph{Brass and Knauer (2003)~\cite{BK}.}
Brass and Knauer's algorithm extends Matou\v{s}ek's idea by fixing
a \emph{triple} of points that contains two closest pairs instead of fixing
a closest pair.
This allows to reduce the dimension in steps of 3
% A triple of points together with the centroid can
%determine a 3-dimensional subspace and its orthogonal space is $(d-3)$-dimensional
unless the triple and the centroid are coplanar.
% By projecting points to a 
%$(d-3)$-dimensional orthogonal subspace, 
The algorithm achieves
time $O(n^{\lceil d/3 \rceil} \log n)$.
The difficulty of the algorithm is the occurrence of
tricky cases, such as a union of two point sets in orthogonal planes.
These cases require special handling.
% , but the actual algorithm
% requires more consideration due to inner symmetries of triples. 

% There are examples of point sets $A \subset \mathbb{R}^4$ where a
% triple of points can be mapped to other $\Omega(n^2)$
% triples in $A$ by isometries (see Figure~\ref{fig:BK}).
% This tricky case, that is, when fixing a triple can have too many
% alternative reductions, can happen only when the point sets lie in orthogonal
% 2-planes and all the triples are coplanar. 
% To deal with this case, the algorithm constructs connected components. Each component starts
% with an antipodal pair of points and components of
% the smallest distance merge if their distance is smaller than
% $\sqrt{2}$. This merging process repeats 
% until either a triple of non-coplanar points in a component is found, or
% the smallest distance between components turns out to be $\sqrt{2}$.
% If such a triple is found, the dimension reduction can be applied.
% Otherwise, the given point set is in orthogonal
% 2-planes. This is because the smallest
% distance between two antipodal pairs in the unit sphere
% can reach the maximum ($\sqrt{2}$) if and only if they are orthogonal 
% pairs. In this case,
% for each component pair, they apply congruence testing algorithm for
% the plane and check if there exists a matching of components such that each
% matched pair is congruent.

%% file: Rotation.tex
\section{Four-Dimensional Rotations}
\label{sec:rot}

A 4-dimensional rotations can be encoded
by a $4 \times 4$ rotation matrix $R$, that it,
a $4\times 4$ orthogonal matrix with determinant $+1$.
It has four eigenvalues of absolute value 1,
whose product is $+1$.
They come in conjugate complex pairs,
 $e^{\pm i\phi}$ and
 $e^{\pm i\psi}$.
The special case of real eigenvalues $1$ or $-1$ is included.
Here, $\phi$ and $\psi$ in the eigenvalues correspond to $\phi$ and
$\psi$ in the angular displacements.

Let us assume first that the four eigenvalues are distinct. This case
is called a non-isoclinic rotation.
The eigenvectors for
 each conjugate complex pair, which are conjugate complex,
span a real 2-plane. We choose an orthonormal basis $v_1,v_2$
for the plane corresponding to
 $e^{\pm i\phi}$,
and an orthonormal basis $v_3,v_4$
for the plane corresponding to
 $e^{\pm i\psi}$.
In the resulting orthonormal basis $v_1,v_2,v_3,v_4$, the matrix $R$
looks as follows:
\begin{equation}
\label{eq:rot_mat}
  R = R_{\phi,\psi} =
  \begin{pmatrix}
    \cos\phi & -\sin\phi  &0&0\\
    \sin\phi & \cos\phi   &0&0\\
0&0&    \cos\psi & -\sin\psi\\
0&0&    \sin\psi & \cos\psi\\
  \end{pmatrix}
\end{equation}
The pair $\{P,Q\}$ of orthogonal planes
$P=\langle v_1,v_2\rangle$
and
$Q=\langle v_3,v_4\rangle$
is uniquely determined by $R$, but the basis
 $v_1,v_2$ and $v_3,v_4$ for each plane is not unique.
Here, $P$ and $Q$ are the only pair of planes that are
invariant under the rotation.
We can require that the
 basis $v_1,v_2,v_3,v_4$ is positively oriented by flipping $v_3$ and
 $v_4$ and changing from
$R_{\phi,\psi}$
to $R_{\phi,-\psi}$ if necessary (or by flipping $v_1$ and $v_2$ and changing to
$R_{-\phi,\psi}$, if we prefer).
The angle pair $\{\phi,\psi\}$ in the range $-\pi< \phi,\psi < \pi$
is then unique up to a simultaneous negation to
$\{-\phi,-\psi\}$.

If the four eigenvalues are not distinct, they come
in equal pairs $e^{i\phi},e^{i\phi},e^{-i\phi},e^{-i\phi}$.
Such rotations are called \emph{isoclinic rotations}.
We exclude the special cases $1,1,1,1$ and $-1,-1,-1,-1$
(the identity and the inversion)
and assume $0<\phi<\pi$.
We can still find a basis $v_1,v_2,v_3,v_4$
for which the matrix has the form $R=R_{\phi,\phi}$, but the
decomposition into two orthogonal planes is not unique.
There are infinitely many pairs of
orthogonal planes
that are invariant under the rotation.
If we insist that the basis is positively oriented,
the rotation can either be written as
$R_{\phi,\phi}$
or
$R_{\phi,-\phi}$.
A rotation of the first class is called a \emph{right
  rotation},
 and the second class a \emph{left rotation}.

%Every rotation can be written as a product of a left rotation and a
%right rotation. This decomposition is unique up to multiplying each
%matrix by $-1$.

%Left rotations and right rotations commute. 

\paragraph{Axis Planes.} 
For a rotation $R$ that is not isoclinic,
there is a pair of orthogonal planes $P= \langle v_1,v_2 \rangle, 
Q= \langle v_3,v_4 \rangle$ 
that is invariant under the
rotation $R$. We call these two planes $P,Q$ axis planes of 
the rotation $R$.
Given a rotation matrix $R$, we can find axis planes $P$ and $Q$ 
by computing $v_1,v_2,v_3,v_4$ as follows.
Let $u_\phi,u_{-\phi},v_\psi,v_{-\psi}$ be eigenvectors of $R$ such
that $u_\theta$ corresponds to the eigenvalue $e^{i\theta}$ where
$\theta \in\{\phi,-\phi,\psi, -\psi \}$. Then, $u_\phi = v_1 + iv_2,
u_{-\phi} = v_1 - iv_2$ and
$u_{\psi} = v_3 +i v_4, u_{-\psi} = v_3 -i v_4$ are conjugate pairs. 
Then, 
$$ v_1 = \frac{u_\phi + u_{-\phi}}{2}, \quad v_2 = \frac{u_\phi -
u_{-\phi}}{2i},$$
$$ v_3 = \frac{u_\psi + u_{-\psi}}{2}, \quad v_4 = \frac{u_\psi -
u_{-\psi}}{2i}.$$

%% file: AngleIsoclinic.tex
\section{Angles between two Planes in 4-Space}
\label{sec:angle}

In classical two- or three-dimensional geometry, an angle (between two
lines, or between two planes in space, or between a line and a plane)
fixes the relative position between the involved object up to
congruence.
In four (and higher) dimensions, we need \emph{two} angles to
fix the relative position between two planes.
(More generally, $k$ angles are defined between $k$-dimensional
subspaces.)

\paragraph{Geometric Definition Using Half Rays.}

\begin{figure}%{r}{0.35\textwidth}
\centering
\includegraphics[width=0.3\textwidth]{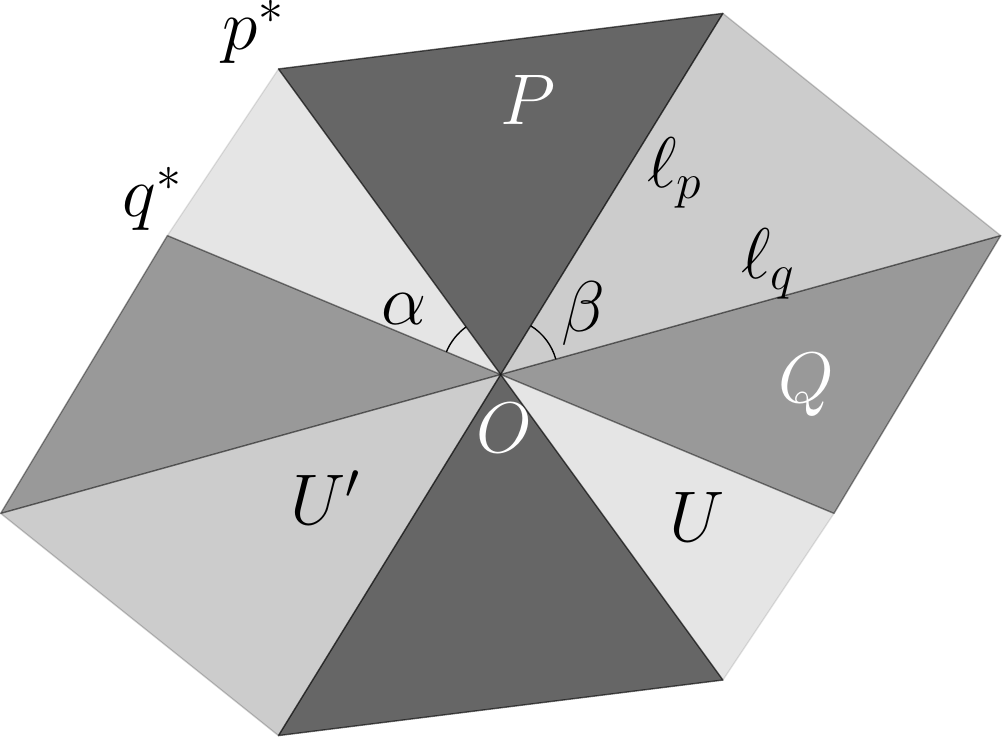}
\caption{Angles between two planes $P$ and $Q$ in 4-space.}
\label{fig:angleplane}
\end{figure}

We first define an angle between two planes in 4-space geometrically
and then show the computation method.
We assume that the planes go through the origin, thus forming
two-dimensional linear subspaces.
The angle between two planes $P,Q$ is defined as a pair
$\alpha,\beta$ of two real numbers in the range $[0,\pi/2]$:
We define $\alpha$ as
the minimum angle between any two half lines $Op$ in $P$ and
$Oq$ in $Q$. Such an $\alpha$
exists~\cite{Mann}. 
We denote such half lines
that make the minimum angle $\alpha$ as $Op^*$ and $Oq^*$. 
Let $U$ be the plane that contains $Op^*$ and $Oq^*$. 
The plane $U$ is perpendicular to the planes $P$
and $Q$, since otherwise $Op^*$ and the projection of $Op^*$ to $Q$
would make an angle smaller than $\alpha$.
Let $U'$ be the orthogonal plane to $U$. Then $U'$ is also
perpendicular to $P$ and $Q$. Let $\ell_p$ be the intersection of $P$
and $U'$. Similarly, let $\ell_q$ be the intersection of $Q$ and $U'$. 
We define
the acute angle made by the lines $\ell_p$ and $\ell_Q$ as $\beta$.
See Figure~\ref{fig:angleplane}.

For later reference we mention the following proposition:
\begin{proposition}
\label{prop-perp}
  There is an orthonormal basis $u_1,u_2 $ of $P$ and
 an orthonormal basis $v_1,v_2 $ of $Q$ such that
 \begin{enumerate}
 \item 
the angle between $u_1,v_1$ is $\alpha$,
 \item 
the angle between $u_2,v_2$ is $\beta$,
\item 
If $\alpha>0$,
the plane spanned by $u_1,v_1$ is perpendicular both to $P$ and to $Q$,
\item 
If $\beta>0$,
the plane spanned by $u_2,v_2$ is perpendicular both to $P$ and to $Q$,
 \end{enumerate}

\end{proposition}

%\label{sec:iso}
\paragraph{Isoclinic Planes.}
The pair of half lines $Op^*$ and $Oq^*$ that defines 
the angle $\alpha$ may not be unique but an infinite number of
such pairs can exist. Accordingly,
there are infinitely many perpendicular
planes $U$ to $P$ and $Q$. In this case, 
an infinite number of planes $U$ cut out $P$ and $Q$ by making all the equal
angles~\cite{Mann}. Thus, $\alpha =\beta$.  We say that a pair of planes
of angle $\alpha,\alpha$ where $0 \leq \alpha \leq \pi/2$ is \emph{isoclinic}.

\paragraph{Computation.}
This can be formulated as follows, see~\cite{BG}.
% Let $L_P$ and $L_Q$ be the vector spaces that span $P$ and $Q$
% respectively. 
%Then,
\begin{align*}
	\cos \alpha = & \max_{\substack{ u_1 \in P \\ \arrowvert u_1
\arrowvert =1}} \phantom{|} 
\max_{\substack{ v_1 \in Q \\ \arrowvert v_1 \arrowvert =1}}
u_1^T v_1 = (u_1^*)^T(v_1^*), \\
\cos \beta = & \max_{\substack{ u_2 \in P \\ \arrowvert u_2
		\arrowvert =1
\\ u_2^T u_1^* = 0 }} \phantom{|} 
\max_{\substack{ v_2 \in Q \\ \arrowvert v_2 \arrowvert =1
		\\ v_2^T v_1^* = 0}}
u_2^T v_2.
\end{align*}
Let $M_P$ and $M_Q$ be the orthonormal basis of $P$ and $Q$. 
The previous equation boils down to 
$$
\max_{\substack{u\in P \\ \arrowvert u\arrowvert =1}} \phantom{|}
\max_{\substack{v\in Q \\ \arrowvert v\arrowvert =1}} u^T v
= 
\max_{\substack{y\in \mathbb{R}^2 \\ \arrowvert y\arrowvert =1}}
\phantom{|}
\max_{\substack{z\in \mathbb{R}^2 \\ \arrowvert z\arrowvert =1}} y^T(M_P^T M_Q) z.
$$
By the minimax characterization of the singular value
decomposition (SVD),
the angle can be computed as the SVD
of the $2\times2$ matrix $M_P^T M_Q$:
$$
 Y^T(M_P^T M_Q)Z
=\begin{pmatrix}\cos \alpha & 0 \\ 0 & \cos \beta\end{pmatrix} ,
$$
where $Y$ and $Z$ are orthogonal $2\times2$ matrices.

\paragraph{Equivalent Definition by Orthogonal Projections.}
Let $P,Q$ be a pair of planes and let $C'$ be an orthogonal projection 
of a unit circle $C$ in $P$ to $Q$. The maximum and the minimum distances $\cos
\alpha$, $\cos \beta$ from a point on $C'$ to the origin define the
same angle $\alpha$ and $\beta$. 
According to this definition, generally the orthogonal
projection $C'$ is an ellipse with the following reasoning.
If $P$ and $Q$ are not isoclinic, there is a
pair of planes $\{U,U'\}$ that are perpendicular to both $P$ and $Q$ and
cut out $P$ and $Q$ at the angle $\alpha$ and $\beta$ respectively.
Then, the intersection of
$Q$ and $U$, (or the line containing the half line $Oq^*$) 
is the major axis (longer axis) of the ellipse because $U$
defines the closest distance between any half lines in $P$ and $Q$.
Similarly, the intersection of
$Q$ and $U'$, (or $\ell_q$)
is the minor axis (shorter axis) of the ellipse. 
However, for a pair of isoclinic planes, the orthogonal projection $C'$
becomes a circle. The argument is shown in
Proposition~\ref{prop:isocli}. 

% The following procedure will be useful in Section~\ref{sec:plane}.
% Let us assume $P$ and $Q$ are not isoclinic and in addition to $C'$,
% let $D'$ be an orthogonal projection of a unit circle in $Q$ to $P$.
% Then, $C'$ and $D'$ are ellipses.
% A pair of points in $P$ can be
% \emph{marked} by the intersection of the unit circle on $P$ and the major axis of $D'$. 
% Similarly, a pair of points in $Q$ can be marked as well.

\paragraph{Left and Right Pairs of Isoclinic Planes.}
Let $P,Q$ %=\langle v_1,v_2 \rangle$, $Q=\langle u_1, u_2 \rangle$ 
be a pair
of 2-planes in 4-space. Let $v_1,v_2$ be orthonormal vectors that
span $P$.
Let $v_1',v_2'$ be the projections of
$v_1,v_2$ to $Q$, and
consider the projections $v''_1$ and $v''_2$ of $v_1'$ and $v_2'$
to the
plane $P^\perp$ orthogonal to~$P$. If $v''_1$ and $v''_2$
are positively oriented together with $v_1, v_2$, 
in other words, if the determinant of $(v_1,v_2,v_1'',v_2'')$ is positive,
we say that $\{P,Q\}$ is a \emph{right pair}, otherwise a \emph{left
pair}.
Whether $P,Q$ is a right or left pair does not
dependent on the order of $P$ and $Q$.
This classification as  right or left is called 
the \emph{chirality} of a pair of planes.
We will use this classification only for isoclinic planes.
When $\alpha=\beta=\pi/2$ (completely
orthogonal planes) or
 $\alpha=\beta=0$ (identical planes), the pair of planes is both
 a left pair and right pair.

%\subsection{Isoclinic Planes}
\paragraph{Relations to Clifford Parallelism.}
A pair $(C,D)$ of great circles in a 3-sphere $\sa$ is called
Clifford parallel if for any point $p$ in $C$, the shortest distance
from $p$ to $D$ is the same. A pair $(C,D)$ of great circles is
Clifford parallel if and only if the pair $(P,Q)$ of planes spanned by $C$
and $D$ is isoclinic. This relation is stated as the following
proposition.

\begin{proposition}
	\label{prop:isocli}
	For a given pair of planes $P$ and $Q$, the following three
	statements are equivalent.
	\begin{compactenum}
	\item\label{item-iso} $P$ and $Q$ are isoclinic.
	\item\label{item-cli} The unit circles $C$ on $P$ and $D$ on
			$Q$ are
			Clifford parallel.
		\item\label{item-proj} The projection $C'$ of a unit circle $C$ on $P$
			to $Q$ is a circle and so is the projection
			$D'$ of a unit circle $D$ on $Q$ to $P$.
	\end{compactenum}
\end{proposition}
\begin{proof}

	[\ref{item-iso} $\Leftrightarrow$~\ref{item-cli}.] The planes $P$ and
	$Q$ are at angle $\alpha,\alpha$ if and only if there are an infinite number
	of pairs of half lines, each pair of which consists of a half
	line $Op$ on $P$ and a half line $Oq$ on $Q$ that form an angle
	$\alpha$. Also, these pairs of lines $\{Op, Oq\}$
	define the pairs of points in $C$ and $D$ with the shortest distance.
	For any pair $\{Op,Oq\}$, the intersection $p'$ of $Op$ and $C$,
	the intersection $q'$ of $Oq$ and $D$, and the origin form
	congruence triangles. Then, the distance between $p'$ and $q'$
	is constant for any points $p'$ on $C$ and $q'$ on $D$. 
	Thus, the planes $P$ and $Q$ are at angle
	at $\alpha, \alpha$ if and only if unit circles $C,D$ on $P$ and $Q$
	are Clifford parallel.

	[\ref{item-iso}$\Leftrightarrow$~\ref{item-proj}.] In the
	previous paragraph, it is explained why the projections $C'$
	and $D'$ generally form ellipses. Similarly to the previous
	argument, points in $Op$ are orthogonally projected to points
	in $Oq$ for the previously defined pairs $\{Op,Oq\}$ 
	since any plane perpendicular to both $P$ and $Q$
	cuts out $P$ and $Q$ with the same angle $\alpha$.
	Therefore, the projections of any points in $C$ to $Q$ are
	equidistant from the origin and so are the
	projections of any points in $D$ to $P$.
	Hence, $P$ and $Q$ are isoclinic
	if and only if the projection $C'$ and $D'$ are circles.
\end{proof}

The terminology for a pair of planes carries over to a pair of great
circles, such as an angle of great circles, 
a right pair of great circles, and
a left pair of great circles. 
For example a pair of great circles in a 3-sphere $\sa$ is 
at angle $(\alpha,\beta)$ 
if the pair of planes spanned by them 
is at angle $(\alpha, \beta)$.
% for $0 \leq \alpha, \beta \leq \pi/2$ as well.

A right/left pair of great circles are
also known as Clifford parallel of the first/second kind~\cite{Ber,Ber2,klein28}.
%For convenience, we simply say that a pair of great circles is a
%right/left pair instead of Clifford parallel of the first/second
%kind.
 Clifford parallelism is originally defined for
 lines in
\emph{elliptic geometry}. Three-dimensional elliptic geometry can be
modeled as the space $\sa/\mathbb{Z}_2$, i.e., the 3-sphere with
opposite points identified. The great circles on the 3-sphere become
the \emph{lines} of elliptic 3-space. 
We are primarily concerned with the geometry of the 3-sphere, and
we will apply
the notion of Clifford parallelism to great circles.

% The new algorithm generates a set of great circles on a 3-sphere $\sa$
% and relates each great circle $C$ to the plane that $C$ spans.
% Later, it will turn out if a pair of great circles are not
% Clifford parallel, we can do the marking procedure, illustrated in
% the previous paragraph ``Equivalent Definition by Orthogonal Projections''.
% Otherwise, this pair defines a unique Hopf
% fibration. 

\paragraph{Generating Clifford-parallel Circles.}
The following useful lemma shows how to parametrize Clifford-parallel circles.
This lemma is used in
Section~\ref{sec:hopfprop}.
\begin{lemma}
\label{lem:paralpha}
Let $v_1,v_2,v_3,v_4$ be a positively-oriented orthonormal basis and
let $C$ be the great circle in a 3-sphere $\sa$
spanned by two orthonormal vectors $v_1$
and $v_2$. If %$\mathcal{D} =\{ D : (C,D)$ 
$C,D$
forms a right pair of great
circles in $\sa$ at angle $(\alpha,\alpha)$,
then $D% \in \mathcal{D}
$ is spanned by the two orthonormal vectors
\begin{align}
%p(\delta)
u_1(\delta)
 &= v_1 \cos \alpha+ (v_3 \cos \delta + v_4 \sin \delta)
	\sin \alpha \label{eq:par}\\
%p'(\delta)
u_2(\delta)
 &= v_2 \cos \alpha+ (v_4 \cos \delta - v_3 \sin \delta)
\nonumber 
\sin \alpha
\end{align}
for some $\delta \in [0,2\pi)$,
and it can be parameterized as
\begin{multline*}
\cos\gamma \cdot u_1(\delta) + \sin\gamma\cdot u_2(\delta)=\\
v_1\cos \alpha \cdot \cos
\gamma
+v_2 \cos \alpha \cdot \sin \gamma
+v_3 \sin \alpha \cdot \cos
(\delta+\gamma)
+v_4 \sin \alpha \cdot \sin (\delta+\gamma)
\end{multline*}
with parameter $0\le \gamma <2\pi$.
 Similarly, for a left pair, $D$ is spanned by 
\begin{align*}
u_1'(\delta) &= v_1 \cos \alpha+ (v_3 \cos \delta + v_4 \sin \delta)
	\sin \alpha \\
u_2'(\delta) &= v_2 \cos \alpha+ (v_3 \sin \delta - v_4 \cos \delta)
\sin \alpha
% u_1(\delta) &= v_1 \cos \alpha- (v_3 \cos \delta + v_4 \sin \delta)
% 	\sin \alpha \\
% u_2(\delta) &= v_2 \cos \alpha+ (v_3 \sin \delta - v_4 \cos \delta)
% \sin \alpha
\end{align*}
for some $\delta \in [0,2\pi)$.
\end{lemma}
\begin{proof}
The circle $D$ can be written as
$\{\,
\cos\gamma \cdot u_1 %(\delta)
 + \sin\gamma\cdot u_2 %(\delta)
\mid 0\le \gamma <2\pi\,\}$,
for any choice of two orthonormal vectors $u_1,u_2\in D$.
In terms of the basis  $v_1,v_2,v_3,v_4$, we may write $u_1 =
\sum_{i=1}^4 a_i v_i$ and $u_2 = \sum_{i=1}^4 b_i v_i$ with
\begin{equation}
  \label{eq:norm-a}
\sum_{i=1}^4 a_i^2 =
\sum_{i=1}^4 b_i^2 = 1,
\qquad
\sum_{i=1}^4 a_ib_i =0.
\end{equation}
The projection of $D$ on the $v_1v_2$-plane containing $C$ is then
the curve
$\cos\gamma \cdot (a_1v_1+a_2v_2)
 + \sin\gamma\cdot (b_1v_1+b_2v_2)$.
 By the definition isoclinic planes of angle $\alpha,\alpha$,
this projection must be a circle of radius $\cos \alpha$.
Thus, the two vectors
$a_1v_1+a_2v_2$ and
$b_1v_1+b_2v_2$ must be perpendicular vectors of length
 $\cos \alpha$.
We may choose $u_1$ in such a way that its projection
$a_1v_1+a_2v_2$
becomes the vector $v_1\cos\alpha$. This implies
$a_1=\cos\alpha$, $a_2=0$,
$b_1=0$, $b_1=\pm\cos\alpha$.
By flipping $u_2$ if necessary, we can achieve that
 $b_1=+\cos\alpha$.

We have
$ a_1^2+a_2^2 = 
b_1^2+b_2^2 = \cos^2 \alpha$
and
$ a_1b_1+a_2b_2 = 0$.
In view of \eqref{eq:norm-a}, this implies
$ a_3^2+a_4^2 = 
b_3^2+b_4^2 = \sin^2 \alpha$
and
$ a_3b_3+a_4b_4 = 0$.
Thus, the two vectors $(a_3,a_4)$ and $(b_3,b_4)$ are orthogonal
vectors of norm $\sin\alpha$.
The general form of such vectors
is 
 $(a_3,a_4)=\sin\alpha\cdot(\cos\delta,\sin\delta)$ 
and $(b_3,b_4)=\pm\sin\alpha\cdot(-\sin\delta,\cos\delta)$.
It can be checked that the choice of the positive sign leads to a
right pair, and the other choice leads to a
left pair.
%
%
%
%  By the definition of an angle, since $C,D$ are
% isoclinic, there exist two orthonormal vectors $u_1,u_2$ such that
% $\angle(v_1,u_1)= \angle(v_2,u_2) = \alpha$ and the planes spanned by
% $\langle u_1,v_1 \rangle$ and $\langle u_2,v_2 \rangle$ are perpendicular
% to both $C$ and $D$.
% Then, $u_1$ and $u_2$ should span $D$.
%
% Since $v_1,v_2,v_3,v_4$ is an orthonormal basis, we may write $u_1 =
% \sum_{i=1}^4 a_i v_i, u_2 = \sum_{i=1}^4 b_i v_i$ with
% % some $a_i$ and $b_i$ for $i= 1, \ldots, 4$ such that
%  $\sum_{i=1}^4 a_i^2 =
% \sum_{i=1}^4 b_i^2 = 1$.
% From $\angle(v_1,u_1)= \angle(v_2,u_2) = \alpha$, $a_1 = b_2 = \cos
% \alpha$.
% From the plane spanned by $\langle u_1,v_1 \rangle$ is perpendicular 
% to both $C$,
% $\langle v_1,v_2 \rangle$, $(u_1-v_1)^T v_2 = 0$ so $a_2=0$. Similarly, $b_1 = 0$.
% From $\sum_{i=1}^4 a_i^2 = \sum_{i=1}^4 b_i^2 = 1$, we may take
% $u_1(\delta) = %u_1 = 
% v_1 \cos \alpha+ (v_3 \cos \delta + v_4 \sin \delta)
% \sin \alpha, u_2(\delta) = %u_2 =
% v_2 \cos \alpha+ (v_4 \cos \delta - v_3 \sin \delta) \sin \alpha$.
\end{proof}

%% file: Pluecker.tex
\section{Pl\"ucker Space and Pl\"ucker Distance}
\label{sec:pluecker}

A classical and effective tool of dealing with linear subspaces in a
vector space is the Grassmannian and the Pl\"ucker embedding. The
\emph{Grassmannian}
$\mathbb{G}(k,V)$ is the collection of all $k$-dimensional linear subspaces
of a vector space $V$. The \emph{Pl\"ucker embedding} embeds
Grassmannians into higher-dimensional projective space.
In this paper, we only need the Grassmannian
$\mathbb{G}(2,\mathbb{R}^4)$. Its Pl\"ucker coordinates are six-dimensional
homogeneous coordinates
in real projective 5-space.

The Grassmannian $\mathbb{G}(2, \mathbb{R}^4)$ is the set of all 
the planes going through the origin in Euclidean
4-space. This is
equivalent to lines in real projective 3-space.
Let $x =(x_0,x_1,x_2,x_3) ,y = (y_0,y_1,y_2,y_3)$ be homogeneous coordinates
of two distinct points in real projective 3-space. 
Then the Pl\"{u}cker vector for a line going through these two points
is the sixtuple $(p_{ij})_{0 \leq i <j \leq 3}$ where $p_{ij} = x_iy_j -x_j
y_i$; in other words, the sixtuple of the $2 \times 2$ determinants of
the $2 \times 4$ matrix with rows $x$ and $y$. 
These coordinates
are determined only up to scaling, and they
 are regarded as homogeneous
coordinates in projective 5-space.
For convenience, we call real projective 5-space with Pl\"ucker
coordinates \emph{Pl\"ucker space}. Each plane in Euclidean 4-space
corresponds to a point in Pl\"ucker space.

To get a metric for 2-planes in Euclidean 4-space, we
normalize the Pl\"ucker coordinates so that the norm of the
 Pl\"ucker vector becomes 1. This represents each
plane by two antipodal points on the unit 5-sphere
$\mathbb{S}^5$, since the normalized 
 Pl\"ucker vector is determined only up to a sign change.
The \emph{Pl\"{u}cker distance}
between two planes $P,Q$ in 4-space
is defined as the Euclidean distance between normalized Pl\"{u}cker coordinates of $P$ and
$Q$, choosing representative points from each antipodal pair such that
the distance becomes smallest.

The Pl\"{u}cker distance is a metric since it is basically a Euclidean
distance.
The next lemma shows that the
Pl\"{u}cker distance is geometrically meaningful 
and does not depend on the choice of a coordinate system.
%property that is
%independent of the choices of bases for $P$ and $Q$.
%and also is invariant over rotations and
%reflections 
%This is shown by relating angles between planes to the Pl\"ucker distance.

\begin{lemma}
\label{lem:pdist}
If a pair of planes $P,Q$ is at angle $(\alpha, \beta)$,
with $0\le\alpha,\beta\le \pi/2$, their
Pl\"{u}cker distance is $\sqrt{2(1- \cos \alpha \cos
\beta)}$.
\end{lemma}

\begin{proof}
By
\cref{prop-perp},
we can choose an orthonormal basis
%Let $P$ and $Q$ be two planes with angle $(\alpha, \beta)$.
 $u=(u_1,u_2,u_3,u_4), v=(v_1,v_2,v_3,v_4)$ 
for %be orthonormal vectors that span
 $P$ %. If $P$ and $Q$ are not orthogonal,
and an orthonormal basis
 $u'=(u'_1,u'_2,u'_3,u'_4),v'=(v'_1,v'_2,v'_3,v'_4)$ 
for $Q$ such that
%be the orthonormal vectors such that 
$u', v'$ are parallel to the projections of $u,v$ to $Q$.
(If $\alpha = \pi/2$, the projection of $u$ to $Q$ is the zero vector,
and in this case we consider any vector $u'$ as parallel to it;
the same holds for $\beta = \pi/2$.)

%
%If $P$ and $Q$ are orthogonal, that is, $\alpha= \beta = \pi/2$,
%let
%$u',v'$ be any two orthonormal vectors
%that span $Q$.

The Plücker coordinates are then the six values
 $(k_{ij})_{0 \leq i <j \leq 3}$,
 where 
$k_{ij}=u_{i}v_{j} - v_{i}u_{j} $.
Since $u,v$ is an orthonormal basis, we can check that
these Plücker coordinates are already normalized, using the
relations  $k_{ij} = - k_{ji}$ and
 $k_{ii} = 0$:
\begin{align*}
	\sum_{0 \leq i < j \leq 3}  k_{ij}^2  
	&	= \frac12  \sum_{0 \leq i < j \leq 3} ( k_{ij}^2 +
	k_{ji}^2 ) 
	= \frac12 \sum_{i=0}^3 \sum_{j=0}^3  k_{ij}^2 \\
	&	= \frac12 \sum_{i=0}^3 \sum_{j=0}^3
(u_{i}v_{j} - v_{i}u_{j})^2 
= \frac12 \sum_{i=0}^3 \sum_{j=0}^3 (u_{i}^2 v_{j}^2 + v_{i}^2 u_{j}^2 
- 2u_{i}u_{j} v_{i} v_{j} ) \\
&= \frac12 \left( \sum_{i=0}^3u_{i}^2 \sum_{j=0}^3 v_{j}^2 - 
  2 \sum_{i=0}^3 u_{i}v_{i} \sum_{j=0}^3 u_{j}v_{j} +
  \sum_{j=0}^3u_{i}^2 \sum_{i=0}^3 v_{j}^2 \right) \\ 
  & =
 \tfrac12 (1-2\cdot 0+1)
 % \frac12 ( (u^T u) (v^T v) - 2(u^T v)^2 + (u^T u)(v^T v)) 
=1  
\end{align*}
Similarly, the Plücker coordinates
$\ell_{ij}=u'_i v'_j - v'_i u'_j $ of $Q$ are normalized:
 $\sum_{0 \leq i <j \leq 3} \ell_{ij}^2 = 1$.
The squared Euclidean distance between these points is
\begin{align*}
%  \sqrt{
\sum_{0 \leq i < j \leq 3} \!\!
	\left( k_{ij}-\ell_{ij}\right)^2
&
=
\sum_{0 \leq i < j \leq 3} 
	k_{ij}^2
+\sum_{0 \leq i < j \leq 3} 
	\ell_{ij}^2
-2
\sum_{0 \leq i < j \leq 3} 
	k_{ij}\ell_{ij}
%\\&
=2\biggl(1-\!\!
\sum_{0 \leq i < j \leq 3} 
	k_{ij}\ell_{ij}
\biggr)
%}
\end{align*}
Let us evaluate the last term.
From the angle between $P$ and $Q$, we get
$u^T u' =\cos \alpha, v^T v' =\cos \beta,$ and
 $u^T v' = v^T u' = 0$, and we have
 $k_{ij} = - k_{ji}$ and
 $\ell_{ij} = - \ell_{ji}$.
%, we get
%$k_{ij}^2 = k_{ji}^2$ and $k_{ij}\ell_{ij} = k_{ji}\ell_{ji}$.
\begin{align*}
	\sum_{0 \leq i <j \leq 3} k_{ij}\ell_{ij} 
	&= \frac12 \sum_{0 \leq i <j \leq 3}( k_{ij}\ell_{ij} +
	k_{ji}\ell_{ji}) 
	= \frac12 \sum_{i=0}^3 \sum_{j=0}^3 k_{ij}\ell_{ij} 
	\\
	&=\frac12 \sum_{i=0}^3 \sum_{j=0}^3 (u_{i}v_{j} - v_{i}u_{j})
	(u'_{i}v'_{j} - v'_{i}u'_{j}) 
	\\& =\frac12 \sum_{i=0}^3 \sum_{j=0}^3
(u_{i}u'_i v_{j}v'_j + v_{i}v'_i u_{j}u'_j - u_{i}v'_iu'_{j}
	v_{j} - v_iu'_iu_jv'_j ) \\
	&=\frac12 \left( \sum_{i=0}^3
 u_{i} u'_i  \sum_{j=0}^3v_{j} v'_j + 
 \sum_{i=0}^3 
 v_{i}v'_i  \sum_{j=0}^3u_{j} u'_j
\right.
\\&\qquad\qquad
 \left.
 - 
 \sum_{i=0}^3 
u_{i}v'_{i}\sum_{j=0}^3u'_{j} v_{j}
 -
 \sum_{i=0}^3 
u'_{i}v_{i} \sum_{j=0}^3u_{j} v'_{j} \right) \\
%
% %
% 	&=\frac12 \left( \sum_{i=0}^3 \sum_{j=0}^3
%  u_{i} u'_i v_{j} v'_j + 
%  \sum_{i=0}^3 \sum_{j=0}^3
%  v_{i}v'_i u_{j} u'_j
% \right.
% \\&\qquad\qquad \left.
%  - 
%  \sum_{i=0}^3 \sum_{j=0}^3
% u_{i}v'_{i}u'_{j} v_{j}
%  -
%  \sum_{i=0}^3 \sum_{j=0}^3
% u'_{i}v_{i} u_{j} v'_{j} \right) \\
% %
& = \tfrac12 \left( (u^T u') (v^T v') + (v^T v')(u^T u') - (u^T v')(u'^T v) -
(u'^Tv)(u^T v')\right)
  \\ & =
\tfrac12(
 \cos \alpha \cos \beta  + \cos \beta   \cos \alpha -0 -0) =
 \cos \alpha \cos \beta  
\end{align*}
Thus, the Euclidean distance is
$
\sqrt{2(1- \cos \alpha \cos \beta)}%^{\frac{1}{2}}
$.
%$$
%\frac{1}{\sum_{0 \leq i<j \leq 3} k_{ij}^2 } +
%\frac{1}{\sum_{0 \leq i<j \leq 3} \ell_{ij}^2 } -
%2 \frac{ \sum_{0 \leq i<j \leq 3} k_{ij} \ell_{ij} }
%{(\sum_{0 \leq i<j \leq 3} k_{ij}^2)(\sum_{i
%\neq j} \ell_{ij}^2) } = 2(1- \cos \alpha \cos \beta).
%$$
Each circle is represented by two antipodal points. Thus we should
consider also negated vectors 
 $(-k_{ij})_{0 \leq i <j \leq 3}$
and  $(-\ell_{ij})_{0 \leq i <j \leq 3}$.
Negation can be achieved by negating one of the vectors
$u,v$ or one of the vectors $u',v'$. 
In the formulas, this changes
$\cos \alpha$ to
$\cos (\pi-\alpha)
=-\cos \alpha$
or
$\cos \beta$ to
$\cos (\pi-\beta)
=-\cos \beta$.
The minimum distance is obviously achieved when choosing the positive
cosine values, i.e., when choosing the angles in the range
 $0\le\alpha,\beta\le \pi/2$.
This finishes the proof.

 This smaller distance value together with the larger value,
$
\sqrt{2(1+ \cos \alpha \cos \beta)}%^{\frac{1}{2}}
$,
form the sides of a rectangle inscribed in the unit circle, as 
expected for two pairs of antipodal points on a sphere.
\end{proof}

Since the angle between two planes is invariant under rotations, we
get the following direct consequence:
\begin{corollary}
	\label{cor:pdist}
The Pl\"{u}cker distance is invariant under rotations and reflections.
\qed
\end{corollary}

\subsection{Other Distances}
\label{Other-Distances}
Conway, Hardin and Sloan~\cite{CHS} considered different distances
in Grassmannian space. 
One of the distances that they considered is called the \emph{chordal
distance}.
For the Grassmannian
$\mathbb{G}(2,\mathbb{R}^4)$, the chordal distance
is defined as $\sqrt{\sin^2 \alpha + \sin^2 \beta}$ for a pair of
planes at angle $\alpha,\beta$.
 The chordal
distance coincides with Pl\"ucker distance if a pair of planes is
isoclinic. 
They also considered
the \emph{geodesic distance}~\cite{Wong2}, 
and \emph{Asimov's distance}~\cite{Asi,GL}.

Those three distances are related to the angles between
$k$-dimensional subspaces.
As mentioned above, two
$k$-dimensional subspaces define $k$ angles
$(\alpha_1,\ldots,\alpha_k)$. (If $k>d/2$, some of these angles are
necessarily 0.)
the geodesic distance, the chordal distance, and
Asimov's distance are defined as the Euclidean norm of
this angle vector,
the Euclidean norm of vector $(\sin \alpha_1,\ldots, \sin \alpha_k)$, and the largest
angle,
 respectively.
The geodesic distance has its name because it represents the true
geodesic distance in the manifold  $\mathbb{G}(k, \mathbb{R}^d)$.

Dense packings of subspaces in $\mathbb{G}(k, \mathbb{R}^d)$
 for these three
distances were experimentally determined in~\cite{CHS}. 

For our algorithm, it is convenient if we can embed $\mathbb{G}(2,
\mathbb{R}^4)$ into some Euclidean space.  Apart from this, the
precise characteristics of the distance don't matter.  The Pl\"ucker
distance is measured in an ambient space of dimension $\binom d k$.
On the other hand, the chordal distance
embeds the Grassmannian $\mathbb{G}(k, \mathbb{R}^d)$ to
an $m$-dimensional sphere, where $m= \binom{d+1} 2 -1$.  (It
represents a subspace by the $d\times d$ symmetric projection matrix
onto that subspace.)  Thus, for large $k$ and $d$, this distance is
preferable over the Pl\"ucker distance, for which the dimension grows
exponentially.  However, for our parameters $d=4$ and $k=2$, the
chordal distance would require a 9-sphere, whereas the Pl\"ucker
distance requires only a 5-sphere. Since the ambient dimension affects
the complexity of the algorithm, the Pl\"ucker distance is preferable
in our case.

%% file: Hopf.tex
\section{The Construction of Hopf Fibrations}
\label{sec:hopf}
For the well-known properties of the Hopf fibration of the 3-sphere,
see for example \cite[Section 4.3.7]{Ber}
and~\cite[Chapter 18.8]{Ber2}.
The goal of this section is to understand the relations
among isoclinic rotations (Section~\ref{sec:rot}),
a left/right pair of great circles (Section~\ref{sec:angle}) 
and Hopf fibrations.
%isoclinic is
%defined as Clifford parallel, and left and right pairs are as
%Clifford parallel of the first and second kinds. Those terms will defined
%later in this section.

Before starting, we skethch the flow of this section without
the definitions of terms:
\begin{compactenum}
\item \label{item:i1} We first construct \emph{the invariant family} 
		for a given great circle $C$ by using only isoclinic
		rotations (Lemmas~\ref{lem:uhopf} and~\ref{lem:chopf}).
	\item Then, we define a \emph{Hopf map} $h$ with respect to
		$C$. This shows that the invariant family for $C$ is equivalent 
		to the \emph{Hopf bundle} for $h$
		(Lemma~\ref{lem:hopf}). 
		This gives a geometric
		perspective of constructing a \emph{Hopf fibration}.
	\item In Section~\ref{sec:hopfprop}, we derive the following
		properties.
		\begin{compactenum}
		\item \label{item:i2} From \ref{item:i1}, we know 
			there is a unique bundle for each great
			circle $C$ (Corollary~\ref{cor:uhopf}). 
			 Also, a pair of Clifford-parallel great circles 
			 belong to a
			 common Hopf bundle (Lemma~\ref{lem:alpha}).
		\item The relation between two great circles defined by
			a right  pair (or a left pair) is
			transitive; thus,
			it is an
			equivalence relation. From~\ref{item:i2}, 
			an equivalence class is
			actually a right Hopf bundle (or a left Hopf
			bundle) (Corollary~\ref{cor:thopf}). 
			The converse is also true.
		\item \label{item:i4} The Hopf fibration maps two great circles at
			angle $\alpha,\alpha$ to two points at
			geodesic distance $2\alpha$ in a
			2-sphere (Lemma~\ref{lem:lrhopf}).
		% \item From \ref{item:i4}, a Hopf fibration is unique up
		% 	to rotations and reflections. This implies 
		% 	independence of choices of a basis
		% 	(Corollary~\ref{cor:rrhopf}).
		\end{compactenum}
\end{compactenum}

To minimize confusion, we provide proofs and explanations only for 
a right Hopf fibration but
parallel lemmas are also valid for a left Hopf fibration.

\subsection{Construction by Isoclinic Rotations}
This construction of a right Hopf fibration provides 
the relation between a right Hopf fibration and a
right rotation.
We will now construct a right Hopf bundle as
the set of orbits of all {right} rotations that map a given circle
to itself.

We mention that the convention of assigning the label left or right
to versions of a structure that come in pairs is somewhat
arbitrary and not uniform in the literature. In fact, right Hopf
bundles are also connected to \emph{left} rotations:
% %According to Proposition~\ref{hopf}, 
a right Hopf bundle is the set of images of a given circle
under all {left} 
rotations.
In the quaternion representation, right rotations according to the
convention in this paper are carried out by
left multiplication with unit quaternions.
Thus, there are good reasons also for alternative choices.

\begin{lemma}
\label{lem:uhopf}
For every great circle $C$ in a 3-sphere and two points $p,q \in C$,
there exists a unique right rotation
$R$ which rotates $C$ in itself and rotates $p$ to $q$.
\end{lemma}
\begin{proof}
Choose an orthonormal basis $v_1,v_2$ of the plane $P$ 
spanned by $C$.
% and arbitrarily label one of the orientations of $C$ as positive and
% the other as negative. 
There exists a unique rotation  
$%\mathbf{R}_\gamma = 
\left( \begin{array}{c c} \cos \gamma & -\sin \gamma \\
\sin \gamma & \cos \gamma \end{array} \right)$ 
%( or $R_{-\gamma} = \left( \begin{array}{c c} 
%\cos \gamma & \sin \gamma \\
%-\sin \gamma & \cos \gamma \end{array} \right)$ )
in the $v_1v_2$-plane that rotates $p$ to $q$. 
Extend $v_1,v_2$
to a positively-oriented basis $v_1,v_2,v_3,v_4$.
It follows that, in this basis, the rotation matrix $R$ must have the form
\begin{align}
\label{eq:rho}
\rho(\gamma) = R_{\gamma,\gamma}.
\end{align}
The reason is as follows:
The first two columns are fixed by the requirement that $P$ maps $p$
to $q$ and leaves the plane through $C$ fixed. The last two columns
are then fixed by the requirement that $R$ is a right rotation.
\end{proof}
Actually, if $p\ne \pm q$, the requirement that $R$ rotates $C$ in
itself is redundant.

The family of right rotations 
$\ifam = \{\rho(\gamma)|\gamma \in [0,2\pi) \}$
rotates a great circle $C$ to itself
where $\rho(\gamma)$ is defined in
(\ref{eq:rho}).
Note that $\ifam$ is a one-dimensional group isomorphic
to the special orthogonal group SO(2). 
We call $\ifam$ the \emph{family of right
rotations} for $C$.

\begin{lemma}
\label{lem:chopf}
Let $\ifam$ be the family of right rotations for some great
circle in a 3-sphere $\sa$.
For every point $p$ in $\sa$, 
the orbit of $p$ under rotations in $\ifam$ is a great circle of $\sa$.
\end{lemma}
\begin{proof}
If $p=(x,y,z,w)$, the orbit of $p$ generated by a
right rotation is
\begin{equation}
\label{eq:param}
\mu(\gamma) = \rho(\gamma) p 
 = u_1 \cos \gamma + u_2 \sin \gamma 
\end{equation}
where
$$
u_1 =p= (x,y,z,w)^T, u_2 = (-y,x,-w,z)^T, \rho(\gamma) \in \ifam
$$
for $ \gamma \in [0, 2\pi)$.
Here, $u_1$ and $u_2$ are orthogonal and $\mu(\gamma)$
is a parametrization of a great circle, so $p$ generates a great
circle as an orbit.
\end{proof}

From Lemma~\ref{lem:chopf}, $\ifam$ partitions $\sa$ into the set 
of great circles that are invariant under rotations in $\ifam$.
Let $\Gamma$ be the set of these great circles and call it the
\emph{invariant family} for $C$.
Eventually, the invariant family $\Gamma$ forms a right \emph{Hopf
bundle}. In the remainder of this section, 
we illustrate what this statement means more precisely.

\subsection{Equivalence of an Invariant Family and a Hopf Bundle.}
A \emph{fibration} is a map that projects a \emph{fiber bundle} to a
\emph{base}, 
by identifying a subspace of the fiber
bundle, called a \emph{fiber}, to a point in a base in a way that any point
in the fiber bundle is contained in exactly one fiber.

Given a great circle $C$, 
by choosing a Cartesian coordinate system $(x,y,z,w)$ such that $C$ lies in the $x,y$-plane,
we define the right Hopf map with respect to $C$ as
\[
h(x,y,z,w) = (2(xw-yz), 2(yw+xz), 1-2(z^2+w^2)) 
\quad \textrm{ for } \quad (x,y,z,w) \in \sa \subset \mathbb{R}^4
\]
For reference, the left Hopf map is
defined as $h'(x,y,z,w) = (2(xw+yz), 2(yw-xz), 1-2(z^2+w^2))$. 

The following lemma shows that a right Hopf map is a fibration.

\begin{lemma}
\label{lem:hopf}
Let $C$ be a great circle in a 3-sphere $\sa$. 
Let $\Gamma$ be the invariant family for $C$.
Let $h$ be the right Hopf map with respect to $C$.
Then, for any $D \in \Gamma$, $h$ maps all points in $D$ 
to one point in the 2-sphere
$\mathbb{S}^2$.
\end{lemma}
\begin{proof}
%Note that the statement is equivalent to showing that 
%each element in $\Gamma$ becomes a fiber. 
The map $h$ indeed maps the 3-sphere to the 2-sphere, since
$(1-2z^2-2w^2)^2 + 4(xw-yz)^2 + 4(yw+xz)^2 = (x^2+y^2+z^2+w^2)^2
=1$. 

As we
have seen in Lemma~\ref{lem:chopf}, for any $D \in \Gamma$, if $p
=(x,y,z,w) \in D$, we can parametrize $D$ as $\mu(\gamma)$ 
in~(\ref{eq:param}) for $\gamma \in [0, 2\pi)$.
Then, 
\begin{align*}
	h(\mu(\gamma)) & =  h( (x,y,z,w) \cos \gamma + (-y,x,-w,z) \sin \gamma )\\
			      & = h((x \cos \gamma - y \sin \gamma, y \cos \gamma+
	x \sin \gamma, z\cos \gamma - w \sin \gamma, w \cos \gamma +
	z \sin \gamma)) \\
	& = \bigl( %\begin{array}{c}
 2 [ (x \cos \gamma -y \sin \gamma)(w \cos \gamma + z \sin
		\gamma) - (y \cos \gamma + x \sin \gamma)(z \cos
	\gamma - w \sin \gamma) ], \\
&\qquad	 2 [(y \cos \gamma+ x \sin \gamma)(w \cos \gamma
	+z \sin \gamma)+(x \cos \gamma - y \sin \gamma)(z \cos \gamma - w \sin
	\gamma)], \\
&\qquad	 1-2[(z \cos \gamma - w \sin \gamma)^2+(w \cos \gamma + z \sin
\gamma)^2] %\end{array} \right
\bigr) \\
& = \bigl( %\begin{array}{c}
 2 [(xw \cos^2 \gamma -yz \sin^2 \gamma + (xz-yw) \cos
\gamma \sin \gamma)  \\ 
&\qquad\qquad -(yz \cos^2 \gamma - xw \sin^2 \gamma +(xz-yw) \cos
\gamma \sin \gamma)], \\ 
&\qquad 2 [(yw \cos^2 \gamma + xz \sin^2 \gamma +
(xw+yz) \cos
\gamma \sin \gamma)   \\ 
&\qquad\qquad +(xz \cos^2 \gamma + yw \sin^2 \gamma -(xw+yz) \cos
\gamma \sin \gamma)], \\ 
&\qquad 1-2[(z^2 \cos^2 \gamma + w^2 \sin^2 \gamma -zw \cos
\gamma \sin \gamma) \\
&\qquad\qquad +(w^2 \cos^2 \gamma + z^2 \sin^2 \gamma +zw \cos
\gamma \sin \gamma)]% \end{array} \right
\bigr) \\ 
		       & = (2(xw-yz), 2(yw+xz), 1-2(z^2+w^2))
\end{align*}
This is independent of $\gamma$, which shows that $D$ is a fiber. 

This also implies that a right Hopf map is a fibration,
since elements in $\Gamma$, or fibers, 
are mutually disjoint and cover the whole $\sa$. 
Remember that in Lemma~\ref{lem:chopf} 
every point in $\sa$ is in a unique orbit, so in a unique element of $\Gamma$,
under rotations in $\ifam$.
\end{proof}

We have shown that the right Hopf map with respect to a great circle $C$
is a fibration that maps an invariant family 
$\Gamma$ for $C$ to a 2-sphere $\mathbb{S}^2$ 
in a way that each great circle in $\Gamma$ is
mapped to a point in $\mathbb{S}^2$.
Hereafter, we are allowed to call a Hopf map a Hopf fibration, 
the fiber bundle $\Gamma$ the Hopf bundle, and each great circle in
$\Gamma$ a Hopf fiber. Also, the 2-sphere which is a base for a Hopf
fibration is called a base sphere. We speak of a right Hopf
fibration/bundle/fiber if a Hopf fibration/bundle/fiber was induced by
a right Hopf map.

\subsection{Properties of the Hopf Fibration}
\label{sec:hopfprop}

Because the invariant family for a great circle $C$ is equivalent to a
right Hopf bundle, we may as well state Lemma~\ref{lem:uhopf} as
follows:

\begin{corollary} %[Corollary of Lemma~\ref{lem:uhopf}]
	\label{cor:uhopf}
	Every great circle belongs to a unique right Hopf bundle.
\end{corollary}

Next, we show the relation between great circles and a Hopf
fibration.

\begin{lemma}
\label{lem:alpha}
Any right pair of great circles belongs to a common
right Hopf bundle.
\end{lemma}
\begin{proof}

Let $C,D$ be a right pair of great circles at angle
$\alpha,\alpha$. Let $h$ be a right Hopf map for $C$. With the
same basis for $h$, $C$ is represented by $\{(\cos \gamma, \sin
	\gamma,0,0)| \gamma \in [0,2\pi)\}$. 
	Then $h$ maps $C$ to the north pole since $h(\cos
	\gamma, \sin \gamma, 0,0) = (0,0,1)$. 
The circle $D$ can be represented by
$
\{(\cos \alpha \cdot \cos \gamma
, \cos \alpha \cdot \sin \gamma, \sin \alpha \cdot \cos
(\delta+\gamma) , \sin \alpha \cdot \sin (\delta+\gamma))|
\gamma \in [0,2\pi) \}
$
%  $\{(\cos \alpha \cdot \cos
% \gamma, \cos \alpha \cdot \sin \gamma, \sin \alpha \cdot \cos
% (\gamma+\gamma_0) , \sin \alpha \cdot \sin (\gamma+\gamma_0))|
% \gamma_0 \in [0,2\pi) \}$ 
for some $\delta \in [0,2\pi)$
by Lemma~\ref{lem:paralpha}. Thus, $h$ maps $D$ to a
point $(\sin 2\alpha \sin \delta, \sin 2 \alpha \cos \delta, \cos
2\alpha)$, since
\begin{align}
	\hbox to 2cm{$
 h(\cos \alpha \, \cos \gamma, \cos \alpha \sin \gamma, \sin
	\alpha \cos (\delta + \gamma), \sin \alpha \sin
	(\delta + \gamma))$\hss} \nonumber \\ 
&= \Bigl( %\begin{array}{c}
 2 (\cos \alpha \sin \alpha \cos \gamma
\sin (\delta + \gamma) - \cos \alpha \sin \alpha \sin \gamma \cos
(\delta + \gamma) ), \nonumber \\
&\qquad
  2 (\cos \alpha \sin \alpha \sin \gamma
\sin (\delta + \gamma) + \cos \alpha \sin \alpha \cos \gamma \cos
(\delta + \gamma) ), \nonumber \\
&\qquad
1- 2 (\sin^2 \alpha \cos^2 (\gamma +\gamma_0) +\sin^2 \alpha \sin^2
(\delta + \gamma))
%\end{array} \right
\Bigr) \nonumber \\ 
\label{eq:hopfba}
&= (\sin 2\alpha \sin \delta, \sin 2 \alpha \cos \delta, \cos
2\alpha).
\end{align} 
This is independent of the parameter $\gamma$.
Therefore, the right Hopf bundle given by $h$
contains both $C$ and $D$ as Hopf fibers. 
\end{proof}

%The converse of Lemma~\ref{lem:alpha} is also true.
This relation of being a right pair is actually transitive.

\begin{corollary}
\label{cor:thopf}
Let us denote $C\parallel_+D$, when
two great circles $C$ and $D$ in $\sa$
form a right pair.
 Then $\parallel_+$ is transitive.  
Each equivalence class of $\parallel_+$ is a right Hopf bundle.
\end{corollary}
\begin{proof}
If $C\parallel_+D$ and $D\parallel_+E$,
 there
is a unique right Hopf fibration containing $C$, $D$ and $E$ 
by Corollary~\ref{cor:uhopf} and Lemma~\ref{lem:alpha}.
Then, $C$, $D$, and $E$
are contained in the same Hopf fibration, so $C \parallel_+ E$.
Observe that $\parallel_+$ is also reflexive and symmetric.
The equivalence class is exactly the invariant family for $C$, or a
right Hopf bundle.
\end{proof}

In our construction, the Hopf map $h$
depends on the
choice of a great circle $C$ and a Cartesian coordinate system
$(x,y,z,w)$, 
However, the
 Hopf fibration resulting from the map is independent of this choice.
To show this, we show how a pair of Hopf fibers is mapped
to a pair of points in a base sphere. A similar statement can be found
in~\cite[Exercise 18.11.18]{Ber2}.

\begin{lemma}
\label{lem:lrhopf}
Two right Hopf fibers at angle $\alpha, \alpha$ are mapped 
to a pair of points of geodesic
distance $2 \alpha$ by the right Hopf map.  
\end{lemma}
\begin{proof}
	\if
	Let $C,D$ be such right Hopf fibers.
In the coordinate system $(x_1,x_2,x_3,x_4)$ where $h$ is computed,
the circle $C$ forms a right pair with the unit circle $C_0$ in
the $x_1x_2$-plane. Therefore, by
 Lemma~\ref{lem:paralpha}, it is generated by
the vectors
$v_1=(\cos \alpha',0,\cos\delta'\sin\alpha',\sin\delta'\sin\alpha')$
and
$v_2=(0,\cos \alpha',-\sin\delta'\sin\alpha',\cos\delta'\sin\alpha')$,
for some $\alpha'$ and $\delta'$.
These orthonormal vectors are of the form
 $v_1= (x,y,z,w)$, $v_2= (-y,x,-w,z)$
with $x^2+y^2+z^2+w^2=1$, and they can be extended to a
positively-oriented 
	orthonormal basis $v_1,v_2,v_3,v_4$
by the
two vectors 
	$v_3= (-z,w,x,-y)$, $v_4 =(-w,-z,y,x)$.
The circle $D$ forms a right-isoclínic pair at angle
%The circles in the fiber bundle of the right Hopf fibration that are
at angle $\alpha,\alpha$ with $C$.
% form one-dimensional family $\mathcal{D}$ 
%and $D \in \mathcal{D}$ as in
Applying Lemma~\ref{lem:paralpha} again, this time to the
basis $v_1,v_2,v_3,v_4$, we see that $D$ contains the point
%We can parametrize the elements of $\mathcal{D}$ by $\delta \in (0,
%2\pi]$ as in \eqref{eq:par}, as taking a representative point  
\begin{align*}
	u_1(\delta) 
%	&= v_1 \cos \alpha + (v_3 \cos \delta  + v_4 \sin \delta) \sin \alpha \\
	&= (x,y,z,w) \cos \alpha + (-z,w,x,-y) \cos \delta \sin \alpha+
	(-w,-z,y,x) \sin \delta \sin \alpha \\
	&= \left( \begin{array}{c} x \cos \alpha -z \sin \alpha \cos
	\delta -w \sin \alpha
\sin \delta \\ y \cos \alpha +w \sin \alpha \cos \delta -z \sin \alpha \sin \delta \\ z \cos
\alpha +x \sin \alpha \cos \delta +y \sin \alpha \sin \delta \\ w \cos
\alpha -y \sin \alpha \cos \delta
+ x \sin \alpha \sin \delta \end{array} \right),
\end{align*}
for some $\delta$.
The Euclidean distance between $h(C)$ 
and $h(D)$
can be evaluated as $\arrowvert h(u_1(\delta)) - h(v_1) \arrowvert $.
After expanding this term, the expression simplifies
to
\begin{align*}
\arrowvert h(u_1(\delta)) - h(v_1) \arrowvert 
& =  2 \sin \alpha\cdot(x^2+y^2+z^2+w^2) = 2 \sin \alpha
\end{align*}
%\todo{calculation \\hopf 2}
independent of $\delta$. %, since $x^2+y^2+z^2+w^2 =1$. It is
% also true for $h(\rho(\gamma)u_1(\delta))$ and $h(\rho(\gamma)v_1)$ for any
% $\gamma$ where $\rho(\gamma)$ is defined in (\ref{eq:rho})
% that 
% \begin{align*}
% \arrowvert h(\rho(\gamma) u_1(\delta)) - h(\rho(\gamma) v_1) \arrowvert 
% & =  2 \sin \alpha(x^2+y^2+z^2+w^2) = 2 \sin \alpha.
% \end{align*}
% %\todo{calculation \\hopf 3}
% The central angle (geodesic distance) is 
% $2 \arcsin (\frac{2 \sin \alpha}{2}) = 2\alpha$.
\fi
By~\eqref{eq:hopfba} in Lemma~\ref{lem:alpha}, the great circle that
is mapped to the point with spherical coordinates $(\gamma,\delta)$,
$
\begin{pmatrix}
 \sin 2\gamma \sin\delta  \\
 \sin 2\gamma \cos\delta  \\
  \cos 2\gamma\\
\end{pmatrix}
$
by the right Hopf map
is spanned by the two orthonormal vectors
\begin{equation*}
u(\gamma,\delta)
=
\begin{pmatrix}
  \cos \gamma\\
  0\\
\cos \delta 
	\sin \gamma\\
\sin \delta
	\sin \gamma\\
\end{pmatrix},
\
v(\gamma,\delta)
=
\begin{pmatrix}
  0\\
  \cos \gamma\\
-\sin \delta
	\sin \gamma\\
\cos \delta 
	\sin \gamma\\
\end{pmatrix}.
\end{equation*}
Let us calculate the angle $\alpha$ between two great circles $C$ and
$D$ with
 spherical coordinates $(\gamma_1,\delta_1)$ and
 $(\gamma_2,\delta_2)$.
Rotating the 2-sphere by varying the parameter $\delta$ corresponds to
a rotation of the $v_3v_4$-plane in $\mathbb{R}^4$, which leaves the
angles between $C$ and $D$ unchanged.
Thus,
  by an appropriate
rotation, we may assume that $\delta_1=0$.

%(Check! Maybe this is noteven necessary.)

The length of the projection of any point of $C$ onto the plane
spanned by $D$ is $\cos \alpha$. We project 
$u(\gamma_1,0 %\delta_1
)$ on the plane with orthonormal basis
$u(\gamma_2,\delta_2)$ and
$v(\gamma_2,\delta_2)$, and get
\begin{align}
\nonumber
  \cos^2\alpha
&=
\langle
u(\gamma_1,0%\delta_1
),
u(\gamma_2,\delta_2)
\rangle^2
+
\langle
u(\gamma_1,0%\delta_1
),
v(\gamma_2,\delta_2)
\rangle^2\\\nonumber
&=
(\cos\gamma_1\cos\gamma_2
+\cos\delta_2\sin\gamma_1\sin\gamma_2)^2
+(-\sin\delta_2\sin\gamma_1\sin\gamma_2)^2
\\\nonumber
&=
\cos^2\gamma_1\cos^2\gamma_2
+
\sin^2\gamma_1\sin^2\gamma_2
+2\cos\delta_2\cos\gamma_1\cos\gamma_2
\sin\gamma_1\sin\gamma_2
\\\nonumber
&=
\cos^2\gamma_1\cos^2\gamma_2
+
(1-\cos^2\gamma_1)(1-\cos^2\gamma_2)
+
\tfrac12
\cos\delta_2
\sin2\gamma_1\sin2\gamma_2
\\\nonumber
&=
2\cos^2\gamma_1\cos^2\gamma_2
+
1-\cos^2\gamma_1-\cos^2\gamma_2
+
\tfrac12
\cos\delta_2
\sin2\gamma_1\sin2\gamma_2
\\\nonumber
&=
\tfrac12
(2\cos^2\gamma_1-1)(2\cos^2\gamma_2-1)
+
\tfrac12
+
\tfrac12
\cos\delta_2
\sin2\gamma_1\sin2\gamma_2
\\\label{cos-gamma}
&=
\tfrac12
(\cos2\gamma_1\cos2\gamma_2
+ 1
+
\cos\delta_2
\sin2\gamma_1\sin2\gamma_2)
\end{align}
We get
\begin{align}
%  \cos c 
%&=
\cos 2\alpha
&=
2\cos^2\alpha-1
%\\&
=
\cos 2\gamma_1 \cos 2\gamma_2 + \sin 2\gamma_1 \sin 2\gamma_2 \cos
\delta_2
\label{cos-2gamma}
\end{align}
from
\eqref{cos-gamma}.

On the other hand, we compute the spherical distance $c$ between
the two points in the base sphere corresponding to
 $(\gamma_1,0)$ and $(\gamma_2,\delta_2)$.
It is the third side of a spherical triangle with angle
$\delta_2-0=\delta_2$ at the north pole and sides
$a=2\gamma_1$ and
$b=2\gamma_2$.
By the spherical cosine law,
\begin{align*}
  \cos c
&= 
\cos a \cos b + \sin a \sin b \cos \delta_2
\\
&= 
\cos 2\gamma_1 \cos 2\gamma_2 + \sin 2\gamma_1 \sin 2\gamma_2 \cos \delta_2
\end{align*}
From this and~\eqref{cos-2gamma}, we conclude that $c=2\alpha$.

Therefore, the angular distance between $C$ and $D$ after applying
the Hopf map is $2 \alpha$.
% for any $C$,$D$ at angle $(\alpha,\alpha)$.
\end{proof}

We have defined the Hopf map with respect to an arbitrarily chosen
great circle $C$ from an invariant family $\Gamma$.
The previous lemma shows that this choice is not essential.
Choosing a different circle $C' \in \Gamma$ will lead to a different
Hopf map, but the two images are related by an isometry of the
2-sphere.

% \begin{corollary}
% \label{cor:rrhopf}
% The images of a set of great circles $\mathcal{S}$ in a 3-sphere 
% under a right Hopf map are independent of choices of a great circle or a
% Cartesian coordinate system. The right Hopf bundle is unique up to 
% rotations and reflections.
% \end{corollary}
% \begin{proof}
% A map from one image of $\mathcal{S}$ to another image  is
% an isometry since it preserves geodesic distances between pairs of points
% in the image of $\mathcal{S}$. 
% \end{proof}

%As a final remark, when we regard two great circles are related
%if two circles in a 3-sphere are fibers of the same right Hopf
%fibration, this relation is transitive. 

The following theorem summarizes the relations of right pairs and
isoclinic rotations. The analogous statement is valid for left pairs
as well.

\begin{theorem}
	\label{thm:right}
	For a pair of great circles $C$ and $D$, 
	$C$ and $D$ is a right-isoclinic pair % at angle $\alpha,\alpha$ 
	if and only if $D$ is an orbit under all right rotations that
			leave $C$ invariant.
\end{theorem}
\begin{proof}

%	[\ref{item-pair} $\Leftrightarrow$ \ref{item-orbit}.]
	By Lemma~\ref{lem:uhopf} and Lemma~\ref{lem:chopf}, there
	exists the family $\mathcal{F}$ of right
	rotations that leave $C$ invariant, and orbits of $\mathcal{F}$ are
	great circles of a 3-sphere. By Lemma~\ref{lem:hopf}, the set of
	these orbits is equivalent to the set of fibers in the (unique)
	right Hopf bundle containing
	$C$. Thus, a circle $D$ is an orbit of $\mathcal{F}$ if and only if
	$C$ and $D$ are in the common right Hopf bundle.
	Lemma~\ref{lem:alpha} implies that $C$ and $D$ are indeed in the
	common right Hopf bundle.
\end{proof}

%% file: ClosestPair.tex
\section{Closest-Pair Graphs}
\label{sec:clograph}
The closest-pair graph $G$ of a given point set $P$ is a graph $G$ such that
each point in $P$ is a vertex of $G$ and a pair of vertices $p$ and
$q$ in $P$ forms an edge if and only if the distance between $p$ and
$q$ achieves the minimum distance among all the pairs in $P$.
The closest-pair graph should not be confused with the nearest-neighbor graph, a
directed graph where an edge from $p$ to $q$ indicates that $q$ is a
nearest-neighbor of $p$.
%  In particular, a nearest-neighbor graph is a
% directed graph, since the nearest-neighbor relation is not symmetric
% and each vertex of a nearest-neighbor graph is of strictly-positive
% outdegree. On the other hand, a closest-pair graph is undirected and some
% vertices of a closest-pair graph may have zero degree.
The closest-pair graph may have isolated vertices, but it must contain
at least one edge.

The closest-pair graph has the following properties:
\begin{enumerate}
\item % (1)
 It can be
constructed in time $O(n \log n)$ by divide and conquer~\cite{SH}
in any fixed dimension,
\item % (2)
 the
degree of each vertex is bounded by the ``kissing number'', 
as described earlier in 
Section~\ref{sec:packing}, and consequently
\item % (3)
 the number of edges is at most 
linear in the number of vertices.
\end{enumerate}

By taking advantage of these characteristics, closest pairs were
already used in previous work~\cite{Atk,BK}, see Section~\ref{sec:3d}. 
We will compute closest-pair graphs not only for the
input point sets in 4-space, but also for sets of planes in Pl\"ucker
space, see Section~\ref{sec:pluecker}.

%% file: Coxeter.tex
\section{The Coxeter Classification of Four-Dimensional Point Groups}
\label{sec:coxeter}

The new algorithm treats the case that a given point set can be
generated by a series of reflections as a special case.
We use the classification of Coxeter groups
to argue that the
cardinalities of point sets that have a high degree of symmetry are bounded.
To this end, this
section provides the classification of such groups in 4-space and
shows the computation of inradii of fundamental regions of such groups. 

%\subsection{Coxeter Groups in 4-space and Fundamental Regions}
The discrete groups generated by mirror reflections
 have been classified in arbitrary dimension,
cf.\
\cite[Table IV on p. 297]{coxeter}.
Each of these groups of rank $d$ can be represented by a collection of mirrors
$r_1, \ldots, r_d$ such that $r_i \cdot r_i =1$ and 
$(r_ir_j)^{m_{ij}}=1$ for $i\neq j$ and $m_{ij} \geq 2$; such a group is called a
Coxeter group. The value of $m_{ij}$ specifies the dihedral angle
between mirrors $r_i$ and $r_j$ as $\pi/m_{ij}$.
The Coxeter diagram, or the Coxeter graph, explicitly encodes these values of $m_{ij}$.

In four dimensions, Coxeter groups were first enumerated by \'Edouard Goursat in 1899:
 There are five \emph{irreducible groups},
which are the symmetry groups of the five regular 4-dimensional
polytopes, called
% COXETER, p. 196, and Table IV on p. 297
% wikipedia
% http://en.wikipedia.org/wiki/Point_groups_in_four_dimensions
% Conway and Smith, On Quaternions and Octonions, 2003 Chapter 4,
% section 4.4 Coxeter's Notations for the Polyhedral Groups
%
% using "alternate symbols" from http://en.wikipedia.org/wiki/Coxeter_group
$A_4, C_4, B_4, F_4$ and $G_4$ according
to~\cite{coxeter}
(or alternatively $A_4$, $BC_4$, $D_4$, $F_4$, and
$H_4$ in today's terms).
Here, $A_m$ denotes a group related to reflectional symmetries
of an $m$-simplex. Similarly, $C_m$ is for an $m$-hypercube and
an $m$-cross-polytope, $B_m$ is for an $m$-demihypercube. $F_4$
represents the reflectional symmetry group for a 24-cell and $G_4$ is
for a 120-cell and a 600-cell. 

In addition, there are the \emph{reducible} groups, direct products of
lower-dimensional reflection groups. They are the groups
$A_3\times A_1$,
$C_3\times A_1$, %$BC_3\times A_1$,
$G_3\times A_1$, %$H_3\times A_1$,
and
$D_2^p\times D_2^q$, for $p,q\ge 2$,
where $G_3$ is the symmetry group of the icosahedron and the
dodecahedron,
$C_3$ is the symmetry group of the cube and the octahedron,
and
$D_2^p$ is the dihedral group of order $2p$
 (alternatively denoted as $I_2(p)$), the symmetry group of the
regular $p$-gon.
Some groups have alternative representations: $D_3 = C_3$,
$D_2^2=A_1\times A_1$, $D_2^3=A_2$, and $D_2^4=C_2$.
For reducible groups, their mirrors fall into two or
more classes such that each mirror in one class is perpendicular to
the mirrors in the other classes.  Since reflections at perpendicular
mirrors commute, this is how reducible groups yield the decomposition into a direct
product of smaller groups.

The arrangement of all mirror hyperplanes of a reflection group
cuts the 3-sphere into equal cells, which can be taken as the
\emph{fundamental regions} of the group. 
These fundamental regions
are not necessarily equal to the Voronoi regions of the point set;
the Voronoi regions are usually cut into smaller cells by
mirrors passing through the centers of the
regions.
It is known that the fundamental region of
a reflection group is a spherical \emph{simplex}
%see
%COXETER
\cite[Theorem 11.23]{coxeter}.
% By a theorem of Cartan (1928)
Thus, in 4-space, the fundamental region is a spherical tetrahedron
$T$, and the group is generated by four independent reflections.

The inradii of fundamental regions of Coxeter groups are of our
particular interest for
Section~\ref{sec:mirror}. 

\subsection{The Radius of an Inscribed Sphere of a Fundamental Region}
\label{sec:inrad}

By looking at the Coxeter diagrams (Table~\ref{tab:coxd} and~\cite[Table IV on p.~297]{coxeter}),
we can read off the dihedral angles between bounding mirrors of fundamental
regions of Coxeter groups by the following conventions.
Each vertex in a Coxeter diagram represents a mirror. A pair of vertices
with no edge means a pair of perpendicular mirrors. An edge with no
number connects a pair of mirrors whose dihedral angle is $\pi/3$.
Edges with number $p$ represent a dihedral angle $\pi/p$.  

\begin{table}[phbt]
\centering
	\begin{tabular}{|| c | c | c || c || }
		\hline
		Groups & Coxeter Diagrams & Normal Vectors& $R_0$ \\
		\hline
		$A_4$ &
		\includegraphics[page=1,scale=0.6]{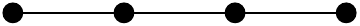}
		& $\begin{array}{c c c c} 
		(1, & 0, & 0, & 0) \\ 
		(-\frac12, & \frac{\sqrt{3}}{2}, & 0, & 0) \\
		(0, & -\frac{1}{\sqrt{3}}, & \frac{\sqrt{2}}{\sqrt{3}}, &
		0) \\
		(0, & 0, & -\frac{\sqrt{3}}{2\sqrt{2}}, &
		\frac{\sqrt{5}}{2\sqrt{2}})
	\end{array}$ & 0.2236067977 \\ \hline	
		$C_4$ &  \includegraphics[page=2,scale=0.6]{diagram}& 
		$\begin{array}{c c c c} 
		(1, & 0, & 0, & 0) \\ 
			(-\frac{1}{\sqrt{2}}, & \frac{1}{\sqrt{2}}, & 0, & 0) \\
		(0, & -\frac{1}{\sqrt{2}}, & \frac{1}{\sqrt{2}}, &
		0) \\
		(0, & 0, & -\frac{1}{\sqrt{2}}, &
		\frac{1}{\sqrt{2}})
		\end{array}$ & 0.1429000737 \\ \hline
		$B_4$ & \begin{tabular}{c} \\
		\includegraphics[page=3,scale=0.6]{diagram}
		\end{tabular} &
		 $\begin{array}{c c c c} 
		(1, & 0, & 0, & 0) \\ 
		(-\frac12, & \frac{\sqrt{3}}{2}, & 0, & 0) \\
		(0, & -\frac{1}{\sqrt{3}}, & \frac{\sqrt{2}}{\sqrt{3}}, &
		0) \\
		(0, & -\frac{1}{\sqrt{3}}, & -\frac{1}{\sqrt{6}}, &
		\frac{1}{\sqrt{2}})
		\end{array}$ & 0.1889822365 
		\\ \hline 
		$F_4$ & \includegraphics[page=4,scale=0.6]{diagram} & 
		 $\begin{array}{c c c c} 
		(1, & 0, & 0, & 0) \\ 
		(-\frac12, & \frac{\sqrt{3}}{2}, & 0, & 0) \\
			 (0, & -\frac{\sqrt{2}}{\sqrt{3}}, & \frac{1}{\sqrt{3}}, &
		0) \\
		(0, & 0, & -\frac{\sqrt{3}}{2}, &
		\frac12)
		\end{array}$ & 0.009671356812 \\ \hline
		$G_4$ & \includegraphics[page=5,scale=0.6]{diagram} & 
		$\begin{array}{c c c c} 
		(1, & 0, & 0, & 0) \\ 
			 (-\frac{1+ \sqrt{5}}{4}, & \frac{\sqrt{10-
	 2\sqrt{5}}}{4}, & 0, & 0) \\
			 (0, & -\frac{2}{\sqrt{10-2\sqrt{5}}}, &
			 \frac{\sqrt{6-2\sqrt{5}}}{\sqrt{10-2\sqrt{5}}}, &
		0) \\
		(0, & 0, &
		-\frac{\sqrt{10-2\sqrt{5}}}{2\sqrt{6-2\sqrt{5}}}, &
		\frac{\sqrt{14-6\sqrt{5}}}{2\sqrt{6-2\sqrt{5}}})
	\end{array}$ 
	& 0.03910328003 \\ \hline	
		$A_3 \times A_1$ & \includegraphics[page=6,scale=0.6]{diagram}& 
		$\begin{array}{c c c c} 
		(1, & 0, & 0, & 0) \\ 
		(-\frac12, & \frac{\sqrt{3}}{2}, & 0, & 0) \\
		(0, & -\frac{1}{\sqrt{3}}, & \frac{\sqrt{2}}{\sqrt{3}}, &
		0) \\
		(0, & 0, & 0, & 1)
	\end{array}$  
		& 0.3015113445
		\\ \hline
		$C_3 \times A_1$ & \includegraphics[page=7,scale=0.6]{diagram}&
		$\begin{array}{c c c c} 
		(1, & 0, & 0, & 0) \\ 
			(-\frac{1}{\sqrt{2}}, & \frac{1}{\sqrt{2}}, & 0, & 0) \\
		(0, & -\frac{1}{\sqrt{2}}, & \frac{1}{\sqrt{2}}, &
		0) \\
		(0, & 0, & 0, & 1)
	\end{array}$ & 0.2108874992 \\ \hline
		$G_3 \times A_1$ & \includegraphics[page=8,scale=0.6]{diagram}& 
		$\begin{array}{c c c c} 
		(1, & 0, & 0, & 0) \\ 
			 (-\frac{1+ \sqrt{5}}{4}, & \frac{\sqrt{10-
	 2\sqrt{5}}}{4}, & 0, & 0) \\
			 (0, & -\frac{2}{\sqrt{10-2\sqrt{5}}}, &
			 \frac{\sqrt{6-2\sqrt{5}}}{\sqrt{10-2\sqrt{5}}}, &
		0) \\
		(0, & 0, &
		0, &
		1 )
	\end{array}$ & 0.1303737577
		\\
		\hline
	\end{tabular}
	\caption{Coxeter diagrams, outward unit normal vectors of fundamental
	regions, %up to congruence, 
and inradii $R_0$ of fundamental regions
of four-dimensional Coxeter groups.} %  of finite families.}
\label{tab:coxd}
\end{table}

Among all four-dimensional Coxeter groups,
we are interested in the finite list, excluding the infinite family
$D_2^p\times D_2^q$; that
is, $A_4, C_4, B_4, F_4, G_4, A_3 \times A_1, C_3 \times A_1$, and $G_3
\times A_1$. Given the dihedral angles, normal vectors of the vertices of
a spherical tetrahedron $T$ can be computed:
 when the dihedral angle between two mirrors $i$ and $j$ is
$\pi/m_{ij}$, $$\hat{n}_i \cdot \hat{n}_j = \cos(\pi - \pi/m_{ij}) =
-\cos(\pi/m_{ij})$$
where $\hat{n}_i, \hat{n}_j$ are unit normal vectors to mirrors $i$ and $j$ and $\cdot$
is an inner product.
% up to congruence mapping.

\begin{figure}%{r}{0.4\textwidth}
	\centering
	\includegraphics[scale=0.6]%[width=0.8\textwidth]
{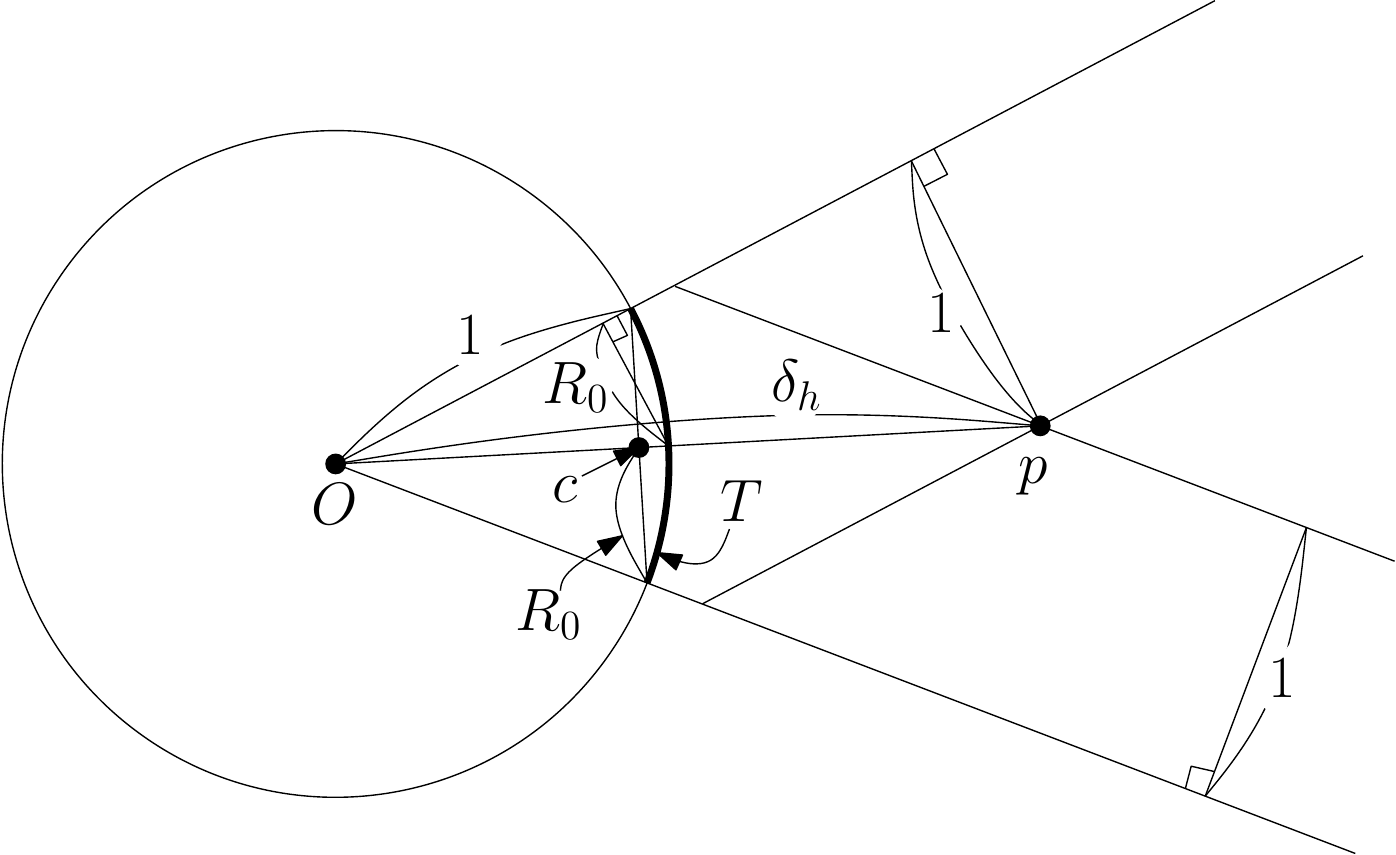}
	\caption{Finding the radius $R_0$ of an
	inscribed 0-sphere in a spherical 1-simplex $T$.
This two-dimensional sketch of the situation holds analogously in
higher dimensions.
 }
	\label{fig:insrad}
\end{figure}
Table~\ref{tab:coxd} enumerates such normal vectors
for the above groups.
The unit normal vector of the $i$-th bounding mirror in a diagram is in the
$i$-th row. For example, for $A_4$, the first two bounding mirrors form an
angle $\pi/3$ and the corresponding unit normal vectors in the outward
directions from the fundamental region form an angle $\pi- \pi/3$.
That is why the inner product of the first two normal vectors for
$A_4$ is
$\cos(2\pi/3) = -\cos(\pi/3)=-1/2$.

We call the distance from the center of the inscribed sphere of a
spherical tetrahedron $T$ to the
mirrors \emph{the Euclidean radius} of $T$. 
We
 compute the Euclidean radius $R_0$ of each
 spherical tetrahedron 
 $T$
that is a fundamental region of
one of above groups as follows.

Each bounding mirror $r_i$ of $T$ spans a hyperplane $H_i$ going through the
origin for $i=1,2,3,4$. After translating each $H_i$ in the orthogonal
direction of $H_i$ toward the interior of the fundamental region by
distance 1, % for $i=1,2,3,4$, 
let $p$ be the intersection point of
these translated hyperplanes. % $H_1,H_2,H_3$ and $H_4$. 
This point has equal distance 1 from all hyperplanes, but
it does not lie on the unit sphere.
Let $\delta_h$ be the distance of $p$ from the
origin $O$.
Then, by rescaling the segment $\overline{Op}$ by $1/\delta_h$ we get the center
$c$ of the
inscribed sphere, and the Euclidean radius of $T$ is
 $R_0 = 1/\delta_h$, see Figure~\ref{fig:insrad}.
%$$ 1: R_0 = \delta_h:1. $$
%Therefore, $R_0 = 1/\delta_h$.
The approximated values of the Euclidean radii $R_0$ of fundamental
regions of 4-dimensional finite Coxeter groups are computed in this
way. These values are
given in
Table~\ref{tab:coxd}.

The smallest Euclidean radius occurs for $G_4$,
and its value is $R_0 \ge 0.039102328.$

We are interested in this radius for the following reason.
Our algorithm will build the closest-pair graph $G$ of a set of points
in $\sa$. One special case arises when every edge $uv$ of $G$ acts as a
mirror, exchanging $u$ and $v$ together with all their neighbors
in $G$. From this it follows that every component of $G$ is an orbit
of some point $u_0$
under the reflection group \emph{generated} by the mirrors perpendicular to
the edges incident to $u_0$. The above considerations allow us to
bound the closest-pair distance of $G$ in such cases:

Thus, the minimum distance $2R_0 \ge 0.0782$ as it is claimed in the
following lemma. 

\begin{lemma}\label{exclude-finite}
  Consider the orbit $I$ of a point $u_0$ on the unit sphere $\sa$
  under a finite four-dimensional symmetry group $H$ which is not
  $D_2^p\times D_2^q$.  Let $\delta$ be the minimum distance $\delta$
  between two points of~$I$, and suppose that the group $H$ is generated
  by the mirror reflections between $u_0$ and each point $v\in I$ at
  distance $\delta$ from $u_0$.
  Then $\delta$ is at least $\delta_1 := 0.07$.
% \textcolor{red}{OLD:}
% If every component of a given geometric graph $G$
%  is
% the orbit of a point under a four-dimensional symmetry group which is
% not  $D_2^p\times D_2^q$, then
% the minimum distance $\delta$ is at least $0.07 > \delta_0$.
% %
% %OLD: then $G$ has at most $n_0$ vertices.
\end{lemma}

\begin{proof}
%The arrangement of all mirror hyperplanes of a reflection group
%cuts the 3-sphere into equal cells, which can be taken as the
%\emph{fundamental regions} of the group. 
%(These fundamental regions
%are not necessarily equal to the Voronoi regions of the point set, as
%the Voronoi regions are usually cut into smaller cells by
%mirrors passing through the centers of the
%regions.)
%It is known that the fundamental region of
%a reflection group is a spherical \emph{simplex},
%see
%COXETER
%\cite[Theorem 11.23]{coxeter}.
% By a theorem of Cartan (1928)
%Thus, in 4-space, the fundamental region is a spherical tetrahedron
%$T$. % and the group is generated from just four reflections.
We have a finite list of possible groups $H$, and for each such group,
we know the shape of its fundamental region.
For each fundamental region $T$, we consider all possibilities
 how the orbit of a point $u_0\in T$ might give rise
the orbit~$I$.
We have to place $u_0$ %in $T$ 
on some facets $F$ of $T$ and
equidistant from the remaining facets of $T$.
%, and generate the orbit of this point.
Otherwise, if the point were not chosen equidistant from the other
faces, the closest-pair graph would not contain edges that
generate all four mirrors.
% span all four dimensions.
The minimum distance is achieved when
 $F=\emptyset$ and $u_0$ lies in the interior of~$T$.
We have analyzed each of the eight groups case by case in
Table~\ref{tab:coxd}, and we have seen that
the minimum distance $\delta$ is $2R_0 \approx 0.0782$.
\end{proof}

%% file: Packing.tex
\section{Packing Arguments and Definitions of Constants}
\label{sec:packing}
We often rely on packing arguments to bound the complexity of certain
configurations in situations where we know that the minimum distance
between points or the minimum Pl\"ucker distance between planes is
bounded from below and we want to show that the number of points or
planes is bounded from above.

%By using volume packing, we can easily establish the following lemma.

\begin{lemma}
	\label{lem:volume}
	If there are $m$ points on a $d$-sphere $\sphere^d$ with minimum
	Euclidan distance $\delta$, then 
	$$m \cdot \kappa_d(\delta/2)^d \leq \omega_d$$
	where $\kappa_d$ is the volume of the Euclidean unit $d$-ball
 and $\omega_d$ is the surface area of the
	unit $d$-sphere. % $\sphere^d$.
\end{lemma}
\begin{proof}
The geodesic distance between the points is at least
$\alpha := 2\arcsin \frac\delta2$.
Thus, if we draw a spherical $d$-ball of radius $\alpha/2$ around the
 $m$ points, these balls form
 a packing on the surface of
the $d$-sphere.
The boundary of such a ball is a $(d-1)$-sphere of Euclidean radius
$\sin \frac\alpha2 = \frac\delta2$
(see Figure~\ref{fig:insrad} for a similar situation).
The $d$-dimensional volume of the spherical ball must be at
least
the volume of the (Euclidean) $d$-ball with the same boundary,
which is $\kappa_d(\delta/2)^d$.
\end{proof}

We will now define a few constants that are used in the new algorithm:

\subsection{Kissing Numbers}

% We use \emph{kissing numbers} in various dimensions. The kissing
% number is the maximum number of non-overlapping equal balls that touch
% another ball of the same size. 

%For packing arguments, 
We denote 
by $K_d$ the kissing number on the $d$-dimensional sphere $\sphere^d$:
 the maximum number of equal interior-disjoint balls on $\sphere^d$ that can
simultaneously touch a ball of the same size.
We use the known bounds
$K_2=5$,
$K_3=12$, 
and $40\le K_5 \le 44$.

These bounds on $K_d$ are derived from the corresponding kissing numbers in
Euclidean space: These kissing number are known
to be 6 in the plane, 12 in 3-space, 24 in 4-space, and 
between 40 and 44 in 5-space. 
 (On the 2-sphere, the kissing number is
smaller than in the plane.)
In the closest-pair graph, the degree of every vertex is obviously
bounded by these kissing numbers.

%  $K_5$ is the kissing number on the surface of a 5-sphere
% 	$\mathbb{S}^5$.
% Since the kissing number of
% 	5-space is between 40 and 44, this number is less than or 
% 	equal to 44. We defining $K_2$ similarly on the 2-sphere, and it is
% 	known that $K_2 \le 5$.
% On the 3-sphere, the 
%  kissing number is $K_3= 12$.
\subsection{The Closeness Threshold}
 We define the constant $\delta_0 := \DELTA0$ for applying 1+3 dimension
	reduction. If the minimum distance of a point set is greater
	than $\delta_0$, the size of the point set is at most $n_0$,
	where $n_0$ is the maximum packing of 3-balls of radius $\delta_0/2$ 
	on a unit 3-sphere $\sa$.
	Then, we can apply the dimension reduction principle in
	Section~\ref{sec:1+d_red} without affecting the time complexity.
	Otherwise, we need to go through other condensing procedures.
	The value of $\delta_0$ will be justified in
	Lemma~\ref{lem:leng} in Section~\ref{sec:orbit}.

	% If a 3-ball of radius $\delta_0$ is
%	projected to a unit 3-ball, its volume is greater than a
%	3-ball of radius $\delta'_0$ where 
%	$$\delta'_0 = 2\sin (\arctan(\delta_0/2)).$$
	Since the surface area $\omega_3$ of $\sa$ is
	$2\pi^2$ and the volume $\kappa_3$ of the unit 3-ball is
	$\frac43 \pi$, by Lemma~\ref{lem:volume},
	a rough estimate of the upper bound
 $n_0$ is 
\begin{equation}
 \label{n0}
n_0 := \lfloor 2 \pi^2 / (\tfrac43 \pi (\delta_0/2)^3) \rfloor <
3.016 \times 10^{11}.  
\end{equation}

% 	\iffalse
% 	is a lower bound on the distance that a closest pair can have
% 	when an orbit path is of length less than 12000. By
% 	Lemma~\ref{lem:leng}, every orbit path is of length $\ell \geq
% 	\pi/\arcsin (\delta/2)$ when $\delta$ is the distance of a
% 	closest pair. From 
% 	$$ 12000 \geq \ell \geq \pi/\arcsin(\delta_0/2), $$
% 	$$ \delta_0 \leq 2 \sin \frac{\pi}{12000} \approx 5.2359 \times
% 	10^{-4}. $$
% 	\fi
% %	This value is less than the side length $s$ of a $p$-gon where
% %	$p< 40 K_5$ and the circumradius of the $p$-gon is
% %	$1/\sqrt{2}$. Note that 
% %	$$ s= \sqrt{2} \sin \frac{\pi}{40 K_5} (> \delta_0).$$

\subsection{The Icosahedral Threshhold}
\label{icosahedral}
In Algorithm M, we will need the maximum number
 $C_1$ of planes at pairwise 
	Pl\"{u}cker distance greater than or equal to a certain distance $\delta_{\min}$.
The distance $\delta_{\min}$ is % defined as follows. 
the Pl\"ucker distance of two isoclinic planes $P,Q$
% is $\delta_{\min}$ if $P,Q$
that
 are mapped to two adjacent vertices of a regular icosahedron by the Hopf
map.

By Lemma~\ref{lem:pdist}, 
the Pl\"ucker distance between two planes of
angle $\alpha,\alpha$ is $\sqrt{2}\sin \alpha$ and by
Lemma~\ref{lem:lrhopf},
such two planes are mapped to a pair of points of geodesic distance
$2\alpha$ by the Hopf fibration.
The edge length of an icosahedron with a circumscribed unit sphere
is $\delta_{\textrm{ico}}=\sqrt{50-10\sqrt{5}}/5$, so the geodesic distance of two vertices
in a sphere is 
$$2\arcsin(\delta_{\textrm{ico}}/2) \approx 1.107148718,$$ 
and the angle between two such
planes is $$ \alpha_{\min} = \arcsin(
\delta_{\textrm{ico}}/2) \approx 0.5535743590.$$
The Pl\"ucker distance of such pair of planes is 
$$\delta_{\min} = \sqrt{2} \sin (\arcsin(
\delta_{\textrm{ico}}/2)) \approx 0.7434960688.$$
Each plane in $\mathbb{G}(2,\mathbb{R}^4)$ is represented by two
antipodal points on the
%with Pl\"{u}cker distances can be
%regarded as a double cover of the 
5-sphere. 
%the upper bound can be estimated by dividing the surface area of
%the unit half 5-sphere by the volume of the 5-ball of radius
%$\delta'_{\min}/2$. %where $\delta'_3$ is the projected radius of the 5-ball of
%radius $\delta_3$ to the unit 5-hemisphere.
%From
%$$ \delta'_3 = 2\sin (\arctan(\delta_3/2)),$$
By Lemma~\ref{lem:volume}, the upper bound can thus be computed as
$$2C_1 \le 2\pi^3 / (\tfrac{8}{15} \pi^2 (\delta_{\min}/2)^5),$$ since the surface
area $\omega_5$ of the unit 5-sphere is $2\pi^3$ and the volume
$\kappa_5$ of a unit 5-ball
is $\frac{8}{15} \pi^2$.
This gives
\begin{equation}
\label{c1}
C_1 \leq \PACK-F. .
\end{equation}
This bound is a rough estimate, and the correct bound is likely to be
much smaller, for two reasons:
(i)  Lemma~\ref{lem:volume} uses only a crude volume argument.
(ii) The Plücker coordinates cannot lie anywhere on the 5-sphere, but
they are restricted to a 4-dimensional manifold, the Plücker quadric.
Unfortunately, this does not directly allow us to apply
 Lemma~\ref{lem:volume} for $d=4$ dimensions, because the Plücker
 surface is negatively curved, and moreover, we don't even know its volume.

%Thus, it would be more appropriate to use the 
 Packings in Grassmannian spaces were
considered by Conway, Hardin and Sloane~\cite{CHS}. They used the
\emph{chordal distance}, defined as $\sqrt{\sin^2 \alpha + \sin^2
\beta}$ for a pair of planes at angle $\{\alpha,\beta\}$
% to provide the experimental bounds
 (see Section~\ref{Other-Distances}).
% for comparisons with Pl\"ucker distance.
%Pl\"ucker distance, $\sqrt{2}(1- \cos \alpha
%\cos \beta)^{1/2}$ for a pair of plane at angle $\{\alpha,\beta\}$, is
%defined differently from the chordal distance but these distances become the same
%when a pair of planes is isoclinic. 
According to their experimental
results, for planes in 4-space, when there are 37 planes, the maximum
chordal
distance that they achieve is about 0.728633689875024.
For isoclinic planes, in which we are interested here,
 the chordal
distance coincides with the Pl\"ucker distance.
This suggests that  the true bound
 $C_1$ is near 37, much smaller than the bound~\eqref {c1}.

%\section{Representation Details}
%\label{sec:rep}

%% file: Canonical.tex
\section{Canonical Axes}
\label{sec:cano}

The {canonical-axes} construction
is
 a well-known procedure for detecting the rotational symmetries in a
planar point configuration~\cite{Man,Ata,Hig}.
We use this construction in Step C11 of Algorithm C in
Section~\ref{sec:iter}.
We can encode labeled-point sets on a circle
as a cyclic string. This can be done by alternating between labels
for a point and angular distances between two adjacent
points~\cite{Man}. 

In addition, by making cyclic shifts in this string so that 
the string becomes lexicographically smallest, but
still starts with labels (not angles), we can get a unique
representation of points on a circle with labels.
These starting points that yield the same string
give rise to a set of $p$ equidistant rays starting from
the origin. We call a collection of these points  \emph{canonical axes}. 
Canonical axes can be found in $O(n \log n)$ time 
when $n$ is the number of labeled points by standard string-processing
techniques~\cite{KMP}.
Canonical axes have the same
rotational symmetries as the
original configuration. 

We assume labels are preserved under
rotations.
Then, if a circle $K$ is mapped to a circle $K'$, canonical axes of $K$ are
mapped to canonical axes of $K'$, so canonical axes are 
equivariant under rotations. 

Moreover, canonical axes are more special
than most other equivariant criteria. If canonical axes of
a circle $K$ are mapped to canonical axes of another circle $K'$ by a
rotation $R$, $K'$ itself is also mapped to $K$ by $R$, so the converse of
equivariance is also true. 

In this context, we define a canonical set and a
canonical set procedure as a special condensing. Refer
to Section~\ref{sec:torus} about a canonical set procedure.
Canonical axes are the result of a canonical set procedure
for a circle and rotational symmetries on a circle.
The new algorithm employs a canonical set procedure 
for a square torus and translational symmetries on a torus. 
See Section~\ref{sec:torus}.

%% file: Congruence_Types.tex
\section{Congruence Types of Vertex and Edge Figures}
\label{sec:congtype}

Assume that a geometric graph $G$ is given.
A vertex figure of a vertex $v$, that is, $v$ together with its neighbors,
is first introduced in Atkinson's algorithm~\cite{Atk}
(see \cref{sec:3d}). Atkinson argued that a vertex figure can be
encoded in a string of length $O(1)$ if the vertex is of bounded
degree, but he omitted the details about representing a vertex figure.
Our new algorithm uses vertex figures and also 
introduces an extended concept called an edge
figure of an edge $uv$, that is, all neighbors of $u$ and $v$ 
together with the edge $uv$.

We now describe the representation details of a vertex figure, 
and of an edge figure.
We assume that a given directed graph $G=(V,E)$ is embedded in the 3-sphere $\sa$
such that all the edge lengths are the same and $G$ has its maximum
degree at most 12. This bound follows from the fact that
$G$ is a (directed) subgraph
 of a closest-pair graph on the 3-sphere, and
the kissing number on the 3-sphere is 12.

Let us begin with a vertex figure.
If a vertex $v$ is of degree 0 or 1, there is only one congruence type
of the neighborhood of $v$.
If a vertex $v$ is of degree 2, the angle between the two incident edges
determines its congruence type.

If a vertex $v$ has degree greater than 2, we can choose two distinct incident edges $va$ and $vb$ such
that the angle $\angle avb$ attains the minimum among all pairs of
such edges.
We order $a$ and $b$ and call these two edges a base pair.
Observe that the vectors $vO$ ($O$ is the origin), $va$, $vb$ are not
coplanar.
Otherwise, $v,a,b$ are in a great circle, $v$ has only $a$ and
$b$ as its neighbors, and $v$ is of degree two.

Let $n_1,n_2,n_3$ be the vectors obtained by
applying the Gram–Schmidt orthogonalization process to the ordered vectors $vO,va,vb$.
Extend these vectors to a positively oriented orthonormal basis
$n_1,n_2,n_3,n_4$, and use this basis
% We use $n_1,n_2,n_3,n_4$ 
as the basis vectors of a Cartesian coordinate system
with origin $v$. Sort
the neighbors
of $v$ lexicographically
by coordinates. Tag each coordinate with direction labels,
$+1$ for outgoing edges, $-1$ for incoming edges,
and $\pm1$ for bidirected edges.
The concatenated sorted string of coordinates with tags yields a string  corresponding to the base pair
$va,vb$.

We can obtain such strings for all possible base pairs.  We use the
lexicographically smallest string to represent the congruence type of
the vertex figure of the vertex $v$.

The congruence type of a directed edge $uv$ is defined similarly by
using a Cartesian coordinate system defined by some base pair.  In
this case, we choose the base pair in a way that it includes $uv$ and
another incident edge $vb$ which forms the minimum angle to $uv$.

These representations of vertex figures and edge figures
can be constructed in time $O(c^2 \log c)$ for
 degree $c$, and the length of the representations is
$O(c)$.
Since the maximum degree $c \le 12$, all these bounds are constant.
%because the maximum degree is bounded by the kissing number $12$ in $3$-space.

%% file: Dimension_Reduction.tex
\section{1+3 Dimension Reduction}
\label{sec:1+d_red}

%!!! NACH FORNE!!

In 4- or higher-dimensional space, Alt et al.~\cite{Alt} were the
first to use
\emph{dimension reduction}. The corresponding dimension reduction method for
our algorithm is more a variant of one in Akutsu~\cite{Aku}.
%although Akutsu's method is also similar to the method by Alt et
%al. Their differences and details was described earlier in Section~\ref{sec:hd}.

We pick an arbitrary point $a_0\in A$.
This point must be mapped to one of the points in $B$. 
For any $b\in B$,
we try to rotate $A$ so that $a_0$ lies on
$b$. To look for rotations $R$ that leave this point fixed,
we project $A$ and $B$ 
on the hyperplane $H$ perpendicular to $Oa_0=Ob$.
To each projected point,
we attach the signed distance from $H$ as a label. 
A rotation in $H$ that preserves labels can be extended to a rotation
$R$ that fixes the point $a_0$.
For each $b\in B$,
the problem is thus
reduced to the one-lower dimensional congruence testing problem for labeled
point sets.  

The new algorithm uses this method for $A$ and $B$ in 4-space 
in the following situation.
Suppose that we know pruned sets $A_0 , B_0$ obtained from $A$ and $B$ such that
$|A_0|\leq n_0$ and $A_0$ is mapped to $B_0$ by the pruning principle
where $n_0$ is the constant defined in~\eqref{n0}.
Then by limiting
the choice of $a_0$ in $A_0$  and the choice of $b$ in $B_0$,
we can check the congruence of $A$ and $B$ in time $O(n_0 n \log n) =
O(n \log n)$,
by known methods for 3-space.

%% file: Pruning.tex
\section{The Pruning and Condensing Principles}
\label{sec:pruning}

The pruning principle plays a very important role in our
algorithm. 
%It was mentioned before as well in
%Section~\ref{sec:model}.
The general scheme of \emph{conventional pruning} is as follows.
\begin{compactenum}
\item Classify a set of points by the value of some function $f$.
%, called a criterion, on a point. % Here, $F$ is called a criterion.
\item Take only one class of points of the same value of a criterion, preferably 
	the smallest class, into consideration and \emph{temporarily} ignore other sets.
\end{compactenum}
We need to make sure that this pruning is
canonical, i.\,e., a criterion $f$ is \emph{equivariant under 
rotations};
this means that
%a criterion $F$ of $x \in A$ is equivariant under a rotation $R$ if 
$f(R\cdot x) =
R\cdot f(x)$
for all rotations $R$.

For example, 
given point sets $A$ and $B$,
if there is a point of non-zero distance to the origin in $A$, 
we can prune $A$ and $B$ by their distances to the origin as follows.
First, classify points by their distances to the origin,
count the number of points in each class, and temporarily focus only on 
the classes of the smallest cardinality $A_s, B_s$.
After pruning by distances to the origin, we may assume that 
all the points are in the same distance to the origin,
which means that
they are on the same sphere.
This is one of the main advantages of pruning. We can assume 
more structure on the point set after pruning.

Another advantage of pruning is that
it reduces the number of points
that we need to consider.
%We say we reduced the number of points by a $c$ factor 
%if the ratio of the number of new pruned points to 
%the number of points before pruning is $c$. 
%If $c \leq \frac12$,
If we can guarantee a size reduction
by a factor $c<1$, we say that we reduced
the number of points successfully. 
Then, pruning can be repeated until it gets stuck, 
without affecting the time
and space complexity. After pruning, we can restart the algorithm
from the beginning with the pruned set
without affecting the complexity of the algorithm,
since $O(n \log n) + O( cn \log cn) + O(c^2n \log c^2n) + \cdots 
= O(n \log n)$.

To assure that each 
pruning reduces the number of points by at
least a half,
we choose the set of the smallest cardinality.
To ensure that the results for $A$ and $B$ are identical, 
we
use some lexicographic rule for tie-breaking.
%We only have to apply the same rule for $B$.
We will
only say that we ``prune by criterion $f$'', without
explicitly mentioning a tie-breaking rule.

We generalize this principle to a more general method, called \emph{condensing}
while maintaining
the advantages of pruning:
We \emph{condense} a finite set $A$ to a nonempty set $A'=
F(A)$ of smaller size, not necessarily a subset of $A$, by using an
equivariant function $F$, that is, $F(R\cdot A) = R\cdot F(A)$
for any rotation matrix~$R$.
The condensing principle is not restricted to picking a subset
according to some criterion, but it can define the function $F$ in a
more general way.
For example,
when we have a perfect matching of the points $A$ that has been
constructed by geometric criteria,
we can replace $A$ by the midpoints of the edges.
Afterwards we only
consider these midpoints instead of the original set~$A$.

%Let $\Phi$ be the set of all the rotations.
%More precisely,
%pruning can be viewed as applying a function $F$ that maps a
%point set $S$ to a point set $S'$ of smaller cardinality
%in an equivariant way under
%rotations:
%	$$F(R\cdot S) = R\cdot S'.$$
%Note that neither $S$ nor $S'$ need to be an input point set.

Any rotational invariants, such as distances and angles,
can be used for designing a function $F$ for condensing. 
For example,
the new algorithm constructs
%also uses criteria that are not rotational invariants ???
% but equivariant mappings: 
canonical axes (Section~\ref{sec:cano}) and edge figures (Section~\ref{sec:congtype}).

At first glance, condensing and pruning looks dangerous because it \emph{throws
  away information}. This could introduce new symmetries, and 
  it might happen that the condensed/pruned sets are congruent, whereas
the original sets are not. 
Therefore, the condensed/pruned set should be kept only
temporarily. 
The prime goal of iterative condensing steps is to eventually
reduce point sets to small enough sets $A''$ and $B''$ 
so that we can afford \emph{dimension reduction}.
% Eventually we want
%to use dimension
%reduction techniques either in
%Section~\ref{sec:1+d_red} or Section~\ref{sec:torus}. 

Pruning and condensing are very powerful and versatile
% flexible
methods, because
we can use them with any %``local'' 
equivariant construction that one might think
of as long as it is not too hard to compute.
 They allow us to concentrate on the cases where condensing makes no progress, and these cases are highly structured and symmetric.  The
difficulty is to pick the right pruning/condensing criteria, and to decide how to
proceed when pruning and condensing gets stuck.

%\emph{%Remark: 
%Condensing and Pruning Convention.}
%We will mostly describe the condensing steps only for the set 
%$A$, but, whenever we apply any condensing on $A$, 
%we will apply the same condensing to $B$ in parallel. If $A$ and
%$B$ are congruent, $B$ will undergo exactly the same sequence of
% steps as $A$. If at any point, a
%difference manifests itself, 
%$A$ and $B$ are not congruent, and we can terminate.

%% file: Overview_Algorithm.tex
\section{Overview of the New Algorithm}
\label{sec:overview}

Finally, we introduce the new optimal algorithm in 4-space.
We first describe the flow of the new algorithm and 
the relation between the modules. 

\begin{figure}[hbt]
\centering
\includegraphics[width=\textwidth]{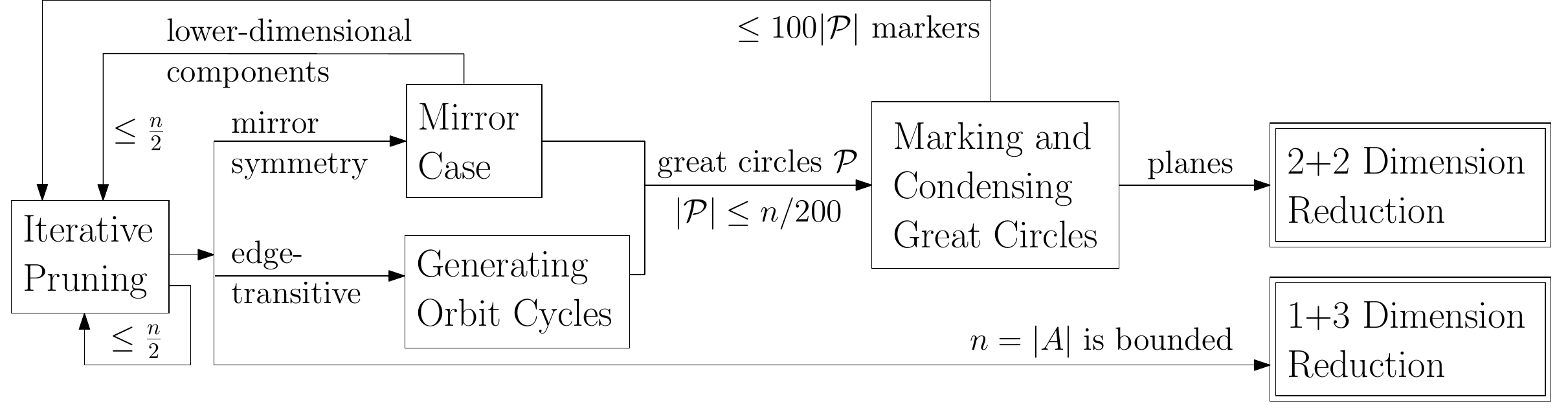}
\caption{The general flow of the algorithm.}
\label{fig:overview}
\end{figure}

As can be seen in Figure~\ref{fig:overview}, the new algorithm
consists of six main modules; iterative pruning, generating orbit cycles, 
marking and pruning great circles, the mirror case, 2+2 dimension
reduction and 1+3 dimension reduction.
The module called ``1+3 dimension reduction'' has already been presented
in Section~\ref{sec:1+d_red}. We explain the other five
modules in Sections~\ref{sec:iter}--\ref{sec:torus}.

\paragraph{Preliminaries.}
%In the problem,
%the translation vector $t$ can be eliminated
%by translating $A$ and
%$B$ so that their centroids become the origin.
%Furthermore,
%we %may
% only consider orthogonal matrices $R$ of determinant $+1$ (direct
%congruences). If we want to allow orthogonal
%matrices of determinant~$-1$ (mirror congruence),
%we repeat the algorithm with a reflected copy of~$B$.

We will declare the problem to be trivially solvable if the
closest-pair distance $\delta$ is large, i.e., $\delta
>\delta_0:=\DELTA0$.
This implies that
the input size $|A|$ is bounded by $n_0 < 3.016\times
10^{11}$,
and hence, by 1+3 dimension reduction, the problem can be reduced to
at most $n_0$ instances of 3-dimensional congruence testing, taking $O(n\log
n)$ time overall.

\emph{Remark.} Whenever we apply some procedure, e.\,g.,
condensing, on $A$, 
we will apply the same procedure to the
other set $B$ in parallel but we will mostly describe those steps only
for the set $A$. If $A$ and
$B$ are congruent, $B$ will undergo exactly the same sequence of
steps as $A$. If  a
difference manifests itself at any point, we know that
$A$ and $B$ are not congruent, and we can terminate.

\paragraph{General Flow.}
We first give a rough overview of our algorithm, omitting details and some
special cases. See Fig.~\ref{fig:overview}.

Our goal is to apply the condensing principle repetitively 
until we can apply either of the two dimension reduction principles to the
original input point sets (not to the pruned sets): 1+3 dimension reduction
(see Section~\ref{sec:1+d_red}) or 2+2 dimension reduction (see
Section~\ref{sec:torus}).  
The 1+3 dimension reduction makes a line and the orthogonal 3-space invariant, 
and the 2+2 dimension reduction makes
 two orthogonal 2-planes invariant. 

Whenever the number of pruned points 
is smaller than some chosen threshold $n_0$,
we can afford 1+3 dimension reduction as mentioned in the preliminaries.
If we finds a set $\plane$ of equivariant great circles from $A$
and $B$, such that $1\le |\plane|  \le C_1$, where $C_1=\PACK-F.$ is defined
in Section~\ref{sec:packing}, 
we can trigger 2+2 dimension reduction.

By applying the 2+2 dimension reduction technique described in Section~\ref{sec:torus},
the problem boils down to congruence testing 
of ``labeled'' points on a two-dimensional torus under
translations. This problem can be solved in $O(n \log n)$ 
by reducing a point set but still preserving symmetries and by using the
Voronoi regions as the neighborhood structure.

 The congruence testing algorithm begins
with pruning by distance from the origin. 
We may thus assume without loss of generality that the resulting set $A'$ lies on the unit 3-sphere 
$\mathbb{S}^3\subset \mathbb{R}^4$. % centered at the origin~$O$.

In iterative pruning (Section~\ref{sec:iter}),
we first check if
the minimum distance $\delta$ between points of $A'$ is bigger than the threshold
$\delta_0$.
If yes, we conclude that  $|A'|\le n_0$ and trigger 1+3 dimension
reduction.
Otherwise,
we construct the closest-pair graph $G$ on $A'$. 
Then we prune the edges of $G$ by congruence type of its edge figures.
%The 
An edge figure % of an edge 
consists of two adjacent vertices 
and all their neighbors.
We apply pruning with other criteria to the set $A'$ until the resulting set
$A_0$ cannot be further reduced.
Then, all edge neighborhoods
 in the graph~$G$ are congruent.
This allows us to find either \emph{orbit cycles} in~$G$,
%(see Section~\ref{sec:orbit})
 or a subset of $A_0$ with mirror
symmetry. % (see Section~\ref{sec:mirror}).
An orbit cycle is a cyclic path $a_1a_2\ldots a_{\ell}a_1$ % on $\sa$ 
with a rotation $R$ such that $Ra_i=a_{(i \bmod \ell)+1}$; in other
words, it is the orbit of the point $a_1$ under a rotation~$R$.

Mirror symmetry is the symmetry that, for each
edge of $G$, swaps the two endpoints and maps the whole
point set onto itself.
It implies that 
the point set $A_0$ must be related to one
of the regular four-dimensional polyhedra,
 or $A_0$ (and $G$) is the
Cartesian product of two regular polygons in orthogonal planes. This
follows from the classical classification of discrete
reflection groups.
The former case can be excluded, since $\delta< \delta_0$,
and in the latter case, we can proceed to 
Section~\ref{sec:plane} for marking and condensing great circles.
See Section~\ref{sec:mirror}.

Let us look at the case when we have found orbit cycles
 (Section~\ref{sec:orbit}) or a set of pairs of regular polygons in orthogonal planes
 (Section~\ref{sec:mirror}).
Geometrically, an orbit
cycle lies on a helix around a great circle $C$ and a pair of
orthogonal planes intersect
a sphere as a pair $C,C'$ of great circles.
Thus, we get a collection $\plane$ of great circles on $\mathbb{S}^3$.
We treat these great circles as objects in their own right, and
we construct the closest-pair graph on $\plane$.
 For this, we use the Pl\"{u}cker embedding of the corresponding
planes into $\mathbb{S}^5/\mathbb{Z}_2 = \mathbb{RP}^5$, mapping each
plane to a pair of antipodal points on the 5-sphere
$\mathbb{S}^5$. 
Then, we construct a closest-pair graph of planes with respect to
the Pl\"{u}cker distance, defined as a normalized Euclidean distance in
the Pl\"{u}cker embedding. 
Refer to Section~\ref{sec:pluecker} for the Pl\"ucker embedding and
Pl\"ucker distance.

For each closest pair $(C,D)$ of great circles in $\plane$, 
if they are not isoclinic (i.e., the projection of $C$ to the plane
containing $D$ is a ellipse but not a circle)
the major axis of this ellipse marks two points of $C$ 
as described in Paragraph ``Equivalent Definition by Orthogonal
Projections'' in Section~\ref{sec:angle}. 
The set of all markers replaces the set  $A_0$.
%and since the number
%of all the markers is a small fraction of $|A_0|$, 
This completes
a successful condensing step, and we restart and continue pruning as before.

Otherwise, all projected ``ellipses'' turn out to be circles. 
In this case, we can find a subfamily of great circles in
a special position: they must be part of a \emph{Hopf bundle} of
circles. The circles of a Hopf bundle can be mapped one-to-one to
points on the 2-sphere $\mathbb{S}^2$. 
See Section~\ref{sec:hopf}.
We can thus use 
the condensing procedures for a 2-sphere in three dimensions as in
Section~\ref{sec:prusphere}.  This
yields a small set $\plane$ of at most 12 great circles.
Then, we can apply the 2+2 dimension reduction technique.

The details for marking or condensing great circles are actually more complicated, 
since we might have a phase
in which $\plane$ is successively pruned, see Section~\ref{sec:plane}.

This concludes the summary of the algorithm.
The steps of the algorithm involve several different operations:
We need closest-pair graphs in 4 and 6 dimensions.
The closest-pair graph of $k$ points
can be calculated in $O(k\log k)$ time
in any fixed dimension,
by a classical divide-and-conquer approach~\cite{SH}.
We also compare a pair of edge figures in $\sa$ but this takes only time $O(1)$
because the maximum degree is at most 12.
We also need Voronoi diagrams in two dimensions and convex
hulls in three dimensions.
Finally, we need to sort lists of numbers lexicographically.
In summary, we will be able to reduce the size of the current point
set $A'$ by a constant factor less than $\frac12$,
 in $O(|A'|\log |A'|)$ time, until dimension reduction is possible.

Hence, we obtain the following theorem.
\begin{theorem}
\label{thm:main}
%Algorithm~\ref{algo:all} 
We can decide if two $n$-point sets $A$ and $B$ in 4-space are
congruent in $O(n \log n)$ time and $O(n)$ space.
%\qed
\end{theorem}
%\begin{theorem}
%\label{thm:main}
%Algorithm~\ref{algo:all} 
%We can decide if two $n$-point sets $A$ and $B$ in 4-space are
%congruent in $O(n \log n)$ time and $O(n)$ space.
%\qed
%\end{theorem}

%% file: Iter.tex
\section{Iterative Condensing Based on the Closest-Pair Graph: Algorithm
C}
\label{sec:iter}
%\label{closest-pair}
After pruning by the distance to the origin, 
we have a set $A\subset \reals^4$ of $n$ points with equal
distance from the origin. This set may already be the result of
some previous condensing steps. 
Without loss of generality, we may assume to be given a set $A$ on the unit sphere $\sphere^3$.
We construct the closest-pair graph $G$ for a point
set $A$ on the 3-sphere and try to condense it.

In this section, we perform an algorithm, which is 
a sequence of pruning and condensing to obtain a new equivariant point
set $A'$
that is in one of the following cases: 
\begin{enumerate}
\item[(i)] the closest distance $\delta$ of $A'$ is greater than $\delta_0$, or 
\item[(ii)] the
closest-pair graph of $A'$ has mirror symmetry, or 
\item[(iii)] the closest-pair graph of $A'$ is in the
edge-transitive case.
\end{enumerate}
If (i) happens, we apply 1+3 dimension reduction.
If (ii) or (iii) happens, we proceed to
Algorithm M in Section~\ref{sec:mirror} or Algorithm O in
Section~\ref{sec:orbit} respectively.
Those sections explain how to reduce the cases to 2+2 dimension reduction.

\begin{figure}[htb]
	\centering
	\includegraphics[width=\textwidth]{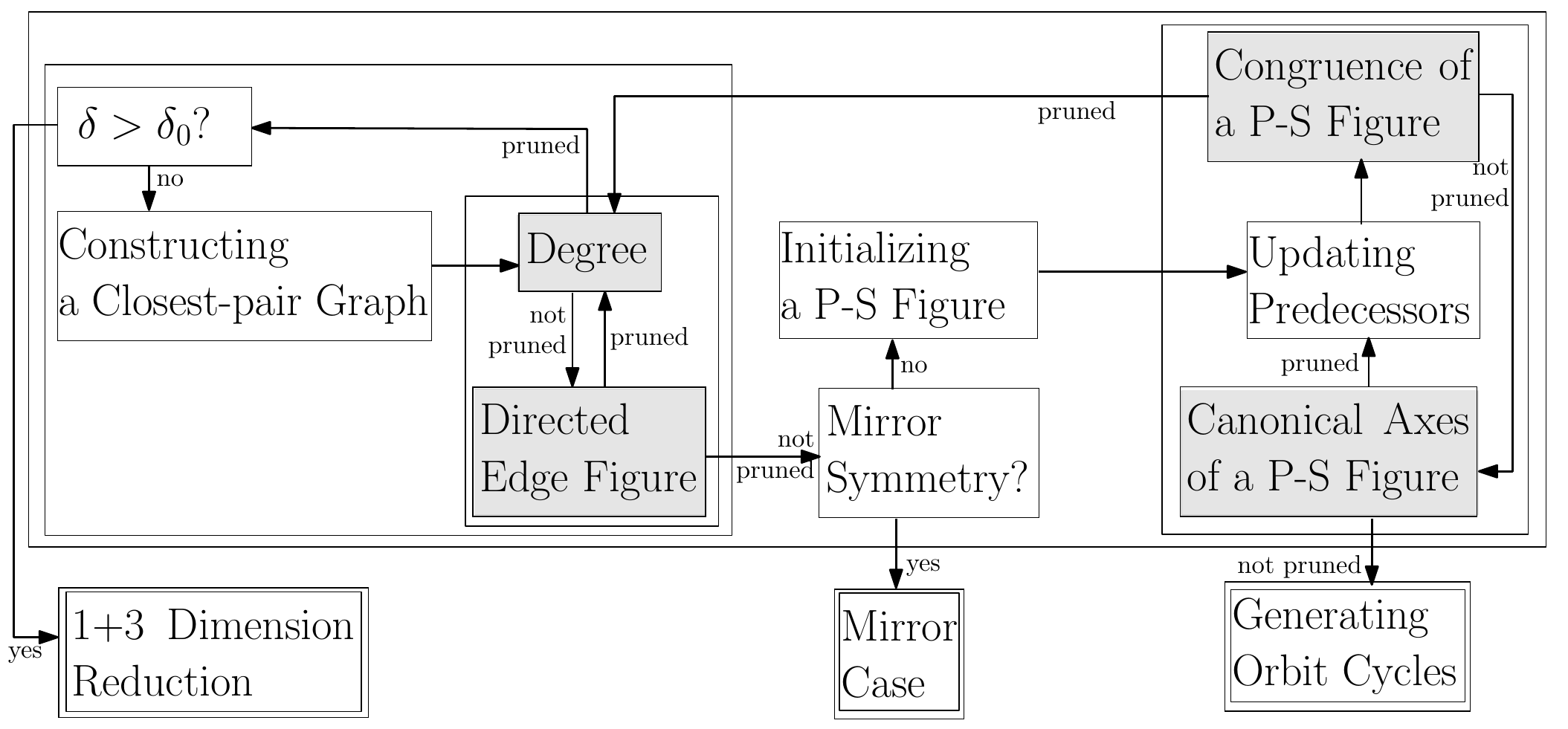}
	\caption{Iterative pruning. The shaded boxes
	represent pruning/condensing steps with pruning criteria or
	condensing methods.
}
	\label{fig:orbit}
\end{figure}

%\subsection{Predecessor-Successor Figures}

\begin{figure}[htb]
  \centering
  \includegraphics{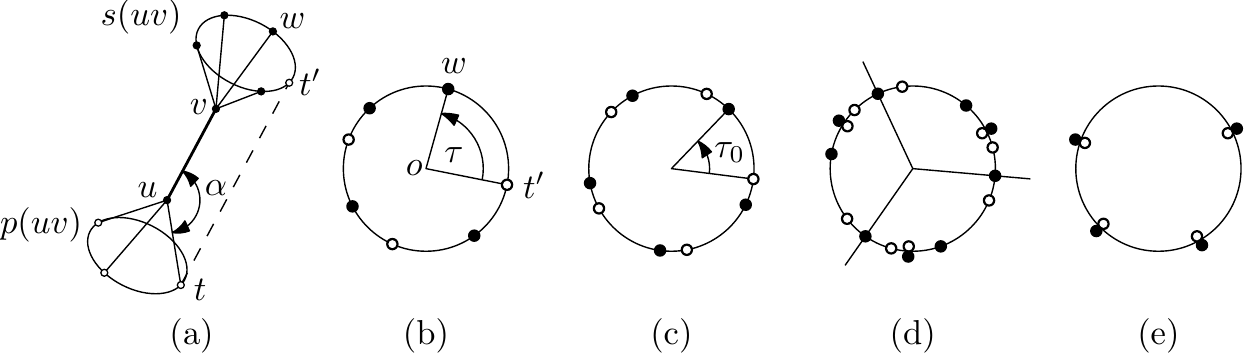}
  \caption{(a) Predecessors and successors of $uv$ at angle
    $\alpha$; $t'$ is the reflected copy of $t$ on the successor circle. (b) The corresponding predecessor-successor figure,
    and the torsion angle $\tau(tuvw)$. Predecessors are drawn white
    and successors black.
(c) An edge-transitive predecessor-successor figure with torsion angle $\tau_0$.
(d) Canonical axes. (e) Mirror symmetry.}
  \label{fig:succ-figure}
\end{figure}

%We are given a set $A$ of points on the 3-sphere.
%This can be the original
%point set, after pruning by distance from the origin, or the result
%of previous reduction steps.
We first explain the notion of a
\emph{\psfig} in $G$.
We have a directed edge $uv$ of $G$ together with 
a set of \emph{predecessor edges} $p(uv)$ incident to $u$
and a set of \emph{successor edges} $s(uv)$ incident to $v$,
as in Fig.~\ref{fig:succ-figure}a.
All edges have the same length, their endpoints lie on the 3-sphere $\sphere^3$, and
all predecessor and successor edges form the same
angle
$\alpha$ with $uv$. Then
 the endpoints of these edges lie on two circles.
If we reflect the predecessor circle at the bisecting hyperplane of $uv$, it
comes to lie on the successor circle. This results in one circle
with a succinct representation of the geometric situation.
% containing
%the geometric information of an arc $uv$,
See Fig.~\ref{fig:succ-figure}a and Fig.~\ref{fig:succ-figure}b.
We refer to the endpoints of the predecessor and successor edges as
\emph{predecessors} and \emph{successors}.
%\footnote{
%(BTW the two circles may degenerate to a point, in case the single
%predecessor or successor lies on a great circle with $uv$ (which is
%good). Is that an issue that we should mention?}
%We call this arrangement of predecessors and successors 
%on a common circle the \emph{\psfig}.
%It may happen that the predecessors and successors coincide
%after the reflection, as in Fig.~\ref{fig:succ-figure}e.
%This means that the involved
%edges
%have a perfect mirror symmetry with respect to the bisecting hyperplane of
%$uv$, a case which will require special treatment.

% ==========================
% [ THIS WAS THE FIRST TRY.
% We have a directed subgraph $D$ of the undirected graph $G$.
% For an arc $uv\in D$, we look at % its predecessors, 
% the \emph{predecessor} arcs $tu$ that enter $u$ and the
% \emph{successor}
%  arcs $vw$ that leave $v$. If we
% restrict our attention only to those
% predecessor and successor
%  arcs which make a specified angle
% $\alpha$ with $uv$, the endpoints of these arcs lie on two circles, as
% shown
% in Fig.~\ref{fig:succ-figure}a. ]
% ==========================

%We can now state the algorithm.
%The following algorithm will successively prune $A$ until the closest distance
%$\delta$ is greater than $\delta_0$, or it will exit to either Algorithm M or
%Algorithm O. 
% These two algorithms will eventually solve the problem
%by 2+2 dimension reduction, or they produce a
%condensed set of points $A'$, and then, this algorithm is resumed with $A'$ taking the
%role of $A$.
%
The following algorithm guarantees that the outcome $A'$ is in one of the promised
cases.

\smallskip
\emphi{Algorithm C} (\emph{Prune the closest-pair graph}).
We are given a set $A\subset \reals^4$ of $n$
 points, equidistant from the origin.
 This set may already be the result of some
previous condensing steps.
 Without loss of generality, we may assume that
$A$ lies on the unit sphere $\sphere^3$.
\begin{itemize}
\labelwidth 32pt
\dimen0=\hsize\advance\dimen0-32pt
\parshape=1 32pt
\dimen0
\item[\bf C1.] [Well-separated points?]
%The closest distance $\delta$ is large?] 
Compute the closest distance $\delta$ between points of $A$. If
$\delta > \delta_0=\DELTA0$, apply 1+3 dimension reduction. ($|A|$ is bounded by a constant.)
\item[\bf C2.] [Construct the closest-pair graph.]
Construct the closest-pair graph $G$, and initialize
 the directed
graph $D$ with two opposite arcs for every edge of % the closest-pair
                                % graph
 $G$.
(This takes $O(n\log n$) time, and the degrees in $G$ are bounded by $K_3=12$.)

\dimen0=\hsize\advance\dimen0-40pt
\parshape=1 40pt
\dimen0
\advance\labelsep8pt

\item[\bf C3.] [Prune by degree.]
If the indegrees and outdegrees in $D$ are not all the same, prune the
vertices by
degree, and return to C1, with the smallest class $A'$ taking the
role of $A$. (Otherwise, we now enter a loop in which we try to prune arcs from $D$.)

\item[\bf C4.] [Prune by directed edge figure.]
The \emph{directed edge figure} of an arc $uv\in D$ consists of
$uv$ together with all arcs of $D$ out of $v$ and all arcs of $D$ that lead into $u$.
If the directed edge figures are not all congruent,
\emph{prune the arcs of $D$} by congruence type of directed edge figures, and return to C3.
(Here we apply the pruning principle not to points but to arcs and
we replace the edge set $D$ by a smaller subset $D'$.
Since the degrees are bounded, we can compare two directed edge
figures in constant time.)

\item[\bf C5.] [Mirror symmetry?]
(Now all arcs have the same directed edge figure.)
If the direct edge figure of $uv$ is symmetric with respect to
the bisecting hyperplane of $uv$, proceed to Algorithm~R
(Section~\ref{sec:mirror}).
\item[\bf C6.] [Choose an angle $\alpha$ with no mirror symmetry.]
Pick an angle $\alpha$ for which the predecessors and successors are
not completely symmetric,
that is, the \psfig\ does not look like
Fig.~\ref{fig:succ-figure}e. 

\item[\bf C7.] [Initialize Successors.]
For every arc $uv\in D$, set
$ s(uv) := \{%\,
 vw : vw \in D, \angle uvw=\alpha%\, 
\}$.
(The size of $s(uv)$ is bounded by $K_2=5$.
We will now enter an inner loop in which we try to prune
 arcs from the sets $s(uv)$.)

\dimen0=\hsize\advance\dimen0-48pt
\parshape=1 48pt
\dimen0
\advance\labelsep8pt

\item[\bf C8.] [Update predecessors.]
Define the predecessor edges by
 $p(uv) := \{\, tu : uv \in s(tu)\, \}$.
Build the \psfig~for each arc, as explained in the text
above.
\item[\bf C9.] [Prune by \psfig s.]
If there are arcs whose \psfig s are not congruent,
{prune the arcs of $D$} accordingly, and return to C3.
\item[\bf C10.] [Check regularity.]
(Now all arcs have the same \psfig. Each figure must
contain the same number $k$ of predecessors and successors,
since the total number of
predecessors in all figures must balance the total number of successors.)
If the predecessor-successor figure consists of two regular $k$-gons,
proceed to Algorithm~O for generating orbit cycles,
see Section~\ref{sec:orbit}. (We call this the \emph{edge-transitive}
case. See Figure~\ref{fig:succ-figure}c.)

\item[\bf C11.] [Prune successors by canonical axes.]  Prune $s(uv)$ to a proper
  nonempty subset by computing \emph{canonical axes} as explained
  in Section~\ref{sec:cano}, and return to C8.
\end{itemize}

% We have to justify the claim in step C10:
%  Since each
% each occurrence of some arc $vw$ as a successor $vw\in s(uv)$ give rise to the
% occurrence
% of $uv$ as a predecessor $uv\in t(vw)$, the total number of
% predecessors  in all \psfig s must equal the total number of successors.

An example of canonical axes is depicted in
Fig.~\ref{fig:succ-figure}d.
If canonical axes consist of $p$ axes, 
we know that $p<k$ because the maximally
symmetric case of two regular $k$-gons
 (the \emph{edge-transitive} case shown in Fig.~\ref{fig:succ-figure}c)
has been excluded in Step~C10.
The loop from C8--C11 maintains
 the following loop invariant on entry to Step~C10:
\begin{equation}
  \label{eq:invariant}
\text{\emph{There is a position of a successor that is not occupied by a predecessor.}}
  \end{equation}
For the reduction in Step~C11,
we rotate the canonical axes counterclockwise until
they hit the first position of type~\eqref{eq:invariant}.
The successors that are intersected
by the canonical axes form a nonempty proper subset $s'\subseteq
s(uv)$.
We thus replace $s(uv)$ for each edge $uv$ by $s'$ and return to Step~C8.
By construction, % $s'$ is nonempty, and moreover, 
we have made sure
that \eqref{eq:invariant} still holds.
After pruning all successor sets, 
the predecessor sets are reduced accordingly in Step~C8, but this cannot invalidate~\eqref{eq:invariant}.
%
%
%  This subset
% % This subset
% %is a proper subset unless $p=k$ and we have a $k$-fold symmetry, where
% %both the black points and the white points form regular $k$-gons, 
% If $s'$ is a proper subset of $s(uv)$, the \psfig~forms an union of
% two regular $k$-gons, as in Fig.~\ref{fig:succ-figure}c.
% We call this an \emph{edge-transitive} case.
% This was already checked in Step~C10 to be treated in Algorithm~O.
% Otherwise, we replace $s(uv)$ for each arc $uv$ by $s'$ and return to Step~C8.
% By construction, $s'$ is nonempty.
% %and moreover, we have made sure
% %that \eqref{eq:invariant} still holds.
% %As a result of reducing all
% %sets
% %of successors, 
% The predecessor sets are also updated, but this does
% not invalidate~\eqref{eq:invariant}.
%
(The invariant~\eqref{eq:invariant} holds
on first entry to the loop because of
 Step~C6.)

%We still need to show that the invariant~\eqref{eq:invariant} holds
%when entering the loop for the first time.
%We know that we have $k$ white points (predecessors) and
%$k$ black points (successors).
%We have ensured
%in Step~C6 that the two sets don't completely coincide. 
%This
%implies~\eqref{eq:invariant}.

The algorithm has three nested loops (indicated by
indentation) and works its way up to higher and higher orders of
regularity.
After C3, all vertices have the same degree. 
%we know that all vertices look the same, at least as far as
%the number of neighbors is concerned. 
After C4, we know that all
\emph{pairs} of adjacent vertices look the same. If we exit to
Algorithm~O in Step~C10, wee will see that certain chains of \emph{four}
points can be found everywhere.
%
%
%
%The reduction of the successor function should not be confused with
%the pruning of arcs of $D$.
%Correspondingly,
%we have successive condensing on three levels:
% (corresponding to the three nested loops): 
%successor functions $s(uv)$, arcs of $D$, and
%points $A$. Each reduction may trigger a pruning operation at the next
%higher level.
%\iflong
% But after pruning $D$, for example, the previous
%(possibly reduced) function is thrown away, and the successor function
%is initialized from scratch in Step~C7.
%\fi
%
%\iflong
%The algorithm consists of several nested loops, which are indicated by
%indentation.

There is the \emph{global loop} that leads back to C1 after each successful
pruning of vertices by degree. Since the size of $A$ is reduced to
less than a half,
we need not count the iterations of this loop.
In addition, there is an outer loop that resumes working at C3
after pruning the edges of $D$,
 and an inner
loop that starts at C8 and is repeated whenever the successor set
$s(uv)$ is pruned.
In these loops, we maintain that $D \neq \emptyset$ and $s(uv) \neq
\emptyset$. 
%we actually need not even be careful about selecting
%the smallest class when we prune.
In Step C3, if we have removed at least one edge from $D$, we will
either be able to prune by degree, or the degree of all
vertices has gone down by at least one. Since the degree is initially
bounded by 12, Step C3 can be visited at most 12 times before exiting
to C1.
Similarly, Step C11 removes at least one element of $s(uv)$, so
this loop is repeated at most 5 times before there
is an exit to the outer loop in step C9.
%(THIS MAY BE SHORTENED IF NECESSARY.)
%\else
%Each inner loop has a fixed bound on the number of repetitions,
%because it decreases the size of $s(uv)$ or the outdegree in $D$ from
%at most 5 or 12 by at least 1.
%Each outer loop reduces $|A|$ to at most $|A|/2$.
%\fi
The most time-consuming step is the construction of the closest-pair
graph in Step C2. 
All other operations take $O(n)$ time, not counting
%for the time being 
the exits to Algorithms M and~O.
Thus, the overall time is $O(|A|\log |A|)$.

%% file: Orbit.tex
\section{Generating Orbit-Cycles: Algorithm O}
\label{sec:orbit}

We now describe how, in the edge-transitive case, the algorithm O
produces a set $\plane$ of at most $|A|/200$ great circles.
All predecessor-successor figures look like
Fig.~\ref{fig:succ-figure}c.
The \emph{torsion angle} $\tau(tuvw)$ 
 between a predecessor edge $tu\in p(uv)$ and a successor edge $vw\in
s(uv)$
%for $tu,uv,vw \in E$ 
is the oriented angle
$\angle(t'ow)$ in the predecessor-successor figure of $uv$.
We define $\tau_0$ as the smallest counterclockwise torsion angle
that appears in the \psfig.
%$\tau(tuvw)$  between a predecessor $t\in p(uv)$ and a successor $w\in
%s(uv)$.
Let $t_0u_0v_0w_0$ be a fixed quadruple with this torsion angle.

\begin{lemma}
\label{orbit-cycle}
  \begin{enumerate}
%\iffalse
  \item \label{continue}
  For every $a_2a_3\in s(a_1a_2)$, there is a
\textup({unique}\textup)
 edge $a_3a_4\in s(a_2a_3)$ such that
$a_1a_2a_3a_4$ is congruent to $t_0u_0v_0w_0$.
\item
\label{4-tuple}
 Moreover, there is a unique rotation $R_0$ that maps
 $t_0u_0v_0w_0$ to
$a_1a_2a_3a_4$
%\fi
\item 
\label{sequence-tuple}
For every triple $a_1a_2a_3$ with
 $a_2a_3\in s(a_1a_2)$, there is a
unique cyclic sequence $a_1a_2\ldots a_\ell$
such that $a_ia_{i+1}a_{i+2}a_{i+3}$
is congruent to $t_0u_0v_0w_0$ for all $i$. \textup{(Indices are taken
modulo}~$\ell$.\textup{)}
\item
\label{unique-rotation}
 Moreover, there is a
unique rotation matrix $R$ such that $a_{i+1}=Ra_i$. In other words,
 $a_1a_2\ldots a_\ell$ is the orbit of $a_1$ under the rotation $R$.
\item 
\label{not-circle}
The points $a_1a_2\ldots a_\ell$ do not lie on a circle.
  \end{enumerate}
\end{lemma}
%
%\begin{proof}
%  (Sketch.)
%The full proof is given in Appendix~\ref{app-O}. It is a straightforward sequence of arguments
%once we have established that
%the three points $t_0,u_0,v_0$ cannot lie on a great circle, and
%therefore a rotation is uniquely determined by the images of these
%three points. If the the points 
% $t_0,u_0,v_0$ would lie on a great circle, this would imply
%a mirror symmetry in the \psfig,
%contradicting the
% filtering steps~C5 and~C6 which have led to the
%invariant~\eqref{eq:invariant}.
%\end{proof}
%
%\iffalse

\begin{proof}% (Sketch)
We first establish two facts about
$t_0u_0v_0w_0$.
\begin{gather}
  \label{eq:no-circle}
%  \vcenter{\begin{minipage}{0.8\linewidth}
\text{%\emph
      {The four points $t_0,u_0,v_0,w_0$ do not lie on a circle.}
}
\\
  \label{eq:no-great-circle}
\text{%\emph
      {The three points $t_0,u_0,v_0$ do not lie on a great circle.}
}
\end{gather}
If $t_0,u_0,v_0,w_0$ would lie on a circle
(not necessarily through the origin),
the \psfig\ of $u_0v_0$
would
have a mirror-symmetric predecessor $t_0u_0\in p(u_0v_0)$ and
successor $v_0w_0\in s(u_0v_0)$, in
contradiction to the choice of $\alpha$ in Step~C6 and to the
invariant~\eqref{eq:invariant}.
If the points $t_0,u_0,v_0$ lie on a great circle~$C$,
then $w_0$ must also lie on
$C$, since $\angle t_0u_0v_0
=\angle u_0v_0w_0 = \alpha$, but this would
contradict~\eqref{eq:no-circle}.

Now, let $a_1,a_2,a_3$ be any three points with
 $a_2a_3\in s(a_1a_2)$. By the definition of predecessors,
% we have
 $a_1a_2\in p(a_2a_3)$. We can thus fit
 $t_0u_0v_0$ to $a_1a_2a_3$ in the \psfig\ of $a_2a_3$.
Since
 the points $t_0,u_0,v_0$ are not
on a great circle, they span a three-dimensional subspace, and
the rotation $R_0$ that maps
$t_0u_0v_0$ to
$a_1a_2a_3$
is uniquely determined.
Since all \psfig s are congruent,
this means that the edge $v_0w_0$ is mapped to
some
 successor edge
 $a_3a_4\in s(a_2a_3)$.
This establishes Properties
\ref{continue}, \ref{4-tuple}, and~\ref{not-circle}.

This process can be continued: 
Since $a_2a_3\in p(a_3a_4)$,
there is a unique edge $a_4a_5\in s(a_3a_4)$ such that
$a_2a_3a_4a_5$ is congruent to $t_0u_0v_0w_0$, and so on.

% $a_1a_2a_3$ to the congruent triple $a_2a_3a_4$. 
By Property~\ref{4-tuple},
there are two unique rotations from
$t_0u_0v_0w_0$ to $a_1a_2a_3a_4$
and  to $a_2a_3a_4a_5$,
and thus there is a unique rotation $R'$ with
$R[a_1a_2a_3a_4]=[a_2a_3a_4a_5]$.
We have seen that the points $a_1a_2a_3$
 are not on a great circle, being congruent to $t_0u_0v_0$ and thus
the rotation $R$ is already uniquely specified by the conditions
$R[a_1a_2a_3]=[a_2a_3a_4]$.
We have therefore established that
 $a_{i+1}=Ra_i$ for $i=1,2,3$ implies
 $a_{i+1}=Ra_i$ for $i=4$. This can be continued by induction, implying that
$a_1a_2\ldots$ is the
orbit of the point $a_1$ under the rotation $R$.

Since the number of points in $A$ is finite, this
orbit must return to $a_1$ and 
form a directed cyclic path $a_1a_2\ldots a_\ell$ 
in the closest-pair graph~$D$.
This establishes Properties
\ref{sequence-tuple} and~\ref{unique-rotation}.
%
%
%Since the backward
%extension to $a_0a_1a_2a_4$ is also unique, by the same arguments, these cycles form a
%decomposition of the triplets $\{\, (u,v,w) : w\in s(u,v)\,\}$ into
%disjoint cycles $\mathcal{C}_3$. Each cycle in
% $\mathcal{C}_3$ is the set of consecutive triplets of
%some directed cyclic path $a_1a_2\ldots a_\ell$ in the closest-pair
%graph.
%
%We denote the set of all these cyclic paths by $\mathcal{C}$. They
%are all congruent.
%  
\end{proof}
% \ref{continue}
% \ref{4-tuple}
% \ref{sequence-tuple}
% \ref{unique-rotation}
% \ref{not-circle}

%\fi

We call the cyclic paths that are constructed in
Lemma~\ref{orbit-cycle} %.\ref{sequence-tuple} 
\emph{orbit cycles}.
The following lemma
gives a bound on the number of orbit cycles in terms of $|A|$ when the
closest-pair distance is small enough.
%
% shows that $|\plane| \le n/200$ where $\plane$ is
%the set of obtained orbit cycles if the closest-pair distance is
%large; $\delta >\delta_0$.
\begin{lemma}
\label{lem:leng}
The number of orbit cycles is at most $|A|/200$ provided that the
closest distance $\delta$ is smaller than $\delta_0=0.0005$.
\end{lemma}
\begin{proof}
  We have $a_{i+1}=R_{\phi,\psi}a_i$ for all $i$
%, with a rotation
%  matrix
%  \begin{equation}
%	  \label{eq:rotation}
%  R_{\phi\psi} =
%  \begin{pmatrix}
%    \cos\phi & -\sin\phi  &0&0\\
%    \sin\phi & \cos\phi   &0&0\\
% 0&0&    \cos\psi & -\sin\psi\\
% 0&0&    \sin\psi & \cos\psi\\
%  \end{pmatrix}
%  \end{equation}
with an appropriate basis
$x_1y_1x_2y_2$
where $R_{\phi,\psi}$ is defined as~\eqref{eq:rot_mat}.
We cannot have $\phi=0$ or $\psi=0$, because otherwise the orbit would
form a regular polygon $a_1a_2a_3a_4\ldots a_\ell$
 in a plane, contradicting %~\eqref{eq:no-circle}.
Lemma~\ref{orbit-cycle}.\ref{not-circle}.
Thus, we know that $\phi,\psi\ne 0$.
 To get a closed loop, we
must have $|\phi|,|\psi|\ge 2\pi/\ell$.
If the projection of $a_i$ to the $x_1y_1$-plane has norm $r_1$
and the projection to the $x_2y_2$-plane has norm $r_2$ with
$r_1^2+r_2^2=1$, then the squared distance is 
$$\delta^2=
\|a_{i+1}-a_i\|^2=(2r_1\sin\tfrac{|\phi|}2)^2+
(2r_2\sin\tfrac{|\psi|}2)^2\ge 4\sin^2 (\pi/\ell).
$$
As a result, 
\begin{equation}
	\label{eq:leng}
\delta \ge 2 \sin \frac{\pi}{\ell}
\end{equation}
We can get the same bound via
 Fenchel's theorem about the length of a closed curve that is not
 contained in a hemisphere~\cite{Fen}.
Thus, if $\delta\le 2\sin(\pi/12000) \approx 0.000523$,
it is guaranteed that
$\ell\ge 12000$, that is, every orbit cycle contains at least 12000
points.
On the other hand,  each point $u\in A$ belongs to a bounded number of orbit cycles:
it has at most $K_3=12$ incoming arcs $tu$, and each each arc
$tu$ has at most $K_2=5$  successor edges $uv\in s(tu)$.
The triple $tuv$ specifies a unique orbit cycle, and thus there are at
most $12\times 5=60$ orbit cycles through $u$. The total number
of orbit cycles is therefore at most $|A|\cdot 60/12000=|A|/200$.
%Every orbit path is of length 
%$\ell \ge \pi/\arcsin (\delta/2)$. 
%By pruning, we may assume the length is the same for all the orbit
%cycles in $\plane$.
%If $\delta \ge \delta_0$, $\ell \ge 60000$.
%
\end{proof}

For each orbit cycle, we can find
an appropriate rotation matrix $R_{\phi,\psi}$. % in~\eqref{eq:rotation}. 
If $\phi=\pm\psi$, then the orbit of any point under this rotation lies
on a great circle, contradicting %~\eqref{eq:no-great-circle}.
Lemma~\ref{orbit-cycle}.\ref{not-circle}.
%Thus, we choose the smaller of $|\phi|$ and $\psi$, and
%
Thus, we get a unique
 invariant plane that rotates by the smaller angle in itself.
We replace each orbit cycle
by the great circle
 in % is the intersection of the 3-sphere and 
this
invariant plane, yielding the
desired set $\plane$ of great circles with
$|\plane| \le |A|/200$.

Section~\ref{sec:mirror} we need a bound on the number of points in a regular
polygon inscribed in a great circle.
\begin{corollary}
	\label{cor:leng}
If the closest distance $\delta < \delta_0$ and there is a regular
$k$-gon inscribed in a great circle of a sphere, then $k> 12000$.
\end{corollary}
\begin{proof}
The same computation as in Lemma~\ref{lem:leng} can be used 
for a great circle.
%but not when $\phi=0$ or $\psi=0$.
From~\eqref{eq:leng} and $2\sin(\pi/12000) > \delta_0$, we find that $k>
12000.$
%\fi
% >>> from math import *
% >>> 2*sin(pi/12000)
% 0.0005235987696171498
% >>> 
\end{proof}

%% file: Mirror.tex
\section{The Mirror Case: Algorithm R}
\label{sec:mirror}
%The following algorithm deal with the case 
%that every arc $uv$ of the directed version of the closest-pair
%graph acts as a mirror that exchanges $u$ and $v$ together with
%their neighborhoods.  Since the reflected mirror hyperplanes act again as
%mirrors, it follows that every component of the graph is the orbit of some
%point $u_0$ under the group generated by reflections perpendicular to
%the edges incident to $u_0$.

\smallskip

\emphi{Algorithm R} (\emph{Treat mirror-symmetric closest-pair
  graphs}).
We are given a nonempty directed subgraph $D$ of the closest-pair
graph on a point set $A\subset\sphere^3$ with
closest-pair distance $\delta\le\delta_0$. %  $|A|\le n_0$. 
 All
directed edges in $D$ have equal edge-figures and exhibit perfect
mirror-symmetry, as established in Steps~C4 and C5 of Algorithm~C.
Algorithm R will produce, in an equivariant way, either
 \begin{enumerate}
  \item [a)] a set $A'$ of at most $|A|/2$ points, or
  \item [b)] a set $\plane$ of at most
 $|A|/200$ great circles.
% for each
% connected
%    component of $G$.
%In this case, every
% connected
%    component of $G$ has at least $...$ vertices.
  \end{enumerate}

\begin{itemize}
\labelwidth 32pt
\dimen0=\hsize\advance\dimen0-32pt
\parshape=1 32pt
\dimen0
\item[\bf R1.] [Make $D$ undirected.]
Construct the undirected version
 of $D$ and call it $G$.
\item[\bf R2.] [Check for eccentric centers of mass.]
Compute the center of mass of each connected 
component of $G$.
If these centers are not in the origin, return the set
$A'$ of these centers.
\item[\bf R3.] [Two-dimensional components?]
If each component is a regular polygon with center at the origin,
return the set $\plane$  of circumcircles of each polygon.
\item[\bf R4.] [Three-dimensional components?]
If each component
spans a
3-dimensional hyperplane $H$,
replace the component by two antipodal
points perpendicular to $H$.
Return the set
$A'$ of these points.

\item[\bf R5.] [Toroidal grid.]
As it was shown in Lemma~\ref{exclude-finite},
now each component of $D$ is the product $P\times Q$ of a regular $p$-gon $P$ and a
regular $q$-gon $Q$ in orthogonal planes, with $p,q\ge 3$.
Pick a vertex $u$ from
each component.
There are four incident edges,
and
the plane spanned by two incident edges is orthogonal to the plane spanned
by the other two edges.
 Represent the component of $G$ by these two orthogonal planes shifted
to the origin. 
Return the set $\plane$ of great circles in these planes (two per component).
\end{itemize}

The algorithm takes linear time.
We still need to show that these cases are exhaustive.
After we make $D$ undirected in Step~R1, 
for every edge
 $uv$ of $G$, the bisecting hyperplane of $uv$
 acts as a mirror that exchanges $u$ and $v$ together with
their neighborhoods.  Since the reflected mirror hyperplanes act again as
mirrors, it follows that every component of the graph is the orbit of some
point $u_0$ under the group generated by reflections perpendicular to
the edges incident to $u_0$.
Thus, the incident edges of a single point determine the geometry of
the whole component.
The graph may have several components, all of which are congruent.

There are infinitely-many groups of the form $D_2^p\times
D_2^q$,
where
$D_2^p$ is the dihedral group of order $2p$, the symmetry group of the
regular $p$-gon. 
These groups are treated in Step R5.
Except this infinite family, there are only eight
other groups, which are excluded by Lemma~\ref{exclude-finite}
%The details have already been described
(see Section~\ref{sec:coxeter}), because the constant $\delta_1=0.07$
from Lemma~\ref{exclude-finite} is (much) larger than~$\delta_0$.

We go through the steps one by one to check if the algorithm
achieves the claimed results.
Step R2 treats the case when a (lower-dimensional) component
does not go through the origin. This includes the cases when $G$
is a matching or a union of ``small'' regular polygons.
Every component contains at least 2 points, and thus $|A'|\le |A|/2$.
Steps R3 and R4 treat the two- and three-dimensional components that
are centered at the origin. (The one-dimensional case of a matching
cannot be centered at the origin, because then we would have $\delta=2$.)
Suppose we have a two-dimensional component and this component is
a regular $k$-gon inscribed in a great circle (Step R3).
From $\delta < \delta_0$ and by Corollary~\ref{cor:leng}, 
we obtain that $k > 12000$. Thus $|C|\le |A|/12000
\le |A|/200$.
A three-dimensional component (Step~R4) contains at least four
points and is reduced to two antipodal points.
Again we have $|A'|\le |A|/2$.

Steps R2--R4 have treated all cases of Coxeter groups which are not
full-dimensional, and Lemma~\ref{exclude-finite}
excludes all full-dimensional groups which are not of the form
 $D_2^p\times D_2^q$.
Thus, in Step R5,
% we have
each
 component of $D$ is the product $P\times Q$ of a regular $p$-gon $P$ and a
regular $q$-gon $Q$ in orthogonal planes, with $p,q\ge 3$.
%\iflong

Note that a regular $p$-gon can be generated as the orbit of a point $u_0$
in  $D_2^p$ or in $D_2^{2p}$, depending on whether we put $u_0$ in the
interior of the fundamental region or on a mirror. But this makes no
difference for the resulting point set.

%\fi
Such a component
 forms a toroidal $p\times q$ grid.  Each
vertex has four neighbors. The two polygons $P$ and $Q$ have
equal side lengths~$\delta$, because otherwise the four neighbors would not be
part of the %pruned
 closest-pair graph $G$. 
The two incident edges of a vertex
$u$ that come from $P$ are orthogonal to the two edges that come from
$Q$, so we can distinguish the two edge classes. (The case when
all four edges are perpendicular is the 4-cube with $p=q=4$ and
 with reflection group
 $D_2^4\times D_2^4$. This case has 16 vertices $(\pm 1/2, \pm 1/2, \pm
 1/2, \pm 1/2)$ and minimum distance $\delta=1$, and it is therefore
excluded.)
%\iflong
Moreover, % this toroidal grid is a square grid:
the grid cells are geometric squares. 
Through
these squares, the classification of edges into the edges from $P$ and
the edges from $Q$ can be transferred consistently to a whole
connected component of $G$.  
%\fi
All copies of $P$ in the grid lie in
parallel planes, and so do the copies of $Q$. 
Accordingly, Step R5 represents
each connected component by two orthogonal planes through the origin.
We need to show that the component is large.
% when the closest distance $d$ is small
As $P\times Q$ lies on the unit 3-sphere, the circumradii $r_P$ and
$r_Q$ of the two polygons satisfy $r_P^2+r_Q^2=1$. 
Thus,
 the larger circumradius, let us say
	$r_P$, is at least $1 /\sqrt{2}$.
% Also, $r_P \ge r_Q$ implies $p \ge q$.
	Since the closest-pair distance is
	$\delta=2r_P \sin \frac{\pi}{p}\le \delta_0=\DELTA0$, % =2 r_Q \sin \frac{\pi}{q}$,
we get $
\sqrt{2} \sin
	\frac{\pi}{p}\le\delta_0$, which implies
$p\ge 8886$.
Each component contains
%$pq\ge3p\ge 1.27 \times 10^5$ 
$pq\ge p$
points and is reduced to two circles.
Thus, the algorithm achieves the claimed reduction.

% >>> pi/asin(0.0005/sqrt(2))*3
% 26657.297073589812
% >>> pi/asin(0.0005/sqrt(2))
% 8885.765691196604
% >>> 

%% file: Algorithm_K.tex
\section{Finding Representative Points from a 2-Sphere: Algorithm~K}
\label{sec:prusphere}

In the following Section~\ref{sec:plane},
we need an auxiliary algorithm to condense a set of great
circles that belong to a common Hopf bundle.
These circles can be mapped topoints on the 2-sphere,
and thus, we adapt
 Atkinson's
algorithm~\cite{Atk} (see Section~\ref{sec:3d})
to condense a finite point set in a 2-sphere.

% Although Atkinson's algorithm preserves rotational symmetries as a
% canonical set procedure (refer to Section~\ref{sec:cano} and
% Section~\ref{sec:cpp}), this algorithm is rather an efficient
% condensing method.
% The algorithm  finds at most 12 
% representative points in $\mathbb{S}^2$ by condensing 
% as in Lemma~\ref{lem:rep}.

\begin{lemma}[Algorithm K]
\label{lem:rep}
%Algorithm~\ref{algo:rep} 
There is a procedure that reduces a set $F$ of points on the 2-sphere
 $\mathbb{S}^2$
to a nonempty set $F'$ of at most
 $\min\{12,|F|\}$
representative
points on $\mathbb{S}^2$,
in $O(|F| \log |F|)$ time.
These points are either
the vertices of a regular tetrahedron, 
 a regular octahedron, % or
 a regular icosahedron, % that is inscribed in the unit sphere,
 a single point, or a pair of antipodal
points.
This mapping from sets $F$ to sets $F'$ is
equivariant under rotations and
reflections.
\end{lemma}

\begin{proof}
  We repeatedly prune the set $F$.
We start by computing the convex hull of $F$.
If it does not contain the origin, 
we can immediately output the vector pointing to the center of gravity
of the points
as a representative.
If the hull is one-dimensional or
two-dimensional, we get
 two antipodal points as representatives.

We can thus assume that the hull is a three-dimensional polytope.
If the vertex degrees are not all the same, we prune the point set and
restart.
We can thus assume that the graph of the polytope is regular, and by
Euler's formula, the degree $d$ can be 3, 4, or 5.
Euler's formula also yields the number $f$ of faces in terms of
the number $n$ of vertices:
$f=(d-2)/2\cdot n + 2\le \frac32 n+2$.
We now try to prune by face degrees. If there are different face
degrees,
we replace $F$ by the centroids of the smallest class of faces,
and restart. We have condensed $|F|$ at least by a factor of 3/4 (the additive term
$+2$ is negligible as long as $|F|$ is large).
The remaining case is when all face degrees and all vertex degrees are
equal.
Then the polytope must have the combinatorics of one of the five
regular polytopes: the tetrahedron, the octahedron, the icosahedron,
the cube, or the dodecahedron.
For the cube and the dodecahedron, we take the centroids of the faces
and thereby reduce the number of points.

We are left with the case of the tetrahedron, the octahedron, and the
icosahedron.  If the edges don't have the same lengths, we condense and
replace $F$ by the midpoints of the smallest edge class. This will
lead to a reduction in the number of points, except for an octahedron and
icosahedron with two edge lengths, each occurring at least
6 times or 12 times, respectively.
We claim that the triangular faces cannot be all congruent in this case,
and we can prune the
8, resp.\
20, faces by congruence type, leading to a reduction to at
most
4 resp.\ 10 triangle centroids. If the triangles were all congruent,
their sides would all have to be long-long-short or
 long-short-short.
 This would mean, for the octahedron, that the 12 edge lengths are divided
 in the proportion $4:8$, and in that case we could have condensed by the
 edges.
The same argument works for the icosahedron, because the 30 edge lengths are divided
 in the proportion $10:20$.

Thus, the only cases where pruning cannot proceed are a
 tetrahedron, an octahedron, or
an icosahedron, with regular triangles as faces. These must be the regular
polytopes.

The most expensive part in
each pruning step is the computation of the convex hull, which
 takes $O(n' \log n')$ time, where $n'$ is the size
of the current point set.
Each step reduces $n'$ by a constant factor of at
most $11/12$. Thus the overall running time is
 $O(|F| \log |F|)$.
\end{proof}

%% file: Plane.tex
\section{Marking and Condensing of Great Circles: Algorithm M}
\label{sec:plane}

We have extracted a set $\plane$ of at most $|A|/200$ 
great circles from the
point set $A$, either from the mirror case (Algorithm~R in
Section~\ref{sec:mirror}) or from orbit cycles (Algorithm~O in
Section~\ref{sec:orbit}).
Algorithm M obtains a small set $A'$ of \emph{marker points} on each
circle so that we can resume Algorithm~C, % in Section~\ref{sec:iter},
or it exits to
Algorithm~T for 2+2 dimension reduction (Section~\ref{sec:torus}).

For this purpose, we look at the angles $\alpha,\beta$ of pairs of
circles $C,D\in \mathcal{P}$.
If $C$ and $D$ are not Clifford parallel,
we can \emph{mark} a pair of points on $C$ and on $D$, as follows:
%The following procedure will be useful in Section~\ref{sec:plane}.
%Let us assume $P$ and $Q$ are not isoclinic and in addition to $C'$,
Let $C'$ be an orthogonal projection of $C$ to the plane through
$D$.
Then $C'$ is an ellipse (not a circle).
A pair of points on $D$ can be
\emph{marked} by the intersection of $D$ and the major axis of $D'$. 
Similarly, two points on $C$ are marked.

 Doing this for all pairs of circles would lead to a quadratic
blowup. Therefore
we construct the closest-pair graph on the \emph{set
  of circles}. We use Pl\"ucker coordinates to map great circles of $\sphere^3$
to points on $\sphere^5$.
We can then compute the closest-pair graph in six
dimensions, and every circle has at most $K_5$
closest neighbors. 

% Pl\"ucker coordinates and Pl\"ucker distance were explained
%  in Section~\ref{sec:pluecker}.
% Lemma~\ref{lem:pdist} and Corollary~\ref{cor:pdist} 
% in Section~\ref{sec:pluecker}
% showed that Pl\"ucker distance is geometrically
% meaningful and does not depend on the choice of a coordinate system.
%\begin{lemma}
%\label{pluecker}
% \begin{enumerate}
% \item 
% Two planes $P,Q$ with angles $\alpha,\beta$ have Plücker
%distance $\sqrt{2(1-\cos\alpha\cos\beta)}$.
%\item A circle can have at most $K_5$ closest neighbors on
% the Plücker sphere $\sphere^5$.
% \end{enumerate}
%\end{lemma}
%\begin{proof}
%  Property 1 is a lengthy calculation, see %which is given in
% Appendix~\ref{sec:a-pluecker}.
%Property 2 is straightforward.
%\iflong 
%(Actually, this bound should be closer to $K_4$
%because the Plücker coordinates satisfy an additional quadratic
%equation and lie on a 4-dimensional surface in
%$\sphere^5$ (the Plücker surface).
%However, this surface is negatively curved, and the kissing
%number for 4 dimensions does not readily apply.)
%\fi
%\end{proof}

The marking approach fails for Clifford-parallel circles.
Clifford-parallel pairs manifest a very rigid
 structure on a sphere.
Section~\ref{sec:hopf} describes
 properties of such pairs.
For convenience, we restate 
these results % in Section~\ref{sec:hopf} 
in the following proposition. 
They are formulated for right pairs of circles, but they hold equally
with left and right exchanged.
\begin{proposition}
\label{hopf}
  \begin{enumerate}
  \item
\label{hopf-transitive}
 The relation of being a right pair is transitive (as well
 as reflexive and symmetric) (Corollary~\ref{cor:thopf}). An equivalence class is called a
    \emph{right Hopf bundle}. 
  \item \label{hopf-map}
For each right Hopf bundle, there is
a \emph{right Hopf map} $h$ that maps the circles
of this bundle
to points on $\sphere^2$ (Lemma~\ref{lem:hopf}).
\item 
\label{distances}
By this map, two Clifford-parallel circles with angle $\alpha,\alpha$ 
%\textup
(and with Euclidean distance
$\sqrt2 \sin\alpha$ on the Plücker sphere $\sphere^5$%\textup
)
are mapped to
 points at
angular distance
$2\alpha$
on % the ``Hopf sphere''
$\sphere^2$ (Lemma~\ref{lem:lrhopf}).
\item
\label{neighbors-on-hopf}
 A circle can have at most $K_2=5$ closest neighbors on
 the Plücker sphere $\sphere^5$
 that form right pairs.
  \end{enumerate}
\end{proposition}
\begin{proof}[Proof of Property~\ref{neighbors-on-hopf}.]
\if
Transitivity
(Property \ref{hopf-transitive}) and the preservation of distances
(Property~\ref{distances})
can be established by calculations, see Section~\ref{sec:hopf}.
The right Hopf map in Property \ref{hopf-map}
 is obtained as follows:
Choose a positively oriented coordinate system
$(x_1,y_1,x_2,y_2)$ for which some circle $C_0$ of the bundle lies in the
$x_1y_1$-plane. Then the map
$h\colon \sphere^3\to \sphere^2$ defined by
\begin{equation*}
  h(x_1,y_1,x_2,y_2)=\bigl(
2(x_1y_2-y_1x_2),\
2(x_1x_2+y_1y_2),\
1-2(x_2^2+y_2^2)
\bigr)
\end{equation*}
maps all points on a circle of the bundle to the same point on
$\sphere^2$.
A difference choice of $C_0$ would lead to a different map, but by
Property~\ref{distances}, the images are related by an isometry
of~$\sphere^2$.
\fi
Property~\ref{neighbors-on-hopf}
is a direct consequence of % follows from
Properties~\ref{hopf-map} and~\ref{distances}.
% by the packing argument on $\sphere^2$.
\end{proof}

\smallskip

\emphi{Algorithm M} (\emph{Mark and condense great circles}).
Given a set $\mathcal{P}$ of great circles on $\sphere^3$,
this algorithm will produce, in an equivariant way, either a nonempty set 
$A'$ of at most $100|\mathcal{P}|$ points
 on $\sphere^3$,
 or a nonempty set $\mathcal{P}'$ of at most \PACK-F.
great circles.

The algorithm updates the set $\mathcal{P}$ and maintains an
equivalence relation $\sim$ on $\mathcal{P}$ such that all circles in
the same equivalence class belong to a common (left or
right) Hopf bundle.  The common chirality of all bundles is indicated
by a variable $\textit{chirality}\in \{ \textit{None}, \textit{Left},
\textit{Right\/}\}$, where \textit{None} is chosen when the equivalence
relation is trivial and all classes are singletons.
The size of the equivalence classes is bounded by 12.

\begin{itemize}
\labelwidth 32pt
\dimen0=\hsize\advance\dimen0-32pt
\parshape=1 32pt
\dimen0

\item[\bf M1.] [Initialize.] % trivial equivalence classes.]
Let every circle $C\in\mathcal{P}$ form a singleton equivalence class.
\item[\bf M2.] [Prune by size.] If equivalence classes are of different
  sizes,
choose the size that occurs least frequently.
Throw away all equivalence classes that are not of this size, together with the circles they
  contain.
\item[\bf M3.] [Few circles?]
If $|\mathcal{P}|\le \PACK-F.$, return the set $\mathcal{P}$.
\item[\bf M4.] [Singletons?] If all equivalence classes are
  singletons,
 set $\textit{chirality} := \textit{None}$.
\item[\bf M5.] [Construct closest pairs.]
Represent each circle $C\in \mathcal{P}$ by two antipodal points on
$\sphere^5$, using Plücker coordinates.
Construct the closest-pair graph $H$ on $\mathcal{P}$ with respect
to the distances on $\sphere^5$.
\item[\bf M6.] [Classify edges.]
Partition the edges of $H$ into
$E_L\cup E_R \cup E_N$, representing
pairs of circles that are left pairs, right pairs, 
and not Clifford parallel.

\item[\bf M7.] [Use non-Clifford-parallel edges.]
If $E_N\ne\emptyset$,
let $\mathcal{N}:= E_N$ and go to Step~M12.

\item[\bf M8.] [Find non-Clifford-parallel pairs from equivalent circles.]
If $E_L\ne\emptyset$ and
$\textit{chirality} = \textit{Right}$,
set $$\mathcal{N}_C:= \bigl\{\{C',D\} \mid
\{C,D\}\in E_L,\ \text{$C'$ is closest to $C$ among the 
circles $C'\parallel_+ C$, $C'\ne C$}
\bigr\}$$ for all $C \in \plane$.
Set $\mathcal{N}:=\bigcup_{C \in \plane}
\mathcal{N}_C$ and go to Step~M12.
%% ?? EXPAND THIS DEF: N_C for each C

\item[\bf M9.] (Same as M8, with left and right exchanged.)

\item[\bf M10.] [Merge classes.]
If $E_L\ne\emptyset$ and
$\textit{chirality} \in \{\textit{Left}, \textit{None}\}$,
merge equivalence classes that contain circles that are connected in
$E_L$.
Use Algorithm~K to
condense each resulting class $F$ to a set $F'$ of at most 12
\emph{representative} circles.
%, as described in below.
Set $\mathcal{P}$ to the set of all representatives circles, and put
them into the same equivalence class if and only if they come from the
same set $F'$.
Set $\textit{chirality} := \textit{Left}$, and return to M2.

\item[\bf M11.] (Same as M10, with left and right exchanged. Since all
  possibilities are exhausted, the only remaining possibility in this
  step is to merge classes and return
  to M2.)

\item[\bf M12.] [Mark points on circles.]
(Each pair in
$\mathcal{N}$ is now a non-Clifford-parallel pair of circles.)
For each pair of circles $\{C,D\} \in\mathcal{N}$, mark the two points
on $C$ that are closest to $D$, and likewise, mark two points on $D$.
Return the set $A'$ of all marked points.
\end{itemize}

%The algorithm contains one loop, when returning from M5 or M6 to M2.
%We shall argue that the number of equivalence classes is reduced
The crucial properties for the correctness and % for % bounding
 the
running time are formulated in a %the following
 lemma:
\begin{lemma}\label{mark-lemma}
  \begin{enumerate}
  \item
\label{reduce-lemma}
After the algorithm returns to Step M2 from step M10 or M11,
the number of equivalence classes is reduced at least by half.
\item Algorithm M terminates in $O(|\mathcal{P}| \log |\mathcal{P}|)$ time.
\item 
In Step M12,
 $\mathcal{N}$ is a nonempty set of
at most $25|\mathcal{P}|$
 non-Clifford-parallel pairs.
%the number of pairs in $\mathcal{N}$,
% containing  pairs.
  \end{enumerate}
\end{lemma}
\begin{proof}
1. The only possibility for the algorithm to stall is that
the edges of $H$ are \emph{within} classes, and no merging takes
place in Step M10 or M11. Each class must be one of the five configurations
listed in Lemma~\ref{lem:rep},
and the smallest possible angular distance
 $2\alpha_{\min}\approx 1.10715$
occurs for two adjacent vertices of the
icosahedron, and a volume packing argument on
 $\sphere^5$ yields that there can be at most \PACK-F. circles with this
 distance. See Section~\ref{sec:packing}.
 Then the algorithm exits
in Step~M3.

2.
The most expensive step is the construction of the closest-pair graph
in Step~M5, which takes
$O(|\mathcal{P}| \log |\mathcal{P}|)$ time.
The algorithm contains one loop, when returning from M10 or M11 to M2.
Since the number of equivalence classes is geometrically decreasing
and each class contains at most 12 circles, the overall time is also
bounded by
$O(|\mathcal{P}| \log |\mathcal{P}|)$.

3. When $\mathcal{N}$ is constructed in Step M7,
 there can be at
most 22$|\plane|$ such pairs, because the degree in $H$ is bounded by $K_5\le 44$.
In Step~M8, %\iflong we first observe that \fi
 the constructed set $\mathcal{N}$ is nonempty:
Every pair  $\{C,D\}\in E_L$ produces at least one element of
 $\mathcal{N}$,
 since
all equivalence classes contain at least two elements, by M2 and M4.
When a pair $\{C',D\}$ in $\mathcal{N}$ is formed,
we have a left pair 
 $\{C,D\}$ 
and a right pair 
 $\{C,C'\}$. If
 $\{C',D\}$ were also Clifford parallel, we would get a contradiction to transitivity
 (Proposition~\ref{hopf}.\ref{hopf-transitive}).
Each circle $C\in \mathcal{P}$ gives rise to at most $5\cdot5=25$ pairs, since
a circle can have at most 5
Clifford-parallel neighbors in~$H$ by Proposition~\ref{hopf}.\ref{neighbors-on-hopf}. 
%It can have at most $K_2=5$ closest circles $C'$
%in its equivalence class, since these must also be closest neighbors
%on
%the ``Hopf sphere'' $\sphere^2$.
\end{proof}

We conclude that the algorithm produces at most
$100|\mathcal{P}|$ points, four points for each pair in~$\mathcal{N}$.

%% file: Cycle.tex
\section{2+2 Dimension Reduction: Algorithm T}
\label{sec:torus}

If we arrive at Algorithm T, we are in the very last step 
so we should restore the initial input point sets $A$ and $B$.
This 2+2 dimension reduction is applied when
we have already identified 
at most $C_1 =\PACK-F. $ pairs of planes $P$ and $Q$ from Algorithm M in
Section~\ref{sec:plane}
such that any candidates $R$ of congruence mapping from $A$ to $B$ has to map
 $P$ to $Q$. 
%, see \eqref{c1}
%
%(If we are given
% $\{P_1,P_2\}$ and
% $\{Q_1,Q_2\}$ as unordered pairs, then we could just try
%two possibilities of matching these pairs.)

We begin by choosing a coordinate system
$(x_1,y_1,x_2,y_2)$ for $A$ so that $P$ becomes the
$x_1y_1$-plane, and similarly for $B$ and $Q$.
We then look for rotations $R$ that leave the
$x_1y_1$-plane invariant. Such rotations have the form
$R = 
\left(
\begin{smallmatrix}
R_1&0\\
0&R_2
\end{smallmatrix}
\right)
$
with two orthogonal $2\times 2$ matrices $R_1$ and $R_2$.
Since $\det R=\det R_1\cdot \det R_2=1$, we try two cases:
 (a) $R_1$ and  $R_2$ are planar rotations;
 (b) $R_1$ and  $R_2$ are planar reflections.
We can reduce (b) to (a) by applying the rotation
%\begin{align}
%\label{eq:reflex}
$%R'=
\left(
\begin{smallmatrix}
0&1&0&0\\
1&0&0&0\\
0&0&0&1\\
0&0&1&0
\end{smallmatrix}
\right)
$%\end{align}
~to~$A$.
Thus, it suffices to describe (a), where
 $R$ has the form
$R_{\phi,\psi}$ in~\eqref{eq:rot_mat},
 i.\,e., a combination of 
 a rotation by $\phi$ in the $x_1y_1$-plane and
 an independent rotation by $\psi$ in the $x_2y_2$-plane.
We use polar
coordinates in these two planes by setting
\begin{equation}
\label{eq:torical}
\begin{pmatrix}
x_1 \\y_1 \\x_2 \\y_2
\end{pmatrix}
=
\begin{pmatrix}
r_1\cos \alpha_1\\
r_1\sin \alpha_1\\
r_2\cos \alpha_2\\
r_2\sin \alpha_2
\end{pmatrix}
\text{\qquad with $r_1=\sqrt{x^2+y^2}$ and $r_2=\sqrt{z^2+w^2}$.}
\end{equation}
The rotation $R_{\phi,\psi}$ 
changes
 $(\alpha_1,\alpha_2)$
by adding
 $(\phi,\psi)$
 modulo $2\pi$, and leaves $r_1$ and $r_2$ unchanged.
In other words,  $R_{\phi,\psi}$ acts as a translation
on the torus $\mt=
[0,2\pi)
\times[0,2\pi)$.
We attach the distances $(r_1,r_2)$ to each point
 $(\alpha_1,\alpha_2)\in\mt$ as a \emph{label}.
The problem becomes therefore a \emph{two-dimensional problem} of testing
\emph{congruence under translation} 
for \emph{labeled} points on
a torus.
We denote the two labeled point sets as $\ta$ and
$\tb$, and a point of $\ta$ can only be mapped to
a point of $\tb$ with the same label.

Points in the $x_1y_1$-plane
should be considered separately, because
% present a difficulty, because
$\alpha_2$ is not unique when $r_2=0$,
and the same problem occurs for the points
 in the $x_2y_2$-plane.
We will defer the treatment of these
points to the end of this section, and start with other points first.
% assume for the time being that there are no such points.

% Points in $P$ and points in $Q$
% should be considered separately, because
% $\alpha_1$ is not unique when $r_1=0$,
% so is $\alpha_2$ when $r_2=0$.
% We defer the treatment of these
% points to the end of this section, and start with other points first.

We now give an algorithm for the following problem:
given two labeled point sets 
$\ta$ and $\tb$ on the torus $\mt$, 
test if $\ta$ and $\tb$ are the same up to {translations}. 
We will find a canonical set of $\ta$ (and  $\tb$)
which is similar to a condensed set. In contrast to a condensed set,
we add no new symmetries to a canonical set, by
updating labels to preserve complete information.
Let $\sym(A)$ for a set $A\subset \mt$ denote
translational symmetry group of $A$, i.e., the set of
translations that map $A$ to itself and preserve labels, if $A$
is a labeled set.

\subsection{A Canonical Set Procedure}
\label{sec:cpp}
Roughly, our goal is to find simplest subsets $A'$ and $B'$ of representative points 
that still preserve the same symmetries as $A$ and $B$ respectively.
In addition, we want to construct $A'$ without arbitrary decisions.
If we can find such subsets,
it is enough to compare an arbitrary point in $A'$ with that in $B'$. 
We construct $A'$ 
for a given point set $A$ in some space $\spaces$ and some group
$\symt$ of symmetries of $\spaces$ in two cases:
(i) $\spaces=\mathbb{S}^1$ (the unit circle),
and
$\symt$ are the rotations of $\mathbb{S}^1$.
(ii)
$\spaces=\mathbb{S}^1\times \mathbb{S}^1$ (the flat torus),
and
$\symt$ are the translations on $\spaces$. 

Now, we formally define a \emph{canonical set procedure} and explain
how to use it.
We denote by $\sym_{\symt}(A)$ the symmetry group of
$A$ within $\symt$:
$$\sym_{\symt}(A) = \{\,
R\in\symt \mid R(A)=A \textrm{ and $R$ preserves labels}\,\}.$$

\begin{definition}\label{canonical}
	A \emph{canonical set procedure} $f_{\textrm{cano}}$ for a space
 $\spaces$ and a subgroup $\symt$ of the symmetries of $\spaces$
 maps every finite set $A\subset \spaces$ to a
set $A'\subset \spaces$ such that the following properties hold.
\begin{compactenum}
\item\label{item:symprev} Symmetries are preserved:
 $\sym_{\symt}(A')=\sym_{\symt}(A)$
\item\label{item:symtrans}
 $\sym_{\symt}(A')$ acts
 transitively on $A'$:
For every $p,q\in A'$, there exists $R\in
\sym_{\symt}(A')$ that maps $p$ to $q$.
\item $A'$ is defined in an equivariant way from $A$:
If $RA=B$
for some $R\in \symt$, and $B$ 
 is mapped to $B'$, then
 $B'=RA'$.
\end{compactenum}
We call $A'$ a \emph{canonical set} of $A$.
\end{definition}
It follows from the definition that the canonical set
$A'$ is nonempty
whenever $A$ is nonempty, provided that $\symt$ is an infinite group.
It is easy to see that condition 3 is implied by the
$\supseteq$ part of
 condition 1. Thus, if the procedure is equivariant, we only have
 to prove that $A'$ does not have more symmetries than~$A$.

A canonical set procedure is used to check congruence 
of two sets $A,B\subset \spaces$ under a
congruence from the class $\symt$ as follows:
We compute their canonical point sets $A'$ and $B'$.
We then pick two arbitrary points $p\in A'$ and $q\in B'$, we
find the unique rotation
 $R\in \symt$ that maps $p$ to $q$, and
finally, we simply have to check whether $RA=B$.
In the above two cases of $\spaces$ and $\symt$,
there is always such a unique $R$.

The correctness of this approach is formulated in the following easy lemma.

\begin{lemma} 
	\label{lem:csetproc}
Suppose that $\spaces$ is a space in which, for any two points $p, q
\in \spaces$, there is a unique symmetry $R = R_{pq} \in \symt$ with
$Rp = q$. 
Let $A'$ and $B'$ be canonical sets of $A$ and $B$ for $\spaces$ and 
$\symt$ respectively.  
There exists a congruence in $\symt$ that maps $A$ to $B$
if and only if
for any $p \in A', q \in B'$, we have
$R_{pq}A=B$.
 \end{lemma}

\begin{proof}%[Proof of Lemma~\ref{lem:csetproc}.]
Suppose there is a congruence $R_0 \in \symt$ such that $R_0(A) =B$
and $p \in A'$ and $q \in B'$ are chosen arbitrarily.
Since $A'$ and $B'$ are obtained by a canonical set procedure,
$B= R_0(A)$ is mapped to $B' = R_0(A')$. Let $p'$ be the pre-image of $q$
under $R_0$. Since $A'$ is a canonical point set, there exists 
a unique $R_1 \in \sym_\symt{A'} \subset \sym_\symt{A}$ 
satisfying $R_1(p) =p'$ and $R_1(A)
=A$. By taking $R = R_0 \cdot R_1$, $R$ is the unique rotation
satisfying $R(p) = R_0(R_1(p)) = R_0(p') =q$ and $R(A) = R_0(R_1(A)) =
R_0(A) = B$. Also, $R \in \symt$ since $R_0, R_1 \in \symt$ and $\symt$
is closed under the multiplication.

The other direction is obvious. 
\end{proof}

In the case~(i) that $\spaces=\mathbb{S}^1$,
 the map from a set of points in $\mathbb{S}^1$ to 
a set of \emph{canonical axes} (refer to Section~\ref{sec:cano})
is the obvious canonical set procedure. 
It yields a set of regularly
spaced directions on~$\mathbb{S}^1$.

A canonical set procedure for case~(ii) %exists as well and 
 is presented in Lemma~\ref{lem:cano-torus} with Algorithm T.

\smallskip
\emphi{Algorithm T} (\emph{A canonical set procedure 
from a labeled point set on the torus}).
The input is a labeled point set $\ta$ on the torus $\mt$. We assume
that two labels can be compared in constant time.
The output is an unlabeled \emph{canonical set} $\hat A$ as in
Definition~\ref{canonical}.
%with the following two properties:
%\begin{compactenum}
%\item [a)]
% $\hat A$ has the same set of translational symmetries as the
%  labeled set $\ta$.
% $\sym \hat A=\sym\ta$.
%\item [b)] The group
%$\sym \hat A$ acts transitively  on $\hat A$:
%For any two points $x,y\in \hat A$,
%the translation from $x$ to $y$ leaves $A$ invariant.
% (In other words, all points of $\hat A$ are isogonal.)
%\end{compactenum}
%\item [c)]
In addition, $\hat A$ should be obtained from $\ta$ 
without making any arbitrary decisions.

\begin{itemize}
\labelwidth 32pt
\dimen0=\hsize\advance\dimen0-32pt
\parshape=1 32pt
\dimen0
\item[\bf T1.] [Prune by labels.]
Choose the label that occurs least frequently in $\ta$, and
% Choose some label that occurs in $\ta$, and
 let $A'$ be the set of points with this label.
(For a while we will now do ordinary pruning,
using only the geometry of the point
set $A'$.)
%(We do ordinary pruning,
%using only the geometry of the point
%set $A'$, until we eventually update labeling.)

\dimen0=\hsize\advance\dimen0-40pt
\parshape=1 40pt
\dimen0
\advance\labelsep8pt

\item[\bf T2.] [Compute the Voronoi diagram.]
 Compute the Voronoi diagram of $A'$ on the torus~$\mt$. 
This can be reduced to a Voronoi diagram in the plane by
replicating the square region representing the torus 
together with the
set $A'$ 9 times in a $3\times 3$ tiling, see for example~\cite{DH}.
Clipping the result to the central tile yields the Voronoi diagram
on the torus in $O(|A'| \log |A'|)$ total time.

\item[\bf T3.] [Prune by shape.]
Translate each point $a\in A'$ to the origin together with its
Voronoi cell.
If the translated cells are not all equal,
%
%Classify the cells by their shape when their defining site is
%translated to the origin. % up to translations.
%If there are different shapes,
%choose the cell shape
%that occurs least frequently, 
%and
 replace $A'$ by the subset of points whose cell shape occurs least
 frequently,
and return to T2.

\dimen0=\hsize\advance\dimen0-32pt
\parshape=1 32pt
\dimen0
\advance\labelsep-8pt

\item[\bf T4.] % [Labeling.] 
[Restore information from the original set $\ta$
 by % attaching
 labeling the points in $A'$.]
 (Now all Voronoi cells are translated
  copies of the same hexagon or rectangle.)
 Assign each point of $\ta$ to its Voronoi cell.  A point on the
boundary is assigned to all incident cells.  Now
for each point % Voronoi site
 $x\in  A'$,  collect the points in its cell
and translate them so that $x$ lies at the origin. Represent each
point as a triple ($\phi$-coordinate,\,$\psi$-coordinate,\,label).
 Concatenate these triples in
 lexicographic order into
a string of numbers that represents
the cell contents, and
 attach the string % of each cell
as a label to the point~$x$. % in $A'$. 
(This string representation is obtained
equivariantly, since two points $x,y\in
A'$ get the same string if and only if the two Voronoi cells
are exact translated copies of each other, including all points in
the cells with their original
labels.
We have thus preserved 
complete information about
the symmetries of $\ta$.)

\item[\bf T5.] [Compress labels.]
Sort %\iflong 
the %label
%\fi
 strings and replace each label with its
rank in sorted order.

\item[\bf T6.] [Finalize.]
If there are at least two labels, return to T1.
Otherwise, 
 return % the set of points
 $A'$ as the {canonical set}
$\hat A$.
\end{itemize}

\begin{lemma}
\label{lem:cano-torus}
Algorithm T computes a canonical set $\hat A$ of a labeled set $\ta$ on the torus in time $O(|\ta|
\log |\ta|)$.
\end{lemma}

\begin{proof}

We first check that $\hat A$ has the claimed properties.
Property~\ref{canonical}.\ref{item:symprev} consists of two inclusions:
We have 
 $\sym (\ta)\subseteq\sym(\hat A)$,
 because $\hat A$ is obtained in a
equivariant way from~$\ta$. None of the operations 
 which are applied to $\ta$ to obtain $\hat A$ destroys any
 translational symmetries.
The other inclusion
 $\sym (\hat A)\subseteq\sym(\ta)$
 is ensured by Step T4. (This would work for any set $A'$.)
%\iflong

%\else
\looseness-1
%\fi
%The symmetries $\sym (\hat A)$ act transitively on $\hat A$ because
% 
To see Property~\ref{canonical}.\ref{item:symtrans}, note that
 the Voronoi cell of a point
 fixes the relative positions of its neighbors.
Starting from any point  $a\in \hat A$
we can reconstruct
 the whole set $\hat A$ if we know the shape of each Voronoi cell.
Step T3 ensures that all Voronoi cells are equal,
and therefore the reconstructed set $\hat A-a$ is the same, no matter from which point $a$ we
start.
%\iflong
The set $\hat A$ forms a lattice on the torus.
%\fi
% not completely happy with this derivation.

Let us analyze the running time. 
Each iteration of the loop T2--T3
takes $O(|A'| \log |A'|)$ time and reduces the size of $A'$ to half
or less. Thus, the total running time of this loop is
 $O(|\ta|\log|\ta|)$, since the initial size of $A'$ is bounded by
 the size of $\ta$.

Steps T4 and T5 involve point location in Voronoi diagrams~\cite{PS} and sorting
of strings of numbers of total length
 $O(|\ta|)$. These operations can be carried out in $O(|\ta|\log|\ta|)$ time.
Thus, each iteration of the whole loop T1--T6
takes $O(|\ta| \log |\ta|)$ time. Step T1 reduces the size of $\ta$ to
a half
or less after each iteration. Thus, the total running time
is $O(|\ta| \log |\ta|)$.
\end{proof}

We are ready for the main result of this section.
\begin{theorem}
\label{thm:ortho}
Given two sets $A,B$ of  $n$ points in $\mathbb{R}^4$, and two
 planes $P$ and $Q$,
 it can be checked in
 $O(n \log n)$ time if there is a congruence that maps $A$ to $B$, and
 $P$ to
 $Q$.
\end{theorem}

\begin{proof}
  We have already seen how to convert input points to points
 $(\alpha_1,\alpha_2)\in \mt$ with labels $(d_1,d_2)$ so that
 congruence testing is reduced to testing congruence by translation on
 the torus.

We still have to deal with
points on coordinate planes.
Let $A_1\subseteq A$ be the points in the $x_1y_1$-plane,
and let $A_2\subseteq A$ be the points in the $x_2y_2$-plane.
%These points are subject only to the $\phi$-component
%of the rotation.
If $A_1\ne\emptyset$, we compute the canonical axes of $A_1$,
as described in Section~\ref{sec:cano}.
They form $k\ge1$ equally spaced angular directions
$\bar\alpha+j\frac{2\pi}k$
modulo $2\pi$, $j\in \mathbb{Z}$.
%We compute $k$ canonical axis directions
%$\beta_1+j\frac{2\pi}k$
% for the corresponding set $B_1$.
 We know that $R$ must map the canonical axes of $A_1$
to the canonical axes for the corresponding set $B_1$.
We incorporate this restriction into
an additional label for each point $u\in A\setminus A_1\setminus A_2$.
If $u$ has a polar angle $\alpha_1$ in the $x_1y_1$-plane, we attach to it
the difference to the nearest smaller canonical angle:
\begin{equation*}
\min \{\, \alpha_1-(\bar\alpha+j\tfrac{2\pi}k)
\mid j\in \mathbb{Z},\ \alpha_1-(\bar\alpha+j\tfrac{2\pi}k)\ge 0\,\}
\end{equation*}
%\iflong
We also have to test if
$A_1$ and $B_1$ are congruent by simply overlaying the canonical axes
and testing if they are equal.
%\else
%We also have to test if $A_1$ and $B_1$ are congruent.
%\fi 

If $A_2\ne\emptyset$, we treat this in the same way and attach an additional
label to the points in $A\setminus A_1\setminus A_2$.

After this, we can just compute the canonical set from Lemma~\ref
{lem:cano-torus}
 and thereby reduce the congruence test to an equality
 test between sets $\hat A$ and $\hat B$, as described before in Lemma~\ref{lem:csetproc}.
If there are no points outside the $xy$-plane and the
$zw$-plane, the problem reduces to two independent congruence tests
in these planes.
The whole algorithm takes $O(n \log n)$ time.
\end{proof}
	
There is an alternative algorithm with the same time complexity
that does not use Lemma~\ref{lem:csetproc}.

\begin{proof}[Alternative Proof of Theorem~\ref{thm:ortho}.]
We can also use Algorithm T for comparing two labeled sets $\ta$ and
$\tb$ on the torus as follows. We run Algorithm~T in parallel
for $\ta$ and $\tb$.  We must make sure that all steps run
identically.
% corresponding labels are the same
In particular, the sorted strings in Step T5
% and the chosen label in Step T1 are identical.
must be the same
for $\ta$ and $\tb$.
If the algorithm runs to the end without finding a difference
between $\ta$ and $\tb$, and if the Voronoi cells of the canonical
sets $\hat A$ and $\hat B$ are equal, we know that
 $\ta$ and $\tb$ are congruent by translation.
There are  $|\hat A|=|\hat B|$ translations that map
 $\ta$ to $\ta$.
Points on coordinate planes can be taken care of by the same way as
the previous proof, and the overall algorithm also takes $O(n \log n)$
time.
\end{proof}

%% file: Conclusion.tex
\section{Concluding Remarks}
\label{chap:conclusion}

We can substitute our algorithm for 4 dimensions as the 
base case in the recursive dimension-reduction algorithms of
Akutsu~\cite{Aku} and
Brass and Knauer~\cite{BK}.
The deterministic algorithm of
Brass and Knauer runs now in
$O(n^{\lceil (d-1)/3 \rceil} \log n)$ time,
and the randomized Monte Carlos algorithm of
Akutsu
 takes time $O(n^{\lfloor (d-1)/2\rfloor /2} \log n)$
for $d \geq 9$ and $O(n^{3/2} \log n)$ for $d\le8$.

It is likely that our algorithm can be simplified.
The constants in the bounds are certainly larger 
than the actual bounds.

\subsection{Practical Implementability}
%There are packages, such as iRRAM for C++, that implement Real-RAM
%models. The four-dimensional algorithm should be implementable in this
%setting.
Instead of using the Real-RAM model,
it also makes sense to  
test congruence with an error
tolerance $\eps$, but this problem
is known to be NP-hard even in the plane~\cite{die,iwa} as mentioned
earlier in Section~\ref{sec:model}.
However, the problem becomes
polynomial if the input points are sufficiently separated compared to
$\eps$.
%, and the NP-hardness results do not apply. % in this case.
  We are confident that our
algorithm, when it is implemented with standard floating-point arithmetic
and with appropriate error margins to shield equality tests from
round-off errors, would decide approximate congruence in the
following weak sense. Given a tolerance $\eps$ that is large compared
to the machine precision, and two sets $A$ and $B$ whose points are
separated by more than, say $10\eps$, the algorithm would correctly
distinguish the instances that are congruent with a tolerance of $\eps$
from the instances that are not even congruent with a tolerance of, say
$100\eps$.  Between $\eps$ and $100\eps$, the algorithm is allowed to
make errors.  Such a result will require a thorough analysis of
the numerical stability of the operations  in our
algorithm.
%\iflong\else\looseness-1\fi

\subsection{Regularity and Related Open Questions}

\paragraph{Does Local regularity imply global regularity?}
The general theme in our four-dimensional algorithms arises the
question whether \emph{local} regularity
implies \emph{global} regularity. 
In other words,
when the neighborhoods of all the points look the same, does it imply
that
a symmetry group acts transitively on the point set or not, if the
neighborhoods are sufficiently large?
It would be interesting to prove this statement quantitatively.

\paragraph{Geometric classification of 4-dimensional point groups.} 
A \emph{point group} is a discrete subgroup $O(d)$, or in other words,
a finite group of orthogonal $d\times d$ matrices.
Hessel's Theorem~\cite{Mar} gives an explicit classification of
the point groups for $d=3$.
A somewhat implicit formulation of Hessel's theorem is stated as follows.
\begin{theorem}
	A three-dimensional point group is one of the following:
	\begin{compactenum}[(a)]
		\item the symmetry group of one of the five three-dimensional platonic solids,
		\item the symmetry group of a prism over a regular
			polygon,
		\item or a subgroup of one of above
			groups.
		%\item The groups $CT, C_{2n}C_n, D_{2n}D_n, D_nC_n$
		%	where $C$ is the symmetry group of cubes and
		%	$T$ is that of tetrahedrons,
		%	$C_n$ and $D_n$ are the cyclic groups and the
		%	dihedral groups of order $n$ and $GH$ is $H
		%	\cup (G-H) \cdot i$ where $i$ is an inversion.
	\end{compactenum}
\end{theorem}
This theorem can be derived without much effort from
Lemma~\ref{lem:rep} (Algorithm~K).

The four-dimensional point groups were already
enumerated~\cite{TS,CS}.
The question is whether the four-dimensional
point group can be enumerated with geometrically interpretable
characteristics.
The following is a conjecture by Rote~\cite{Ro}.
\begin{conjecture}
	A four-dimensional point group is either
	\begin{compactenum}[(a)]
	\item the symmetry group of one of the five four-dimensional
	regular polytope,
	\item a direct product of lower-dimensional points groups,
	\item or a subgroup of one of the above groups.
\end{compactenum}
\end{conjecture}
Since the list of four-dimensional point groups is known, it should in
principle be easy to settle this conjecture. However, it would be nice
to established it directly, without resorting to
the previous classification.
Felix Klein, in his book on non-Euclidean geometry from 
1928, writes that the determination of
the four-dimensional point groups should be ``easy''
\cite[pp.~240--241]{klein28}\footnote
{%\selectlanguage{german}
\emph{Die Kugeldrehungen der euklidischen Geometrie lassen sich
  eineindeutig den wesentlich voneinander verschiedenen Quaternionen
  sowie den Schiebungen jeder der beiden Arten so zuordnen, dass die
  zugehörigen Gruppen isomorph sind.} [\dots]\ Die Kenntnis dieser
Isomorphie ist sehr nützlich; denn mit ihrer Hilfe ist es z.~B.\
leicht, \emph{die endlichen}, d.~h.\ die nur aus endliche vielen
Bewegungen bestehenden \emph{Bewegungsgruppen} der räumlichen
ellipti\-schen Geometrie zu bestimmen. [Emphasis in the original]}.

\paragraph{Tilings on a sphere.} Another interesting question is the classification of
regular and semiregular tilings on
spheres. It is possible that our techniques can shed light on
symmetries on the 3-sphere,
%for
%$ example the classification of finite subgroups of the orthogonal
% group $O(4)$ of $4\times4$ orthogonal matrices, 
or on regular and
 semiregular tilings of~$\sa$. %\footnote{References?}